\documentclass[%
 	aip,
 	amsmath,amssymb,
preprint,%
preprintnumbers, 
]{revtex4-1}
\usepackage{graphicx}
\usepackage{dcolumn}
\usepackage{bm}

\usepackage[utf8]{inputenc}
\usepackage[T1]{fontenc}
\usepackage{mathptmx}
\usepackage{etoolbox}

\makeatletter
\def\@email#1#2{%
 \endgroup
 \patchcmd{\titleblock@produce}
  {\frontmatter@RRAPformat}
  {\frontmatter@RRAPformat{\produce@RRAP{*#1\href{mailto:#2}{#2}}}\frontmatter@RRAPformat}
  {}{}
}%
\makeatother

\usepackage{amsfonts}
\usepackage{graphicx}
\usepackage[table]{xcolor} 
\usepackage{setspace}
\usepackage[hidelinks]{hyperref} 
\usepackage{mathtools}
\usepackage{commath}
\usepackage{gensymb}
\usepackage[thinc]{esdiff}
\usepackage[title]{appendix}
\usepackage{numprint}
\usepackage{diagbox}
\usepackage{makecell}

\npthousandsep{,}\npthousandthpartsep{}\npdecimalsign{.}

\usepackage[export]{adjustbox}

\graphicspath {{figures/}}

\DeclareMathOperator{\sech}{sech}

\newcommand{\cartCoeff}{\mathrm{F}}
\newcommand{\scalarPhi}{\Phi}
\newcommand{\cylAzi}{\phi}

\begin{document}
\preprint{FERMILAB-PUB-26-0539-PPD-TD}

\title{Precise Modeling of a Complex Solenoidal Magnetic Field Using a Combination of Analytic Functions and a PINN}
\author{Cole Kampa}
\email{kampa@caltech.edu}
\affiliation{ 
Department of Physics, Mathematics and Astronomy, California Institute of Technology, Pasadena, CA, 91125, USA
}
\author{Susan Dittmer}
\affiliation{ 
Department of Physics and Astronomy, Northwestern University, Evanston, IL, 60208, USA
}

\author{Henry Glass}
\affiliation{
Fermi National Accelerator Laboratory, Batavia, IL, 60510, USA
}

\author{Michael Schmitt}%
\affiliation{ 
Department of Physics and Astronomy, Northwestern University, Evanston, IL, 60208, USA
}

\date{21 August 2026} 

\begin{abstract}    
We demonstrate an iterative approach to modeling a sparsely measured magnetic field in a large-bore solenoid. This approach uses a hybrid of traditional and machine learning techniques. The traditional technique is a linear least-squares fit using a series solution to Laplace's equation, while the machine learning technique involves the training of a physics-informed neural network (PINN) on the least-squares fit residuals. We use a newly defined activation function ``DELTAsnake,'' a modification to the snake activation function proposed by Ziyin et al. that allows for stronger curvature and non-monotonicity. The combined model approximately obeys Maxwell's equations to a level sufficient for producing high quality physics simulations and analysis. Our approach is applied to a highly realistic calculation of the expected magnetic field in the Mu2e experiment's Detector Solenoid which includes a simple model for the expected statistical measurement uncertainties. Using ten toy measurement simulations, we demonstrate the capabilities of our model in comparison to the least-squares method alone; the least-squares method alone results in a reduced chi-squared statistic of ${2.15 \pm 0.01}$, while our approach improves the reduced chi-square to ${1.034 \pm 0.005}$. Furthermore, for an average toy simulation, we show that the range of the RMS of the three field component residuals reduces from ${0.07-0.37}$~Gauss to ${0.05-0.07}$~Gauss.
We find that this novel method is robust against a realistic systematic uncertainty deriving from Hall probe calibration bias and can be used to significantly reduce the number of measurements required to achieve an accurate model.
\end{abstract}

\maketitle

\section{Introduction}
\label{sec:introduction}

Magnetic fields play a central role in a wide range of scientific and industrial applications~\cite{muon_g2_2025,ongena_fusion,grover_MRI}. In particle physics experiments like Mu2e~\cite{mu2e_sensitivity2023}, accurate knowledge of the field is essential for reconstructing charged particle trajectories and measuring momenta of charged particles. Errors in the field model directly propagate to the momentum measurement and can degrade the sensitivity of the experiment.

To achieve a high quality model of the field, experimentalists often measure the field directly and create a data-driven model for the field throughout the experimental volume. A common approach is to fit an analytic model -- typically a truncated series expansion that satisfies Maxwell’s equations -- to a sparse set of measurements~\cite{CMS_BField_2008,ATLAS_BField_2008,Frank_2012}. This method has the advantage of being fast, interpretable, and free of assumptions about the underlying conductor geometries. Experience has shown, however, that such models can struggle to capture subtle, complex, or asymmetric features. In recent particle physics experiments, for example, this technique has been used to model magnetic fields with a $10^{-3}$ relative accuracy, whereas Mu2e requires a field modeling accuracy of $10^{-4}$ (see Ref.~\cite{FMS2018}). The limitations in the analytic method are due primarily to basis function selection. One can often choose a convenient coordinate system that results in basis functions that respect some symmetry, e.g. cylindrical symmetry in the case of Mu2e. These symmetries are typically only approximate in the magnetic environment, which results in some features that the basis functions cannot easily model. While the complete (infinite) set of basis functions would be able to model said features, in practice truncating to a finite set of terms limits this capability; such a truncation is required due to the finite set of measurement positions.

An alternative approach is to use flexible machine learning models, such as neural networks (NNs), trained directly on the measured values. In addition to being more flexible than the analytical functions, NNs do not rely on any assumptions about the conductor geometries. When physical constraints are included, these models can learn complex structure without requiring large numbers of hand-engineered basis terms that rely on approximate symmetries and a reasonable selection of boundary conditions. Physics-Informed Neural Networks (PINNs)~\cite{PINN2019}, in particular, allow soft enforcement of physical laws such as Maxwell’s equations via an extension to the loss function. We find that a PINN alone is not sufficient for modeling the field of the Mu2e Detector Solenoid (DS) at a level required for producing high quality physics results. Instead, the best magnetic magnetic field model is obtained when applying a PINN to capture sub-dominant features that remain after fitting with the analytic model.

In this work, we present a hybrid modeling strategy that combines these two approaches. We begin by fitting a cylindrical harmonic least-squares model to the measured field. Then, we train a PINN to learn the residual field components that the analytic model fails to capture. The PINN is designed to use both hard and soft constraints from Maxwell’s equations to ensure physical consistency. Finally, we iterate the least-squares fit on the PINN-subtracted data to recover any remaining structure that the least-squares model struggled to fit in the first iteration. In this manner, we are able to attain our accuracy goals.

Our approach is tested using a high-fidelity simulation of the Mu2e Detector Solenoid (DS) magnetic field, including detailed representations of both active and passive magnetic elements. The simulation is used to generate realistic test data; it allows us to benchmark the model performance under controlled conditions. We demonstrate that the hybrid model substantially improves reconstruction quality, generalizes well between the points of the measurement grid, and is robust against noise and realistic systematic effects. We also explore the ability of the PINN to recover key field features even when trained on a dramatically reduced measurement set, highlighting its potential in scenarios where mapping time or access is limited.

The remainder of this paper is organized as follows. In Section~\ref{sec:methods}, we briefly describe the Mu2e solenoid system and outline the hybrid modeling approach. The modeling description includes the formulation of both the least-squares fit and the PINN. In Section~\ref{sec:results}, we present results from the nominal field scenario as well as a few robustness tests, including a simulated Hall probe calibration systematic error and a case with dramatically reduced measurement density. We conclude in Section~\ref{sec:summary} with a summary of the results and their implications for future field modeling efforts.

\section{Methods}
\label{sec:methods}

We apply a hybrid modeling approach comprised of linear least-squares fitting and a novel PINN scheme. This approach is applied to a realistic calculation of the Mu2e DS magnetic field.

\subsection{Magnetic Field Calculation and Data Preparation}
\label{subsec:methods_fieldcalc}

The Mu2e experiment~\cite{Mu2eBernstein} at Fermilab will search for the Charged Lepton Flavor Violating (CLFV) process muon to electron conversion, which is forbidden in the Standard Model of particle physics. The experiment comprises a proton beam and a set of three superconducting solenoid systems to produce and manipulate a muon beam. The muons are stopped on an aluminum stopping target, and the detector systems measure electrons emitted from the target to identify the signal process.

The Mu2e magnetic environment is complex. Three large superconducting solenoid systems, the Production Solenoid (PS), Transport Solenoid (TS) and Detector Solenoid (DS), establish the magnetic field throughout the experiment. In the PS, the field captures decay products from proton-nucleon interactions and directs them to the TS. The TS field is toroidal which allows for efficient charge and momentum collimation and transport to the DS. By the time the beam arrives at the DS it is comprised primarily of low-momentum ($p \lesssim 80$~MeV/c) negative muons,
which then stop in the aluminum target and yield the Mu2e initial state, a muon bound to an aluminum nucleus.
Under the signal process, neutrinoless muon-to-electron conversion in the field of a nucleus, an electron with $p \approx 105$~MeV/c is emitted from the target. The DS field, which has a negative axial gradient near the TS-DS interface, guides this electron toward the downstream detector systems. To accurately measure the momentum and energy of the electron using the detectors, the magnetic field in the downstream DS region is designed to be as uniform as possible (approximately 1$\%$). The Field Mapping System (FMS) apparatus, the Detector Solenoid Field Mapper (DSFM), sparsely samples the DS magnetic field using Hall probes~\cite{FMS2018}. Modeling procedures are applied to these measurements to create a continuous model for the field.

A Cartesian coordinate system is defined in Mu2e with an origin in the center of the TS in $y$ and $z$ and $x=0$ defined along the DS axis. The ordinal directions are ${(\hat{i}, \hat{j}, \hat{k})}$: $+\hat{j}$ points from the floor to the ceiling, $+\hat{k}$ is parallel to the DS solenoidal axis and points from the upstream to downstream direction of the beam in the DS, and $+\hat{i}$ is defined to complete the right-handed system.

In the DS, where we apply our modeling technique, the coils have radii of approximately $1$~m and span an axial length of approximately $10$~m. We map a region of the bore that encompasses the sensitive tracking volume which extends to a radius of $0.8$~m. This results in a large mapping volume of approximately $20$~m$^3$.

We use a combination of custom and commercial software packages to calculate the expected magnetic field in the Mu2e DS. The \texttt{helicalc} package, described in the supplementary material, is used to calculate active contributions to the magnetic field due to the superconducting network in the PS, TS, and DS. The commercial three-dimensional Finite Element Analysis (FEA) magnetostatics simulation software suite Opera from Dassault Syst\`{e}mes is used to calculate the contribution from the passive magnetic materials~\cite{opera}.

\begin{figure*}
\centering
\includegraphics[trim=2cm 4cm 0.0cm 0.0cm, width=0.8\textwidth]{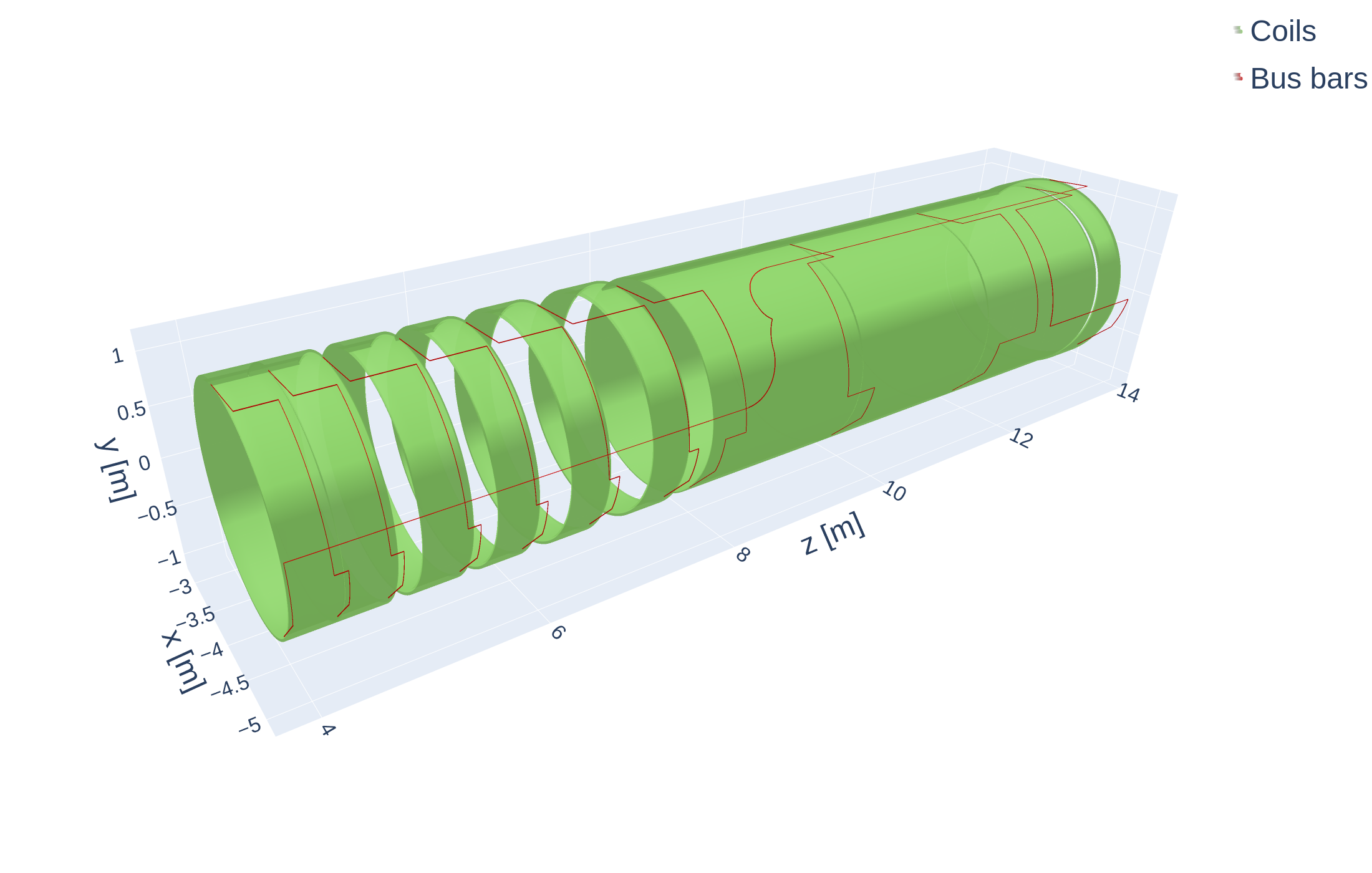}
\caption{A 3D representation of the DS coils and bus bars. While the visualization represents the coils as simple cylindrical shells \texttt{helicalc} represents the coils with helical windings in the field calculation.}
\label{fig:DS_3D}
\end{figure*}

The superconductors in each solenoid system are treated in different ways, given proximity to the region of interest, the DS bore. The PS and TS coils are represented as ideal solenoids, i.e. thick cylindrical shells of current, following the prescription of the SolCalc MATLAB routine \cite{SOLCALC2018}. On the other hand, a much more detailed geometric representation of the DS coils is used. Each coil is represented as one or more helical windings of thick superconducting cable with a rectangular cross-section. Coils with multiple layers are wound with opposite helicity in subsequent layers, and layers are spliced together with a thick circular arc of superconductor. Finally, a complex network of superconducting bus bars connect the DS coils in series. The bus bars are composed of thick circular arcs and thick straight bars. A 3D visualization of the DS conductor network is shown in Figure~\ref{fig:DS_3D}. All of these superconductor field contributions are computed in \texttt{helicalc}.

\begin{figure*}
\centering
\includegraphics[trim=0cm 0cm 0.0cm 0.0cm, width=0.8\textwidth]{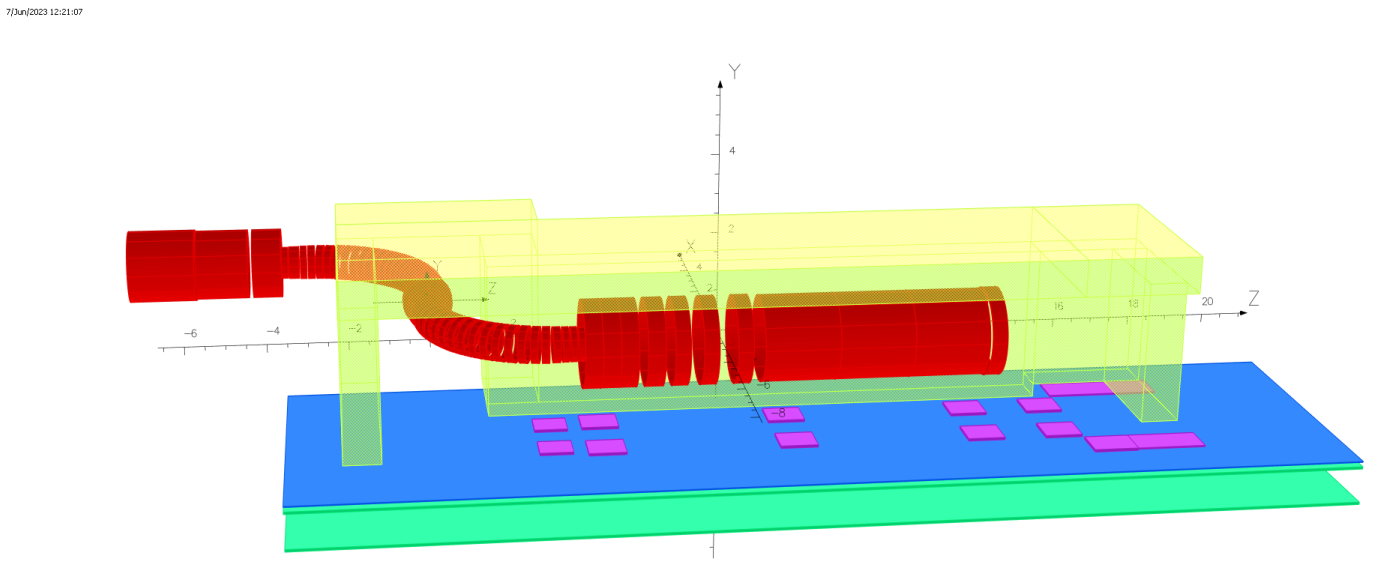}
\caption{A 3D representation of the Mu2e coils and passive magnetic materials included in the Opera calculation. The coils (red) will be welded via a support structure to the steel floor plates (magenta), which are embedded in the concrete floor. The floor contains several layers of rebar (green and blue), and the DES blocks (translucent yellow) contain rebar. The $-x$ wall of the concrete shielding blocks has been removed only in the visualization, for clarity of the coils and floor plates.}
\label{fig:rebar_coils}
\end{figure*}

The magnetic material contributions to the magnetic field are calculated using the Opera magnetostatics package. A set of ideal solenoids represents the PS, TS, and DS coils. This representation is practically identical to the representation used in \texttt{helicalc} as confirmed through validation studies. In addition to the coils, there are several sources of magnetic material: reinforcement bar (rebar) in the concrete shielding blocks, rebar in the floor, and steel plates embedded in the floor.

To ease the computational burden of including passive magnetic materials in Opera, we use simple geometric blocks with a homogeneous mixture of steel and concrete to approximate the magnetic properties of concrete with rebar. In reality a complex grid of cylindrical steel rods is embedded in the concrete. The homogeneous model accounts for the relative density of rebar material as well as the efficiency of flux transport given the expected layout of the cylindrical rods. A full technical description of the material modeling is provided in the supplementary material.

The Opera calculation is completed twice, once with all of the magnetic material included ($\vec{B}_{\mathrm{w}}$) and once without any magnetic material contributions ($\vec{B}_{\mathrm{w/o}}$). Figure~\ref{fig:rebar_coils} visualizes the Opera model that includes the passive magnetic materials.

We use \texttt{helicalc} for the main contributions and the difference ${\vec{B}_{\mathrm{w}} - \vec{B}_{\mathrm{w/o}}}$ from OPERA for the impact of the passive magnetic materials. The final field calculation at each field point is the sum of the field contributions from each superconducting element $i$ as calculated by \texttt{helicalc} along with the field contributions from the passive magnetic materials as calculated by Opera:
\begin{align}
\vec{B}_{\mathrm{true}}(x, y, z) &= \bigg( \sum_i \vec{B}_i(x, y, z) \bigg)_{\mathtt{helicalc}} \notag \\
&+ \bigg( \vec{B}_{\mathrm{w}}(x,y,z) - \vec{B}_{\mathrm{w/o}}(x, y, z) \bigg)_{\mathrm{Opera}}
\end{align}

\begin{figure}
\centering
\includegraphics[width=0.45\textwidth]{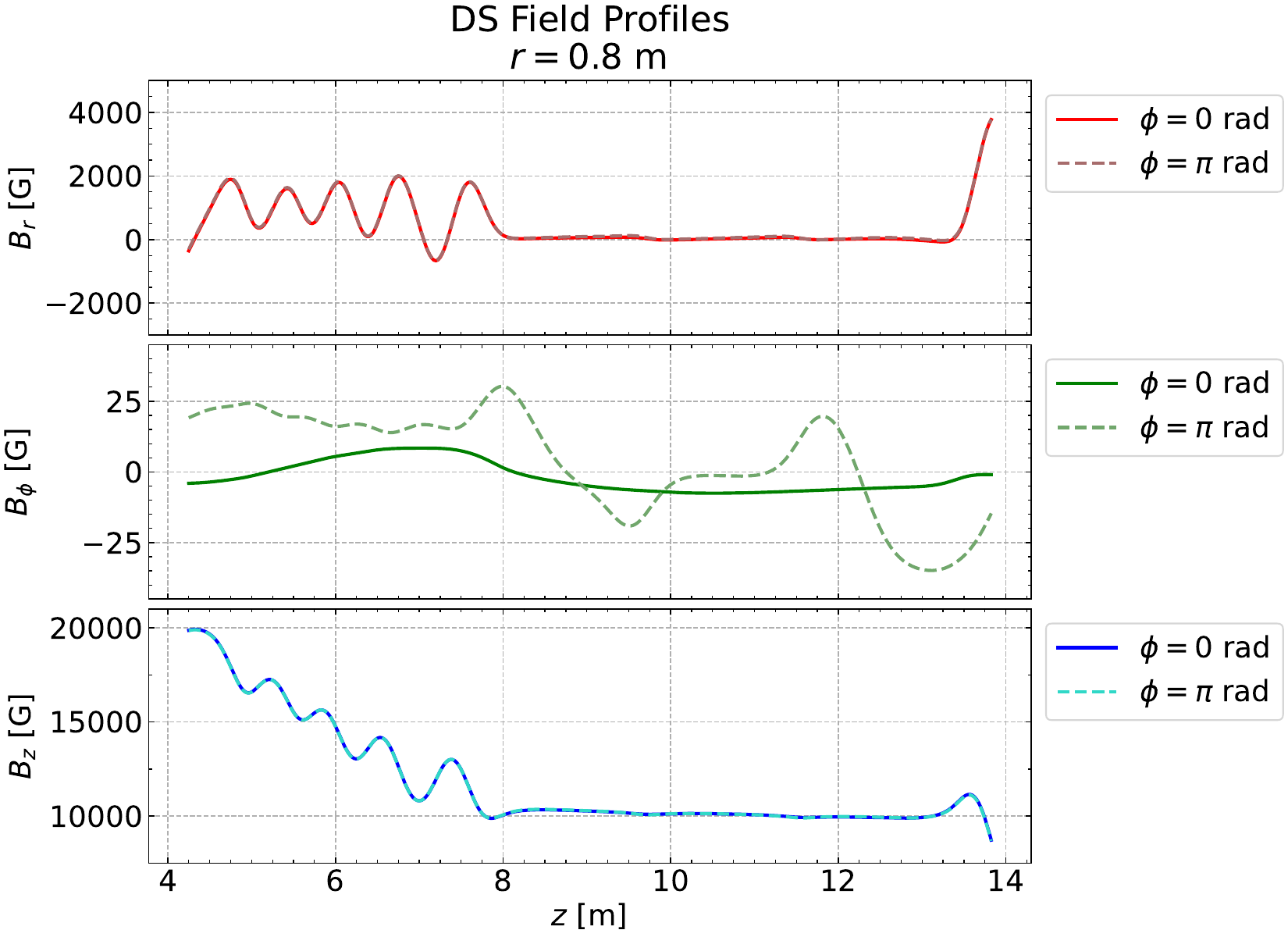}
\caption{Axial field profiles in the Mu2e DS. The top panel shows $B_z$ vs. $z$. The middle panel shows $B_r$ vs. $z$. The bottom panel shows $B_\phi$ vs. $z$. In each plot, the solid line denotes the $\phi=0$~rad profile while the dashed line denotes the $\phi=\pi$~rad profile.}
\label{fig:field_profiles}
\end{figure}

As a simple demonstration of the complexity of the magnetic field in the DS, Figure~\ref{fig:field_profiles} shows 1D profiles of the three field components along the axial direction ($z$) at a radial distance of $0.8$~m from the solenoid axis. Along with being complex, the magnitudes of the field components vary widely.

For an evaluation of our methods we prepare two field point samples. The first sample represents the expected measurement values of the DS field and comprises a sparse cylindrical grid of approximately ${N_{\mathrm{meas}}=2.4 \times 10^4}$ field points throughout the DS volume ($r \leq 0.8$~m, $z \in [4.25, 13.85]$~m). This sample is denoted as the \textit{measurement sample}. The grid of points in the measurement sample is defined as the expected sampling locations of the DSFM, which will measure the DS field with eight 3D Hall probes spaced radially along the DSFM propellers~\cite{FMS2018}. The propellers are rotated azimuthally in $\phi$ ($\Delta \phi = \pi/8$~rad) and translated axially in $z$ ($\Delta z = 50$~mm) to create the sparse grid of measurements.
Each Hall probe contains three orthogonally mounted GaAs Hall elements which display a characteristic thermal noise with a standard deviation of approximately $0.3$~Gauss. This thermal noise is modeled and used to represent a statistical measurement error applied to each field component; a multivariate Gaussian distribution ($\mathcal{N}_3$) with mean vector of zeros and variances of $(0.3$~Gauss$)^2$ is sampled and summed with each ideal field point. The thermal noise in each field component is uncorrelated. The field in the measurement sample is given by:
\begin{equation}
\vec{B}_{\mathrm{meas}}(x, y, z) = \vec{B}_{\mathrm{true}}(x,y,z) + \mathcal{N}_3(\vec{\mu}, \bm{\Sigma})
\end{equation}
where $\vec{\mu} = (0, 0, 0)$ and $\bm{\Sigma} = (0.3 \mathrm{G})^2 \times \mathbb{I}_3$. For the nominal configuration we generate ten toy measurements by resampling the thermal noise distribution.

The second field point sample is used strictly for evaluating the modeling results and is denoted as the \textit{test sample}. It is defined on a 3D Cartesian grid with a step size of $\Delta x = \Delta y = \Delta z = 25$~mm such that the volume of the DS bore is uniformly sampled. A radial cut of \mbox{$r \leq 0.8$~m} is applied to the test sample to match the extent of the measurement sample. The test sample contains approximately \mbox{$N_{\mathrm{test}} = 1.2\times 10^6$~points}. Measurement error is not included in the test sample. The model fitting and training procedures are performed on the measurement sample alone, while the evaluation of the results are performed on the test sample. The field evaluated at the test sample points is $\vec{B}_{\mathrm{true}}$. In other words, comparing the test sample to the prediction of the fitted model evaluates the modeling accuracy relative to the underlying (or ``true'') magnetic field. In contrast, comparing the measurement sample to the model prediction characterizes the residuals expected in a real measurement campaign, incorporating the effects of measurement noise, potential systematic effects, and the spatial sampling density used during fitting and training.

\subsection{Least-Squares Approach}
\label{subsec:methods_leastsquares}

To model the magnetic field throughout the Mu2e DS volume from a sparse set of measured points, we begin with an analytic approach. We construct functional forms that satisfy Maxwell’s equations in a source-free region and fit their free parameters to the measured points using linear least-squares minimization. This method is widely used in particle physics~\cite{CMS_BField_2008,ATLAS_BField_2008,Frank_2012} and has the advantage of requiring no detailed knowledge of the conductor geometry. In the case of the Mu2e DS, more than $\numprint{140}$ conductors contribute to the field, and their precise positions after construction and cooldown are not easily determined. The analytic method bypasses this complexity, but at the cost of requiring many basis functions to model detailed features of the field.

To reduce this burden, we design basis functions that reflect the approximate cylindrical symmetry of the Mu2e DS. The set of basis functions includes constant and linear functions of the Cartesian coordinates.

The specific functional forms used are presented in the remainder of this section. They build upon prior work~\cite{pollack2020}, with small extensions to broaden the expressiveness of the model.

A static magnetic field in a source-free region obeys Maxwell's equations:
\begin{equation}\label{eq:div_B_LSQ}
\nabla \cdot \vec{B} = 0
\end{equation}
and
\begin{equation}\label{eq:curl_B_LSQ}
\nabla \times \vec{B} = 0\ .
\end{equation}
With these constraints, the range of allowed functional forms that can be used to describe $\vec{B}$ is restricted. We choose to define a scalar magnetic potential $\scalarPhi$ as
\begin{equation}\label{eq:B_defn_Phi_LSQ}
\vec{B} = - \nabla \scalarPhi
\end{equation}
which obeys Equation~\ref{eq:curl_B_LSQ} by construction. Equation~\ref{eq:div_B_LSQ} can be recast for $\scalarPhi$:
\begin{equation}\label{eq:laplace_LSQ}
\nabla^2 \scalarPhi = 0\ .
\end{equation}
In other words, any scalar function $\scalarPhi$ that has a Laplacian of zero is allowed by Equations~\ref{eq:div_B_LSQ}~and~\ref{eq:curl_B_LSQ}. With a set of $M$ basis functions $\scalarPhi_i$, a series solution to Laplace's equation is constructed as a sum of basis functions:
\begin{equation}
\scalarPhi_{\mathrm{LSQ}} = \sum_i^M \scalarPhi_i
\end{equation}
which is used as an analytic model function.

In Cartesian coordinates, one can trivially construct terms up to first-order in the spatial components and Equation~\ref{eq:laplace_LSQ} will be satisfied:
\begin{align}
\scalarPhi_{\mathrm{cart}}(x, y, z) = -\big(& \cartCoeff_0 + \cartCoeff_1 x + \cartCoeff_2 y + \cartCoeff_3 z + \cartCoeff_4  x y \notag \\
&+ \cartCoeff_5 x z + \cartCoeff_6 y z + \cartCoeff_7 x y z\big)
\end{align}
which leads to the Cartesian terms in the analytical model
\begin{align}\label{eq:B_cart_LSQ}
B_{x, \mathrm{cart}}(x, y, z) &= \cartCoeff_1 + \cartCoeff_4 y + \cartCoeff_5 z + \cartCoeff_7 y z \\
B_{y, \mathrm{cart}}(x, y, z) &= \cartCoeff_2 + \cartCoeff_4 x + \cartCoeff_6 z + \cartCoeff_7 x z \\
B_{z, \mathrm{cart}}(x, y, z) &= \cartCoeff_3 + \cartCoeff_5 x + \cartCoeff_6 y + \cartCoeff_7 x y
\end{align}
which has seven free parameters $\cartCoeff_i$.

To solve Laplace's equation in a cylindrical coordinate system, we assume that a separable solution exists, and enforce boundary conditions to derive an appropriate set of basis functions.
We define the cylindrical coordinate system in terms of the Cartesian system: the radial coordinate $r=\sqrt{x^2 + y^2}$, the azimuthal coordinate $\cylAzi = \tan^{-1}(y/x)$, and the axial coordinate $z$, which remains the same as the Cartesian $z$.
In this coordinate system, Laplace's equation reads:
\begin{equation}\label{eq:laplace_cyl}
0 = \frac{\partial^2 \scalarPhi}{\partial r^2} + \frac{1}{r} \frac{\partial \scalarPhi}{\partial r} + \frac{1}{r^2} \frac{\partial^2 \scalarPhi}{\partial \cylAzi^2} + \frac{\partial^2 \scalarPhi}{\partial z^2}\ .
\end{equation}
The $\phi$ component necessarily results in oscillatory solutions due to the $2\pi$~periodicity of the azimuthal coordinate. While both exponential and oscillatory solutions are permitted in $z$, we choose the oscillatory form, which better reflects the structure of the solenoidal fields. Given the known forms of the $\phi$ and $z$ solutions, Equation~\ref{eq:laplace_cyl} reduces to a Bessel differential equation in $r$. A detailed derivation of such a cylindrical harmonic solution appears, for example, in Pollack et al.~\cite{pollack2020}
A prototypical term reads:
\begin{equation}\label{eq:phi_n_k_partial}
\scalarPhi_{n,k} (r, \cylAzi, z) = I_n(k r) \sin(n \cylAzi) \sin(k z)
\end{equation}
where $n$ is a non-negative integer, $k$ is a real constant, and $I_n$ are modified Bessel functions of the first kind.
The $\sin$ terms are selected for illustration; $\cos$ terms are equally valid and included in the series. The full functional form, including $\sin$ and $\cos$ terms is:
\begin{align}
\scalarPhi_{n,k}(r, \cylAzi, z) = I_n(k r) \big( &A_{n,k} \cos(k z) \cos(n \cylAzi) \notag \\
+ &B_{n,k} \sin(k z) \cos(n \cylAzi)  \notag \\
+ &C_{n,k} \cos(k z) \sin(n \cylAzi) \notag \\
+ &D_{n,k} \sin(k z) \sin(n \cylAzi) \big)
\end{align}
where $A_{n,k}$, $B_{n,k}$,  $C_{n,k}$, and  $D_{n,k}$ are unconstrained coefficients that modify the overall normalization and sign of each term.

To restrict the axial solutions to a set of countable, orthogonal functions we set a periodic boundary condition on $\scalarPhi$ in $z$ over some length $L = z_1 - z_0$ for start and end points $z_0$ and $z_1$. We make a simplifying translation of the axial coordinate $z$ such that $z=0$ is in the center of these endpoints, or $z_1=-z_0=L/2$. The boundary condition is imposed at each point in the $r,\cylAzi$ plane:
\begin{equation}\label{eq:axial_BC_LSQ}
f(r, \cylAzi) = \scalarPhi(r, \cylAzi, -L/2) = \scalarPhi(r, \cylAzi, +L/2)
\end{equation}
for some unknown $f(r, \cylAzi)$. After fitting the model to the data, we can extract $f(r, \cylAzi)$ although it is not the primary quantity of interest.
It should be understood that while there is no explicit boundary condition on the exterior radial surface of the bounding cylinder, when fitting this series solution to measurements the resulting model function can only be trusted within the radial extent of the measurements. 
For the Mu2e DS, the radial extent of the measurements is $r=0.8$~m; we do not attempt to use the model beyond this radius. Furthermore, for $r\geq1$~m we encounter the superconductors in the solenoid coils, so the underlying assumption that we are in a source-free region breaks down.

The magnetic field in the Mu2e DS is far from axially periodic, but functional forms with a periodic boundary condition in $z$ can describe the measurement sample. This can be understood by considering that the magnetic field is the gradient of the scalar field; while the scalar potential is fixed to match at the axial ends, the gradients are not. Some additional flexibility to fit the points is gained by tuning the hyperparameter $L$. There is a subtle trade-off when selecting $L$. As $L$ increases, the total number of terms required to match the data will increase. On the other hand, if $L$ is too small and the boundaries are near the data there is little flexibility in the functions to match the data at the axial ends, as they must conform to the condition of equal potential $\scalarPhi$ on the two surfaces.

By imposing the boundary condition in Equation~\ref{eq:axial_BC_LSQ},
we discretize $k$:
\begin{equation}
k_m = \frac{2 \pi (m+1)}{L}
\end{equation}
where $m$ is a non-negative integer. We choose $(m+1)$ to discard the $k_m=0$ terms which define $\scalarPhi_{n,k}$ that are independent of the spatial coordinates and do not contribute to the field. With the set of $k_m$, we construct the series solution to $\scalarPhi_{\mathrm{cyl}}$ in cylindrical coordinates by summing $\scalarPhi_{n,k}$ up to some maxima of $m$ and $n$:
\begin{align}\label{eq:phi_cyl_final}
\scalarPhi_{\mathrm{cyl}} = \sum_{m=0}^{m_{\mathrm{max}}} \sum_{n=0}^{n_{\mathrm{max}}} I_n(k_m r) \big( &A_{m,n} \cos(k_m z) \cos(n \cylAzi) \notag \\
+ &B_{m,n} \sin(k_m z) \cos(n \cylAzi)  \notag \\
+ &C_{m,n} \cos(k_m z) \sin(n \cylAzi) \notag \\
+ &D_{m,n} \sin(k_m z) \sin(n \cylAzi) \big)\ .
\end{align}

By increasing $m_{\mathrm{max}}$, smaller spatial features in $z$ are captured. Additionally, due to the coupled nature of $z$ and $r$, terms with increasing $m$ (i.e. increasing $k_m$) scale the radial dependence in the term $I_n(k_m r)$. Similarly, increasing $n_{\mathrm{max}}$ includes terms with smaller spatial features in $\cylAzi$ while modifying the radial dependence via the order of the Bessel function. Setting $z=\pm L/2$ elucidates the functional form of boundary condition $f(r, \cylAzi)$:
\begin{align}\label{eq:axial_BC_LSQ_full}
f(r, \phi) = \sum_{m=0}^{m_{\mathrm{max}}} \sum_{n=0}^{n_{\mathrm{max}}} &(-1)^{m+1} I_n(k_m r) \notag \\
&\times \big( A_{m,n} \cos(n \cylAzi) + C_{m,n} \sin(n \cylAzi) \big) \ .
\end{align}

We apply the gradient to $\scalarPhi_{\mathrm{cyl}}$ (Equation~\ref{eq:phi_cyl_final}) to determine the magnetic field defined by this cylindrical harmonic model:
\begin{widetext}
\begin{align}\label{eq:B_cyl_final}
B_{r, \mathrm{cyl}} = \sum_{m=0}^{m_{\mathrm{max}}} \sum_{n=0}^{n_{\mathrm{max}}} k_m I'_n(k_m r) \big[ &A_{m,n} \cos(k_m z) \cos(n \cylAzi) + B_{m, n} \sin(k_m z) \cos(n \cylAzi) \notag \\
+ &C_{m,n} \cos(k_m z) \sin(n \cylAzi) + D_{m, n} \sin(k_m z) \sin(n \cylAzi) \big] \notag \\
B_{\cylAzi, \mathrm{cyl}} = \sum_{m=0}^{m_{\mathrm{max}}} \sum_{n=0}^{n_{\mathrm{max}}} \frac{n}{r} I_n(k_m r) \big[ - &A_{m,n} \cos(k_m z) \sin(n \cylAzi) - B_{m, n} \sin(k_m z) \sin(n \cylAzi) \notag \\
+ &C_{m,n} \cos(k_m z) \cos(n \cylAzi) + D_{m, n} \sin(k_m z) \cos(n \cylAzi) \big] \notag \\
B_{z, \mathrm{cyl}} = \sum_{m=0}^{m_{\mathrm{max}}} \sum_{n=0}^{n_{\mathrm{max}}} k_m I_n(k_m r) \big[ - &A_{m,n} \sin(k_m z) \cos(n \cylAzi) + B_{m, n} \cos(k_m z) \cos(n \cylAzi) \notag \\
- &C_{m,n} \sin(k_m z) \sin(n \cylAzi) + D_{m, n} \cos(k_m z) \sin(n \cylAzi) \big]\ .
\end{align}
\end{widetext}
where $I'_n(k_m r)$ is the first derivative of $I_n(k_m r)$.

As the least-squares analytical model is not the focus of this paper, we do not describe some of the subtler points of the model or the hyperparameter optimizations (see the supplementary material). The optimized model uses the following hyperparameter values: $L=12.5$~m, $m_{\mathrm{max}} = 54$, and $n_{\mathrm{max}} = 6$. When including $N_\mathrm{cart}$~($=7$ with all terms) of the trivial Cartesian solutions the least-squares model has $N_{\mathrm{free}}$ free parameters:
\begin{equation}\label{eq:Nfree}
N_{\mathrm{free}} = 4\ n_{\mathrm{max}}\ (m_{\mathrm{max}}+1) + 2\ (m_{\mathrm{max}} + 1) + N_\mathrm{cart}
\end{equation}
which is slightly complicated by the removal of $C_{m,0}$ and $D_{m,0}$ terms, which are otherwise unconstrained given ${\sin(n\cylAzi) = 0}$ when $n=0$. $N_{\mathrm{free}} = \numprint{1436}$ for the nominal model.

After fitting, we refer to the prediction from the least-squares model on the measurement sample as $\vec{B}^{(\mathrm{meas})}_{\mathrm{pred}, i}$ which is used to define the residual at field $i$ ($\vec{r}_i = (x_i, y_i, z_i)$):
\begin{equation}\label{eq:res_defn_LSQ}
\vec{B}^{(\mathrm{meas})}_{\mathrm{res}, i} = \vec{B}_{\mathrm{meas}, i} - \vec{B}^{(\mathrm{meas})}_{\mathrm{pred}, i}
\end{equation}
noting that an equivalent definition is used for calculating the residuals on the test sample.

\subsection{Physics-Informed Neural Network}
\label{subsec:methods_pinn}

A PINN is trained on the residuals on the measurement sample (Equation~\ref{eq:res_defn_LSQ}) from the least-squares fit described in Section~\ref{subsec:methods_leastsquares} to improve the modeling accuracy. As with the least-squares model function, we desire a trained PINN that obeys Maxwell's equations in a source-free region. In contrast to the expansion in Bessel functions, there is no clear analytical path to ensuring these constraints in a NN. The flexibility of NNs, however, allows for soft constraints in the form of additional loss terms~\cite{PINN2019}. We define a PINN with three inputs describing the field point location in Cartesian coordinates and one output describing a residual scalar magnetic potential $\scalarPhi_\mathrm{NN}$. A simple hidden layer structure is used: $N_{\mathrm{layers}}$ hidden layers each contain $N_{\mathrm{nodes}}$ nodes, all using the same activation function (described below). The network employs a fully connected feed-forward topology. The field described by the PINN is defined in an analogous way to Equation~\ref{eq:B_defn_Phi_LSQ}:
\begin{equation}
\vec{B}_{\mathrm{NN}} = -\nabla \scalarPhi_{\mathrm{NN}}\ .
\end{equation}

This formulation satisfies $\nabla \times \vec{B}=0$ by construction (i.e. a hard constraint) and naturally follows the derivation of the functional forms used in the least-squares model. Ultimately, $\scalarPhi_{\mathrm{NN}}$ is added to $\scalarPhi_{\mathrm{LSQ}}$ so that $\scalarPhi_{\mathrm{model}} = \scalarPhi_{\mathrm{LSQ}} + \scalarPhi_{\mathrm{NN}}$ describes the best-fit model to the data. To constrain $\nabla \cdot \vec{B}$, we include an additional loss term defined as the mean squared divergence evaluated at a set of $N_c$ collocation points randomly sampled from the DS mapping volume.
The total loss function minimized during the PINN training is:

\begin{widetext}
\begin{equation}
L_{\mathrm{PINN}} = L_B + \lambda L_{\nabla \cdot B} = \frac{1}{N_{\mathrm{meas}}} \sum_{i=1}^{N_{\mathrm{meas}}} ||\vec{B}^{(\mathrm{meas})}_{\mathrm{res}, i} - \vec{B}_{\mathrm{NN}, i}||^2 + \frac{\lambda}{N_{c}} \sum_{j=1}^{N_{c}} (\nabla \cdot \vec{B}_{\mathrm{NN}, j})^2
\end{equation}
\end{widetext}
where $\lambda \geq 0$ is a hyperparameter that regulates the relative weight of the divergence loss and $\vec{B}_{\mathrm{NN}, i}$ is the residual field vector predicted by the PINN at field point $i$. $\vec{B}_{\mathrm{NN}, j}$ is the predicted residual field vector at collocation point $j$. Ideally, minimizing $L_{\mathrm{PINN}}$ should lead to minimal $||\vec{B}^{(\mathrm{meas})}_{\mathrm{res}, i} - \vec{B}_{\mathrm{NN}, i}||^2$ consistent with $\nabla \cdot \vec{B} = 0$. This PINN structure is motivated by the work done by Coskun et al.~\cite{coskun2022} but notably differs given the novel blend of hard and soft physical constraints. Similar methods have been applied successfully by Scheinker and Pokharel to time dependent electromagnetic fields for modeling beam dynamics~\cite{scheinker2023}.

\begin{figure*}
\centering
\includegraphics[trim=0cm 0cm 0.0cm 0.0cm, width=0.65\textwidth]{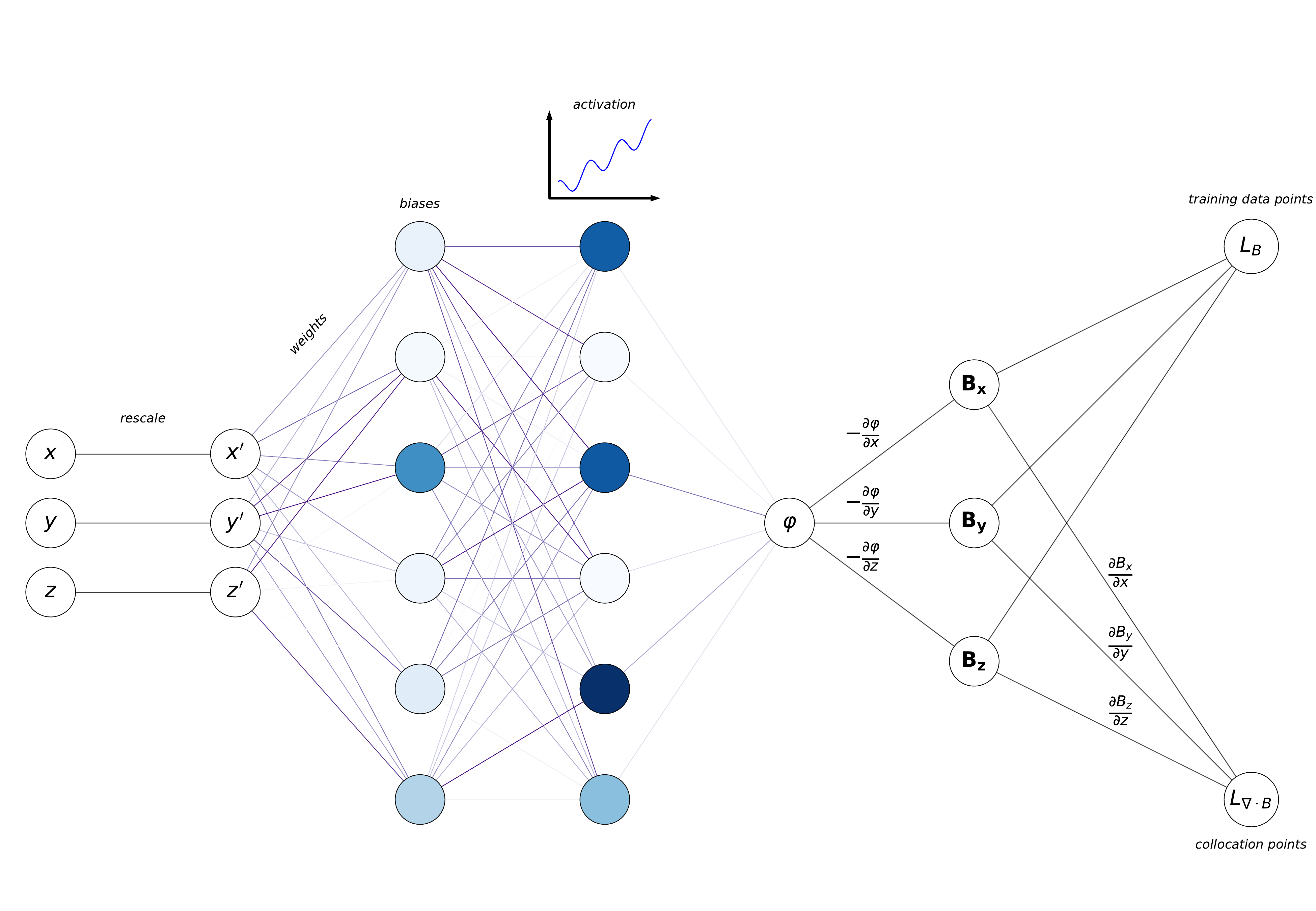}
\caption{A network diagram for the PINN. The Cartesian coordinate inputs are rescaled and fed into a fully-connected feed-forward PINN, which uses the DELTAsnake activation function at every hidden layer node. For clarity this diagram includes two hidden layers with six nodes per hidden layer. The colors of the edges (nodes) represent weight (bias) values after training. The scalar output $\scalarPhi$ of the NN is differentiated, which yields the field vector. Whether feeding in training data or collocation points, the final step is either to calculate $L_B$ by comparing the prediction to the residual field vectors in the training set or to calculate $L_{\nabla \cdot B}$, respectively. When the PINN is used to predict the field vectors after training, the loss terms are not calculated and the field vector is directly returned.}
\label{fig:network_diagram}
\end{figure*}

All derivatives are calculated using automatic differentiation, which is implemented and optimized in modern machine learning packages like TensorFlow~\cite{tensorflow2015-whitepaper}. While automatic differentiation is already being used to find the gradient of the loss function with respect to the weights and biases for backpropagation, we additionally use it to differentiate the output with respect to the input. Conveniently, automatic differentiation yields exact derivatives, so the soft constraint in the loss function is unbiased.

Given the large variation in the domain of each coordinate in the DS bore (recall \mbox{$z \in [2.9, 14.9]$~m} and \mbox{$r \in [0, 0.8]$~m}), each Cartesian coordinate is rescaled such that the values are zero centered and have a range of $[-1, 1]$. The rescaled coordinates are dimensionless.
For example, the rescaled $x$ coordinate is
\begin{equation}
x' = \frac{x - \overline{x}_{\mathrm{meas}}}{(\max(x_{\mathrm{meas}}) - \min(x_{\mathrm{meas}}))/2}\ .
\end{equation}

The measurement and test samples use an identical rescaling. The spatial derivatives used in the calculation of ${\vec{B}_{\mathrm{NN}}}$ and ${\nabla \cdot \vec{B}_{\mathrm{NN}}}$ are correspondingly rescaled for physical consistency:
\begin{equation}
\diffp{}{x} =  \frac{2}{\max(x_{\mathrm{meas}}) - \min(x_{\mathrm{meas}})} \times \diffp{}{{x'}}\ .
\end{equation}
A pictorial representation of this network structure is shown in Figure~\ref{fig:network_diagram}.

The choice of optimal network size and activation function for our test case derives from a number of hyperparameter optimizations, which are described in the supplementary material.
The optimal network structure determined from the hyperparameter optimization is $N_{\mathrm{hidden}} = 8$ and $N_{\mathrm{nodes}} = 64$. The appropriate choice of activation function is subtle; the least-squares model is periodic axially and azimuthally but monotonic radially. 
While the least-squares model captures dominant axial and azimuthal components, its limited representational power (i.e. finite number of terms) may leave behind residual structures that contain periodic features, particularly at shorter wavelengths than those used in the model, e.g. set by $m_{\mathrm{max}}$. This is analogous to underfitting in polynomial regression; this behavior is documented in classical regression literature~\cite{montgomery2012linear}. With this in mind, we want the PINN to be able to capture oscillatory features.
As demonstrated by Ziyin et al.~\cite{snake2020}, common activation functions struggle to model even basic oscillatory behaviors. Instead, the so-called ``snake" activation function can successfully model oscillatory behavior while avoiding a vanishing gradient which plagues many common activation functions (e.g. $\tanh$). The snake activation is defined as:
\begin{equation}
y(x; a) = x + \frac{1}{a} \sin^2 (a x)
\end{equation}
where $a$ is a hyperparameter that determines the spatial frequency of the sinusoidal term.

Through a series of robustness tests we noticed sub-optimal training when applying the standard snake activation. Given this observation, we developed a modified version of the snake activation function: the Derivative-Enhanced Local Tuning Activation modification to snake, or ``DELTAsnake''. We define the DELTAsnake activation function:
\begin{equation}\label{eq:snake}
y(x; f, a, G) = G \big(a x \frac{1-f}{1+f} + \sin^2(a x) \big)
\end{equation}
The hyperparameter $f$ controls the ratio of the maximum negative derivative to the maximum positive derivative and is set in the range $[0, 1]$.
The contribution of the $\sin^2$ term is tuned with hyperparameters $f$ and $a$. Hyperparameter $G$ modifies the curvature of the function.
This ability to make the function non-monotonic and tune the curvature results in faster and more consistent convergence during training. By setting $f=0$ and $G=1/a$, one can recover the original snake activation function.

\begin{figure*}
\centering
\includegraphics[trim=0cm 0cm 0.0cm 0.0cm, width=0.8\textwidth]{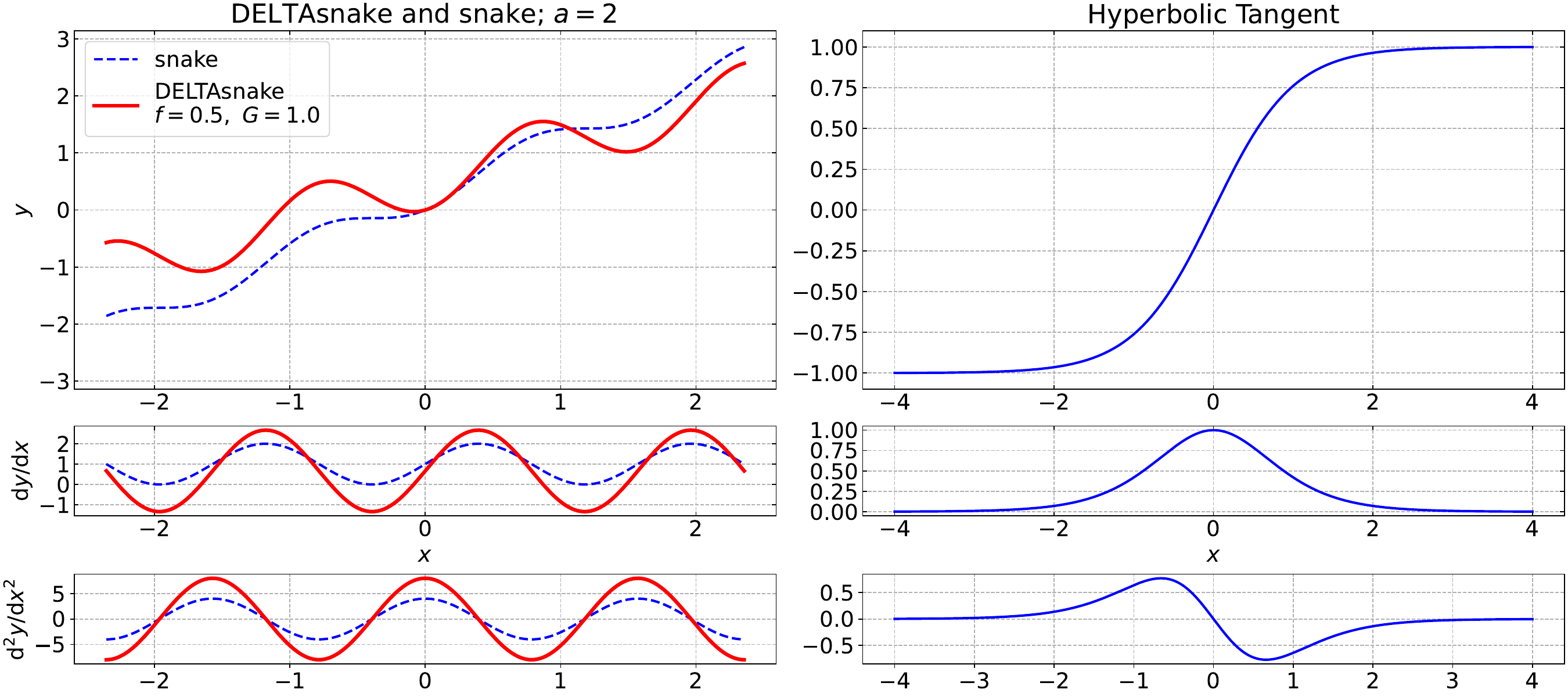}
\caption{The DELTAsnake and snake activation functions (left) both outperform common activation functions like hyperbolic tangent (right) given the periodic nature of the data. Additionally, the snake-like functions do not suffer from vanishing gradients as suggested in the middle panel of the plots. DELTAsnake has additional flexibility in monotonicity and curvature.}
\label{fig:snake_activ}
\end{figure*}

To illustrate the impact of these modifications, we compare the first and second derivatives of DELTAsnake, snake, and $\tanh$ in Table~\ref{tab:DELTAsnake_derivs} and provide plots comparing the functions and their derivatives in Figure~\ref{fig:snake_activ}.

\begin{table*}
\center
\caption{A comparison of functional forms and first and second derivatives of DELTAsnake, snake, and $\tanh$.}
\bgroup
\def\arraystretch{2.0}
\begin{tabular}{l|c|c|c|c}
\textbf{Activation} & $\mathbf{y}$ & $\mathbf{y'}$ & $\mathbf{y'}$ \textbf{range} & $\mathbf{y''}$ \\
\hline
DELTAsnake & $\ \ G \big(a x \frac{1-f}{1+f} + \sin^2(a x) \big)\ \ $ & $\ \ G \big(a \frac{1-f}{1+f} + a \sin(2 a x) \big)\ \ $ & $\ \ [\frac{-2fG a}{(1+f)}, \frac{2Ga}{(1+f)}]\ \ $ & $2 G a^2 \cos(2 a x)$ \\
\hline
snake & $x + \frac{1}{a}\sin^2(a x)$ & $1 + \sin(2 a x)$ & $[0, 2]$ & $2 a \cos(2 a x)$ \\
\hline
$\tanh$ & $\tanh x$ & $1 - \tanh^2 x$ & $[0, 1]$ & $-2 \tanh x \sech^2 x$
\end{tabular}
\egroup
\label{tab:DELTAsnake_derivs}
\end{table*}

A detailed analysis of the properties of DELTAsnake is currently under way and will be discussed in a later work. In this particular application, we find values \mbox{$f=0.5$}, \mbox{$G=1$}, and \mbox{$a=5$} work well. We show concretely in the supplementary materials that a network using the DELTAsnake activation outperforms an equivalent network using the standard snake activation function as well as the hyperbolic tangent activation function.

As recommended in Ref.~\cite{snake2020}, when using the snake activation, the biases ($b$) are initialized to zero and the weights ($w$) are initialized from a uniform distribution in the range:
\begin{equation}
w \in [-\sqrt{3/N_{\mathrm{nodes}}}, \sqrt{3/N_{\mathrm{nodes}}}]\ .
\end{equation}
When using DELTAsnake, due to the larger curvature, we find stabler results with a smaller initialization range. From trial and error we developed a heuristic for the range of the weight initializations for DELTAsnake that depends on $a$, $G$, and $N_{\mathrm{nodes}}$ and relates to the period of DELTAsnake oscillations:
\begin{equation}
w \in \big[ - \kappa \frac{\pi}{G a \sqrt{N_{\mathrm{nodes}}}}, \kappa \frac{\pi}{G a \sqrt{N_{\mathrm{nodes}}}} \big]
\end{equation}
with a hand-tunable parameter $\kappa$ that should be $\mathcal{O}(1)$. When $\kappa=1$ the initialization range holds $w$ to within one period of the DELTAsnake oscillation for the case of $G=1$. An additional suppression factor of $1/G$ is included to balance the increased curvature. If $\kappa$ is too large, the training may not converge. If $\kappa$ is too small, the network gets stuck in uninteresting solutions or false minima. For the PINN applied to the Mu2e DS magnetic field, we find a value of $\kappa=0.7$ works well. For the best network structure, which has $N_{\mathrm{nodes}} = 64$, this results in an initialization range of approximately $w \in [-0.055, 0.055]$.

To prepare the data for training, the measurement sample is randomly split into \textit{training data} and \textit{validation data} with an $80\%$/$20\%$ training/validation split. The network is trained using the ADAM optimizer~\cite{Adam2017} for a nominal training period of $\numprint{10000}$ epochs. In each epoch $N_c=\numprint{50000}$ collocation points are sampled and passed through the network to calculate $L_{\nabla \cdot B}$. An initial learning rate ($\mathrm{LR}$) of $\mathrm{LR} = 2 \times 10^{-3}$ is used, with a learning rate reduction scheme that monitors the training loss for plateaus. The learning rate is reduced by a factor of $0.5$ with a patience of $300$ epochs. The minimum allowed $\mathrm{LR}$ is set to $5 \times 10^{-5}$. To encourage early training towards a convergence of $L_B$ before applying a harsh physics constraint, $\lambda$ is initialized to a small value and adjusted throughout the training period. We evaluated several annealing strategies and found that a gradual linear ramp yields the most consistent results and is straightforward to tune. A pre-training period of $\numprint{1250}$ epochs is completed with $\lambda=0.001$. We then apply the linear ramp until $\lambda$ reaches its final value $\lambda_{\mathrm{max}}$. Every $3$ epochs we apply a constant additive increase:
\begin{equation}
\lambda_{i+1} = \min(\lambda_i + 0.0001, \lambda_{\mathrm{max}})\ .
\end{equation}
Annealing $\lambda$ in this way provides the benefit of being able to pre-train with a small $\lambda$ value, without shocking the loss function with a large jump in $\lambda$ in few epochs.

An early stopping routine monitors the validation loss and stops the training with a patience of $\numprint{5000}$.

\subsection{Fit Quality Statistic}
\label{subsec:methods_fitquality}

To evaluate the quality of the fits we compare the value of the reduced chi-square ($\chi_{\mathrm{reduced}}^2$) with one; a value close to one is indicative of a good fit based on the measurement uncertainties. The reduced chi-square value is equal to the chi-square value, $\chi^2$, divided by the number of degrees of freedom, ${N_{\mathrm{DOF}} = N_{\mathrm{meas}} - N_{\mathrm{LSQ}} - N_{\mathrm{PINN}}}$, where $N_{\mathrm{LSQ}}$ denotes the number of free parameters in the least-squares fit and $N_{\mathrm{PINN}}$ is the generalized degrees of freedom (GDF) which we estimate using a Monte Carlo method proposed by Ye~\cite{YeGDF}. In the initial least-squares fit, which does not include the PINN, we set $N_{\mathrm{PINN}}=0$. Details of the estimation of $N_{\mathrm{PINN}}$ are included in the supplementary material. The full definition of the reduced chi-square is:
\begin{align}\label{eq:chi2}
\chi_{\mathrm{reduced}}^2 &= \frac{\chi^2}{N_{\mathrm{DOF}}} \notag \\
&= \frac{1}{N_{\mathrm{DOF}}} \sum_i^{N_\mathrm{meas}} \frac{(\vec{B}_{\mathrm{meas},i} - \vec{B}_{\mathrm{pred},i}^{(\mathrm{meas})})^2}{\sigma_i^2}
\end{align}
where $\sigma_i = 0.3$~Gauss is the injected measurement uncertainty. While Equation~\ref{eq:chi2} refers to the measured and predicted values of from the initial least-squares fit, corresponding definitions of the measurement and prediction values are used to determine $\chi_{\mathrm{reduced}}^2$ in the downstream modelings stages.

\section{Results}
\label{sec:results}

Our modeling technique is applied to the ten toy measurement simulations for the Mu2e DS described in Section~\ref{subsec:methods_fieldcalc}. We provide a summary of the reduced chi-square values from the toys and plots to further highlight the results for one of the toys. In the subsequent sections we describe robustness tests of the model that include a test of the impact of a systematic shift in the measurements and a test demonstrating the power of the PINN when using a significantly reduced set of measurement points.

\subsection{Nominal Results}
\label{subsec:results_nominal}
\subsubsection{Least-Squares Approach}
\label{subsubsec:results_leastsquares}

All field points in a given toy simulated measurement sample are included in the least-squares fit. We set the length parameter $L=12.5$~m and include terms in the series to orders $m_\mathrm{max} = 54$ and $n_{\mathrm{max}} = 6$, based on a hyperparameter optimization. Using intuition from the Nyquist-Shannon sampling theorem, one can alternatively select a reasonable value for $m_\mathrm{max}$ and $n_{\mathrm{max}}$;
the model should not include frequencies above half the sampling frequency. In $z$, for example, the measurement sampling step size is $\Delta z = 0.05$~m, therefore $m$ should not exceed a value of $124$. The azimuthal step size is $\Delta \cylAzi = \pi/8$~rad, so $n$ should not exceed a value of $7$. Other considerations include consistency of convergence and fitting time, so we ultimately use a model with a more conservative truncation in axial frequencies. The least-squares fits result in a fit quality of ${\chi^2_{\mathrm{reduced}} = 2.15\pm0.01}$. We provide plots below for a toy with ${\chi^2_{\mathrm{reduced}} = 2.14}$. Running a single fit requires approximately 20~minutes of wall-clock time when executed on an AMD Ryzen Threadripper 3970X 32-Core CPU.

\begin{figure}
    \centering
    \begin{minipage}[t]{0.45\textwidth}
        \centering
        \includegraphics[width=\linewidth]{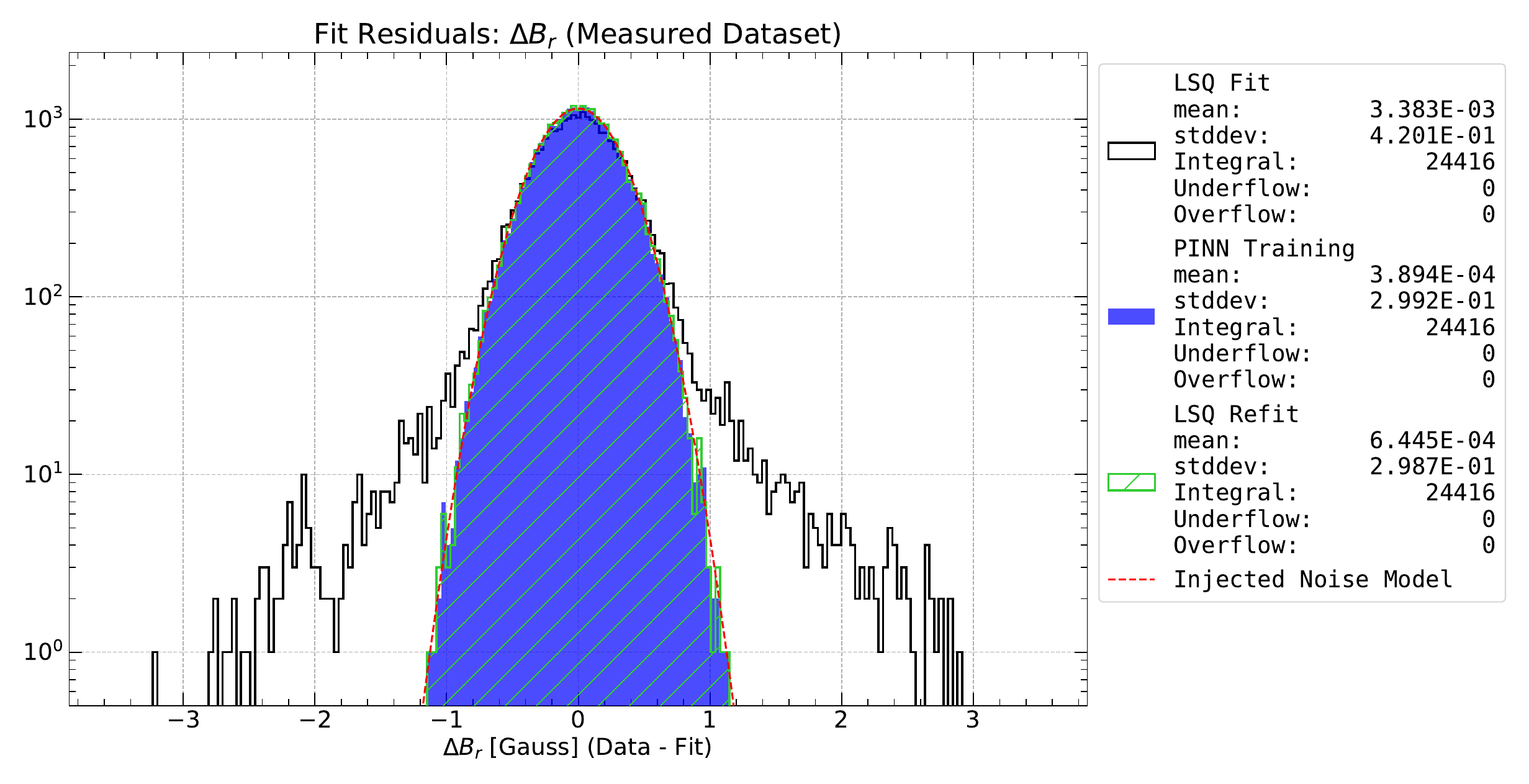}
    \end{minipage}
    
    \vspace{-0.5em} 
    
		\begin{minipage}[t]{0.45\textwidth}
        \centering
        \includegraphics[width=\linewidth]{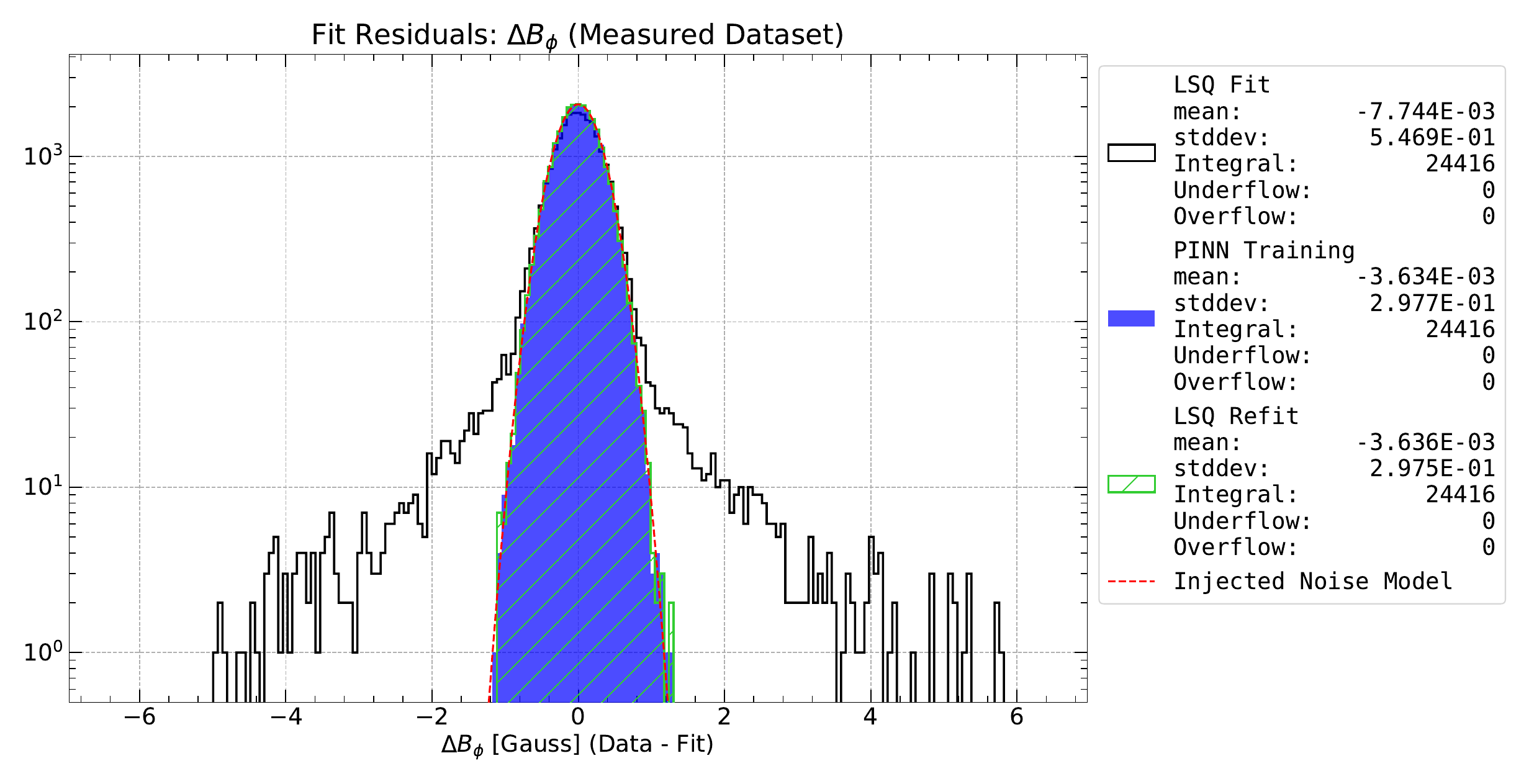}
    \end{minipage}
    
    \vspace{-0.5em} 

    \begin{minipage}[t]{0.45\textwidth}
        \centering
        \includegraphics[width=\linewidth]{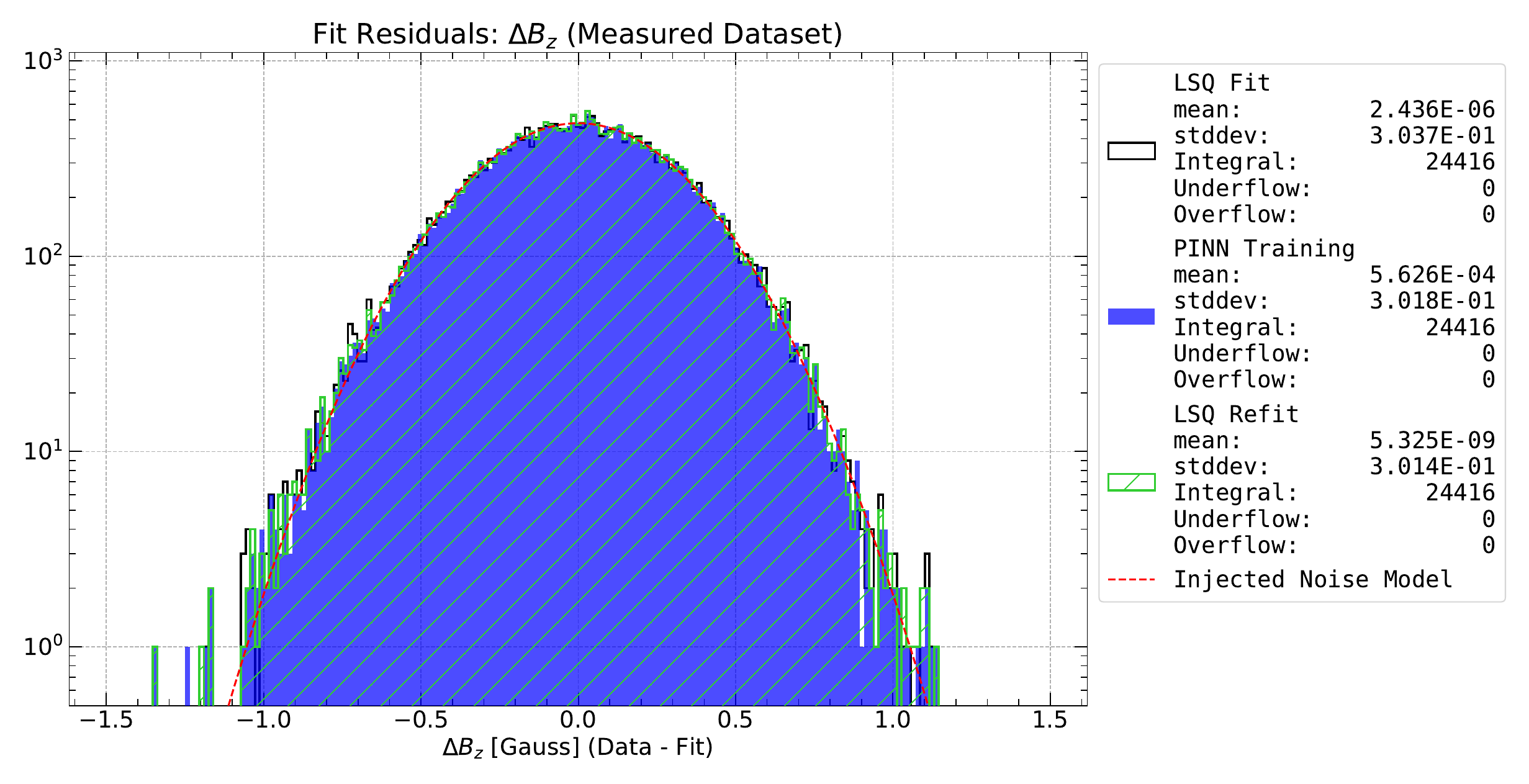}
    \end{minipage}
    \caption{Residual plots, data - fit, for the three field components $B_r$ (top), $B_\cylAzi$ (middle), $B_z$ (bottom) evaluated at all points in the measurement sample. The residuals after each step of the modeling are included: least-squares fit (black outline), PINN training (blue filled), and least-squares iteration (green hatch). An overlay of the injected noise model is included as a red dashed line.}
    \label{fig:residuals_df_meas_nominal}
\end{figure}

\begin{figure}
    \centering
    \begin{minipage}[t]{0.45\textwidth}
        \centering
        \includegraphics[width=\linewidth]{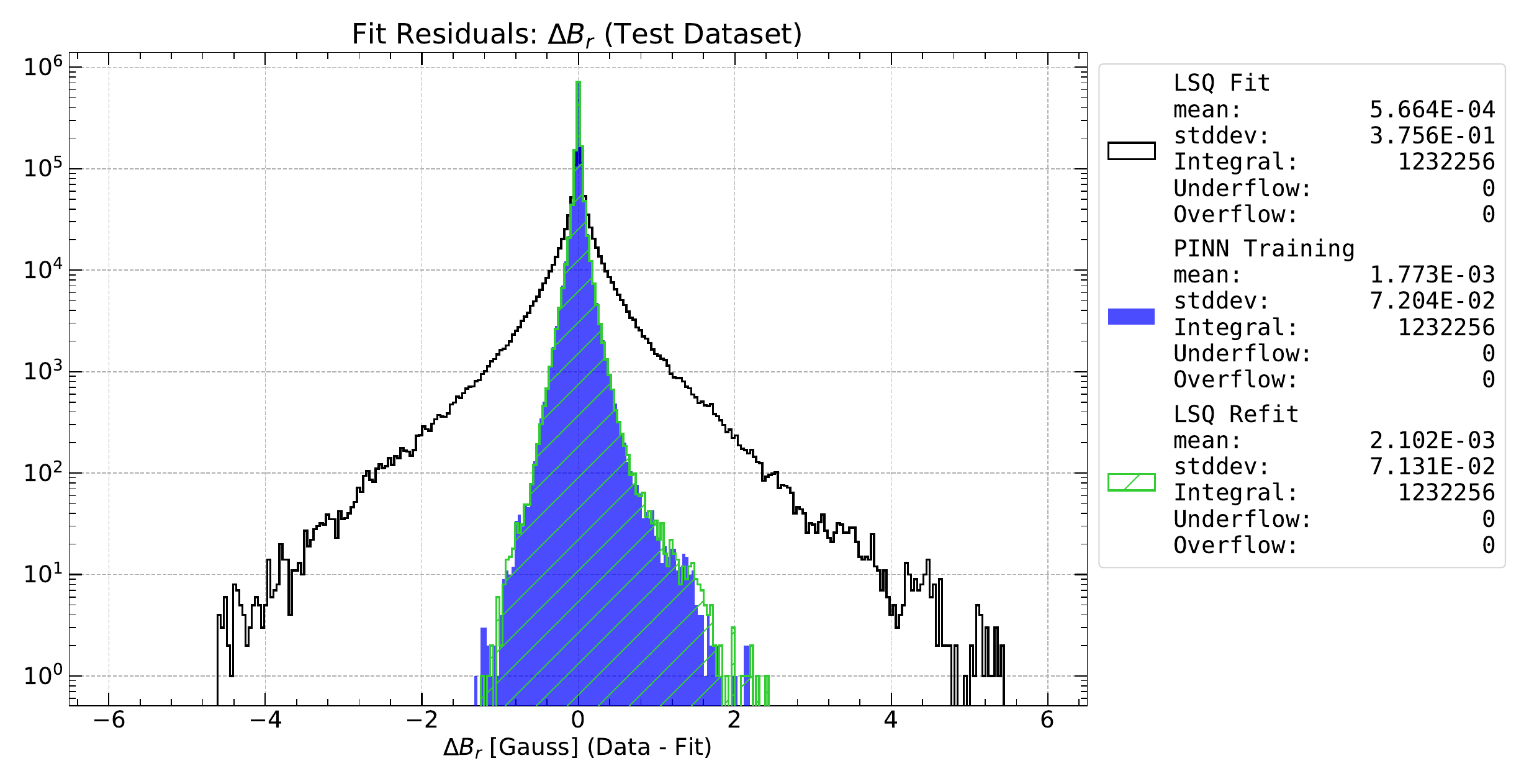}
    \end{minipage}
    
    \vspace{-0.5em} 
    
		\begin{minipage}[t]{0.45\textwidth}
        \centering
        \includegraphics[width=\linewidth]{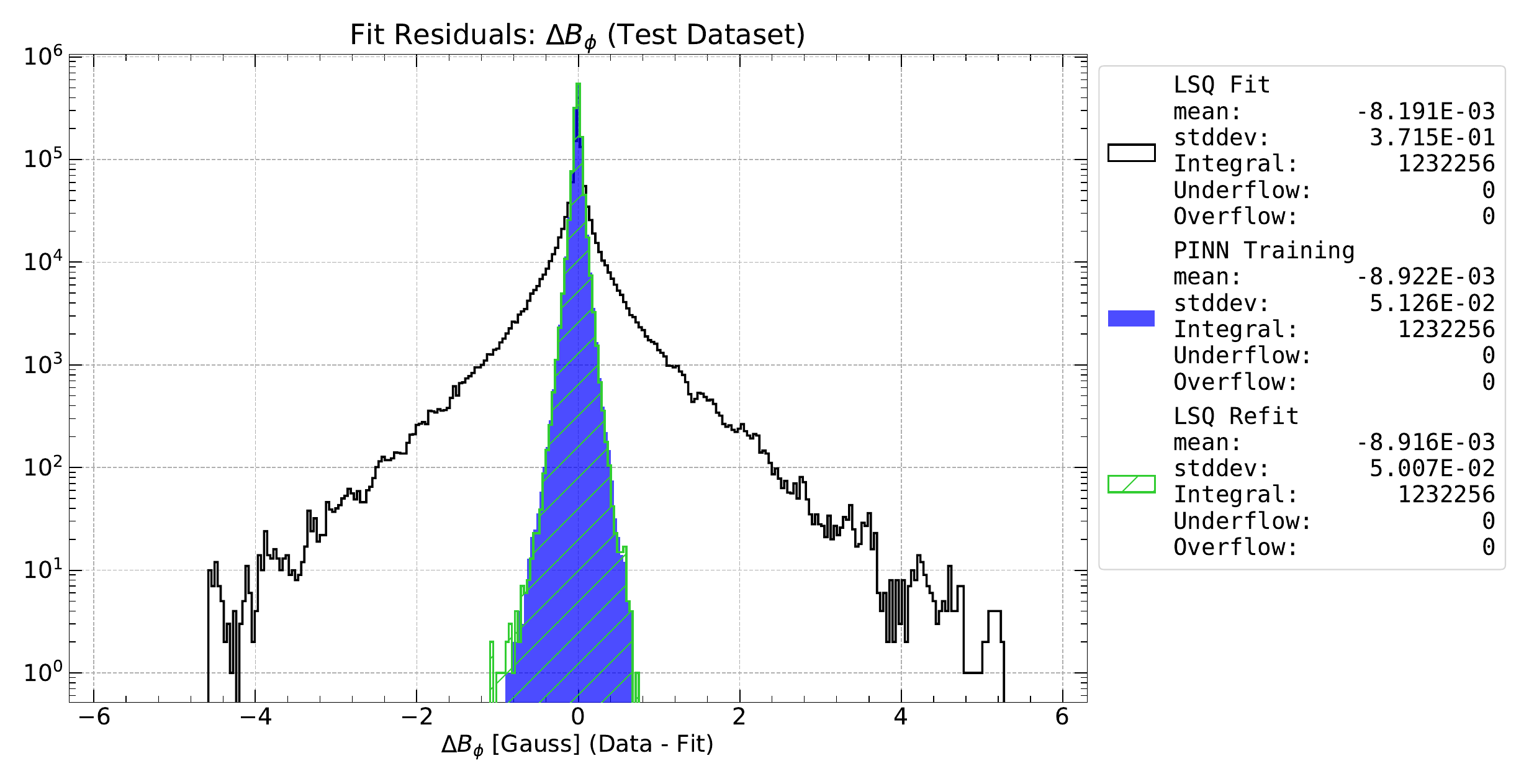}
    \end{minipage}
    
    \vspace{-0.5em} 

    \begin{minipage}[t]{0.45\textwidth}
        \centering
        \includegraphics[width=\linewidth]{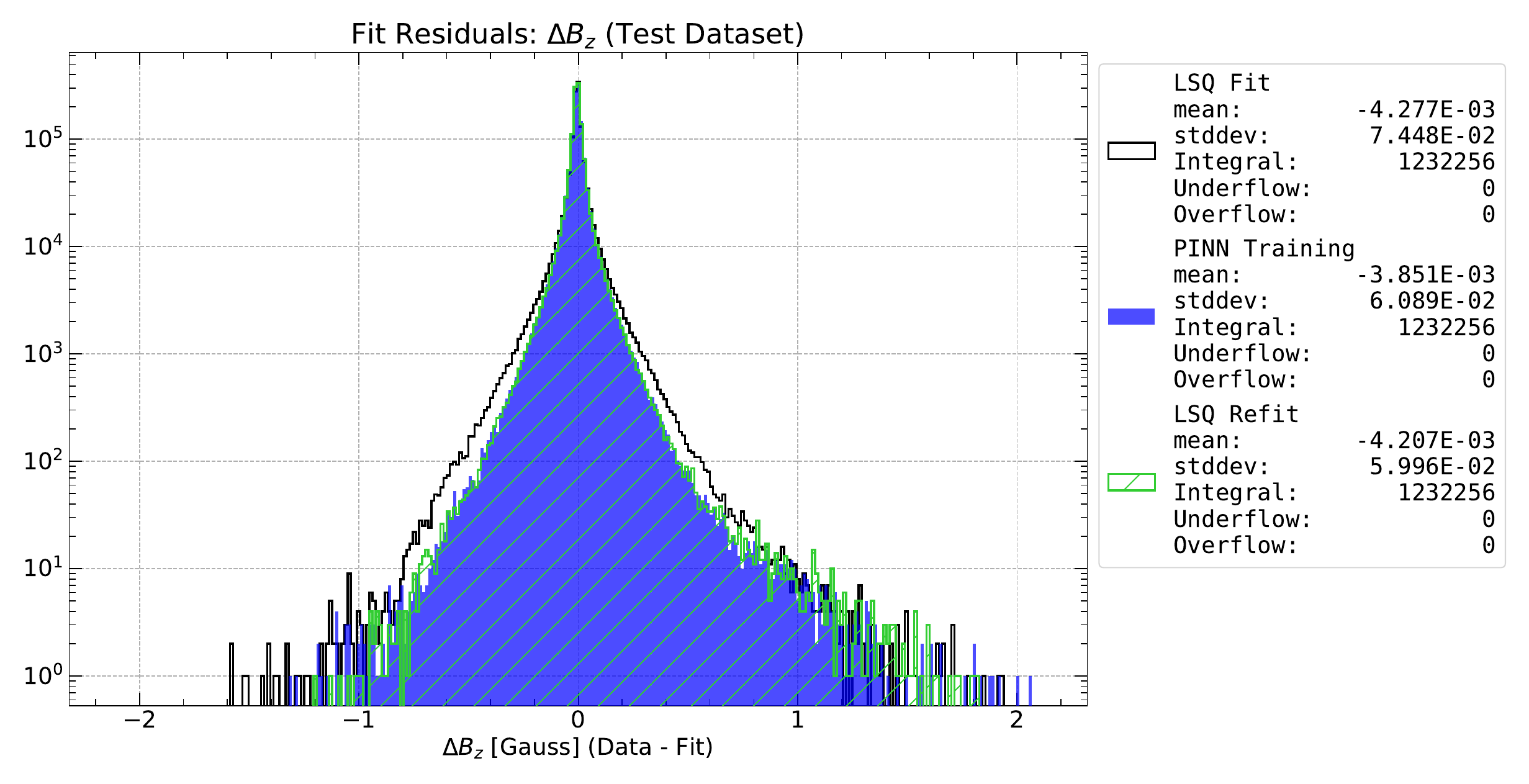}
    \end{minipage}
    \caption{Residual plots, data - fit, for the three field components $B_r$ (top), $B_\cylAzi$ (middle), $B_z$ (bottom) evaluated at all points in the test sample. The residuals after each step of the modeling are included: least-squares fit (black outline), PINN training (blue filled), and least-squares iteration (green hatch).}
    \label{fig:residuals_df_test_nominal}
\end{figure}

Plots of residuals in the following sections (Figures~\ref{fig:residuals_df_meas_nominal}~and~\ref{fig:residuals_df_test_nominal}) clearly indicate the least-squares model accurately reconstructs $B_z$ but struggles to model sub-dominant features in $B_r$ and $B_\cylAzi$.

\subsubsection{Physics-Informed Neural Network}
\label{subsubsec:results_pinn}

The PINN is applied to the residuals of the least-squares fit for each toy. In other words, the PINN learns a relic field contribution that can be included in the final model.
The choice of network structure and activation function is not obvious, so we complete a number of hyperparameter optimizations to build an optimal network. This optimal network has $N_{\mathrm{layers}} = 8$, $N_{\mathrm{nodes}} = 64$, and a DELTAsnake function with $f=0.5$, $a=5$, and $G=1$ at every node of every hidden layer. A choice of $\lambda_{\mathrm{max}} = 0.1$ is sufficient to produce a physically consistent model that retains enough flexibility to model the relevant features of the field. The fit quality after training the PINN is ${\chi^2_{\mathrm{reduced}} = 1.037\pm0.005}$, and ${\chi^2_{\mathrm{reduced}} = 1.034}$ for the toy included in the figures. The training procedure requires approximately 30~minutes of wall-clock time per toy when executed on an NVIDIA GeForce RTX$^{\mathrm{TM}}$ 2080 Ti GPU with 11GB GDDR6 memory.

Using a Monte Carlo method~\cite{YeGDF} with twenty trials, wherein we randomly perturb the input residuals by a small amount and retrain the PINN, we estimate the GDF to be ${N_{\mathrm{PINN}} = \widehat{\mathrm{GDF}} = \numprint{1156}\pm57}$. Given the uncertainty on $N_{\mathrm{PINN}}$ is small relative to the total degrees of freedom, we use the mean value when computing $\chi^2_{\mathrm{reduced}}$. We note that $N_{\mathrm{PINN}}$ is an order of magnitude fewer than the number of free parameters in the PINN ($=\numprint{29441}$), which is consistent with the findings of Ye~\cite{YeGDF} and Gao and Jojic~\cite{GaoGDFNN}. In particular, Gao and Jojic demonstrate that the GDF is significantly reduced in neural networks that are deep and regularized. Our PINN is both deep (${N_{\mathrm{layers}} = 8}$) and is regularized with the physics constraint.

\begin{figure}
\centering
\begin{minipage}{0.45\textwidth}
\centering
\includegraphics[trim=0cm 0cm 0.0cm 0.0cm, width=1.0\textwidth, left]{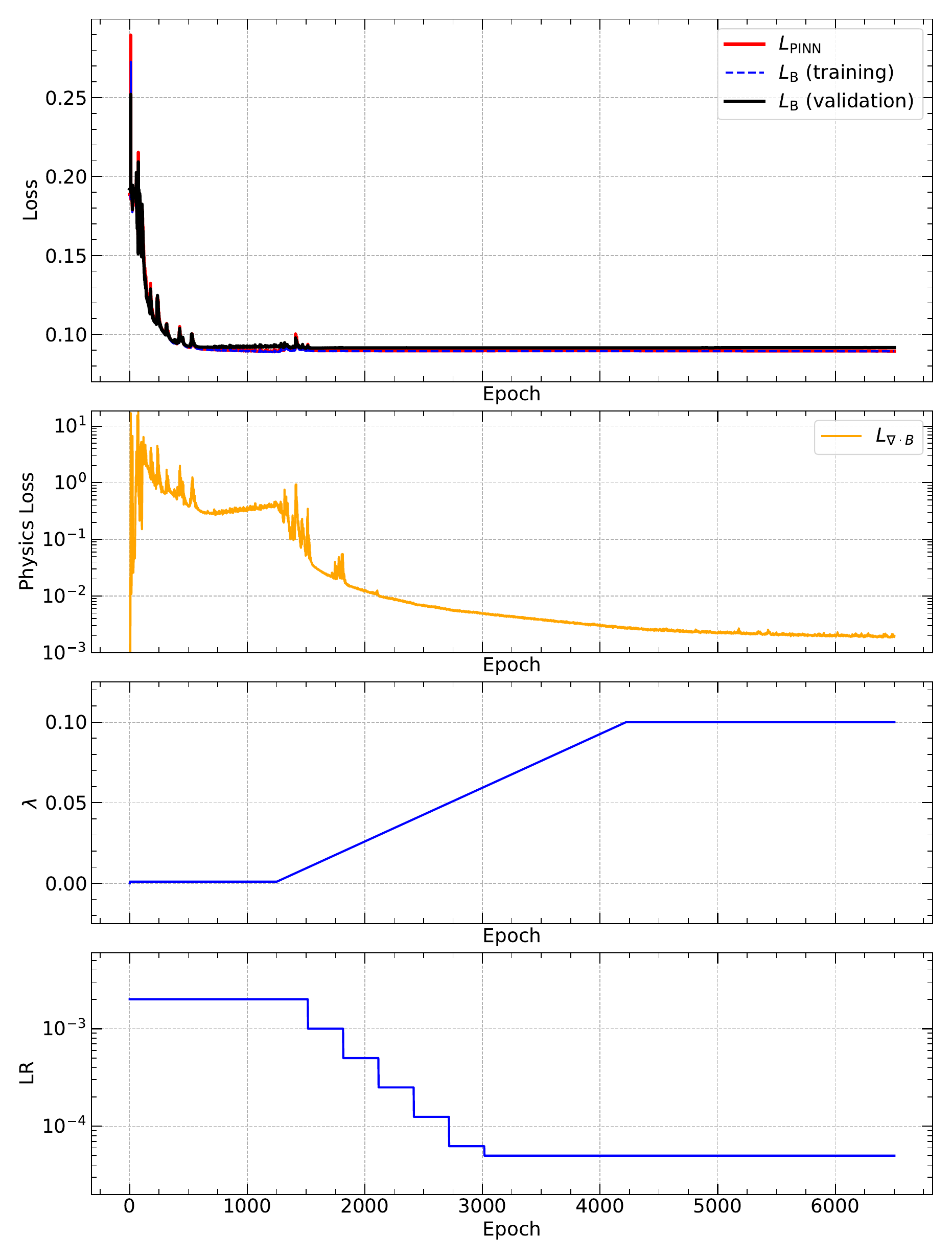}
\end{minipage}
\caption{Plots of losses relating to $\vec{B}$ accuracy (top row), physics loss (second row), physics constraint hyperparameter $\lambda$ (third row), and LR (fourth row) as a function of epoch throughout the training period.}
\label{fig:loss_lambda_LR_vs_epoch}
\end{figure}

After training the PINN, we study the features of the trained network in detail. The total loss function and its components, the hyperparameter $\lambda$, and the LR throughout the training are shown in Figure~\ref{fig:loss_lambda_LR_vs_epoch}.
The loss exhibits some small instabilities, particularly early on during the $\lambda$ annealing. Nonetheless, the overall trend is clearly convergent. Interestingly, during the pre-training $L_{\nabla \cdot B}$ initially increases rapidly while the network settles towards a small MSE loss. Before the pre-training concludes $L_{\nabla \cdot B}$ naturally begins to decrease given the very small non-zero value of $\lambda$ during pre-training. The network consistently improves the loss using the initial LR throughout pre-training and for the initial stage of $\lambda$ annealing. The LR then rapidly falls to the minimum value. Note that during the rapid decrease in LR the MSE loss is quite stable, while $L_{\nabla \cdot B}$ continues to be reduced. This is exactly the desired behavior of the $\lambda$ annealing: an initial training period that focuses on MSE loss, followed by an extended period of annealing and final training once $\lambda$ has reached $\lambda_{\mathrm{max}}$ that focuses on reducing unphysical features in the trained PINN.

\begin{figure}
\centering
\includegraphics[trim=0cm 0cm 0.0cm 0.0cm, width=0.45\textwidth]{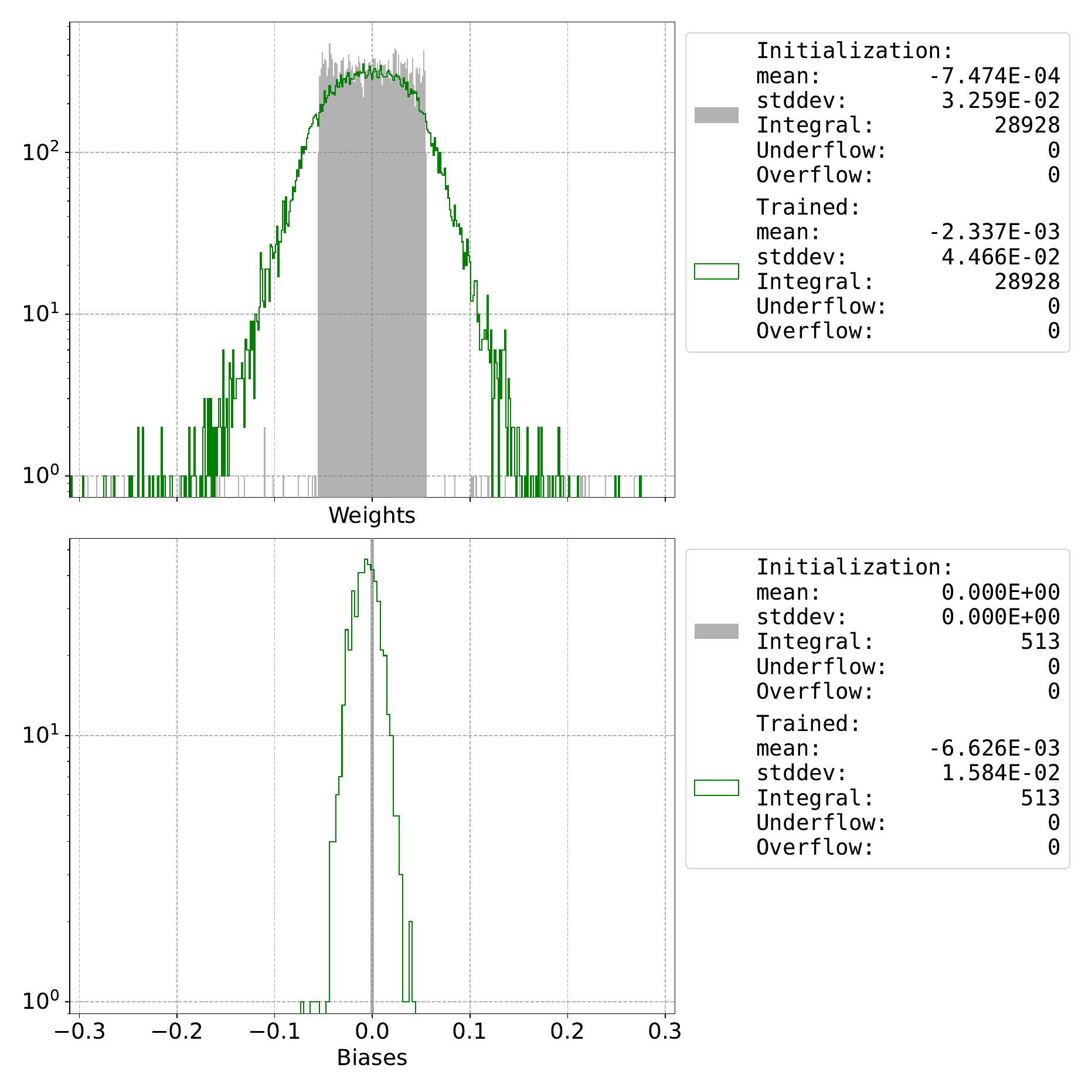}
\caption{Plots of weights (top) and biases (bottom) of the PINN. The initialization is represented by a gray shaded histogram while the values after training are represented by a green outline histogram.}
\label{fig:weights_biases}
\end{figure}

By inspection, we can ensure the weights and biases of the trained network are under control. Figure~\ref{fig:weights_biases} displays the weights and biases at initialization and after training. The weights are approximately Gaussian and are restricted to within approximately half a period of the DELTAsnake oscillation. Similarly, the biases are approximately Gaussian. Both the weights and biases display means that are non-zero but small.

\begin{figure}
\centering
\begin{minipage}[t]{0.45\textwidth}
\centering
\includegraphics[trim=0cm 0cm 0.0cm 0.0cm, width=1.0\textwidth]{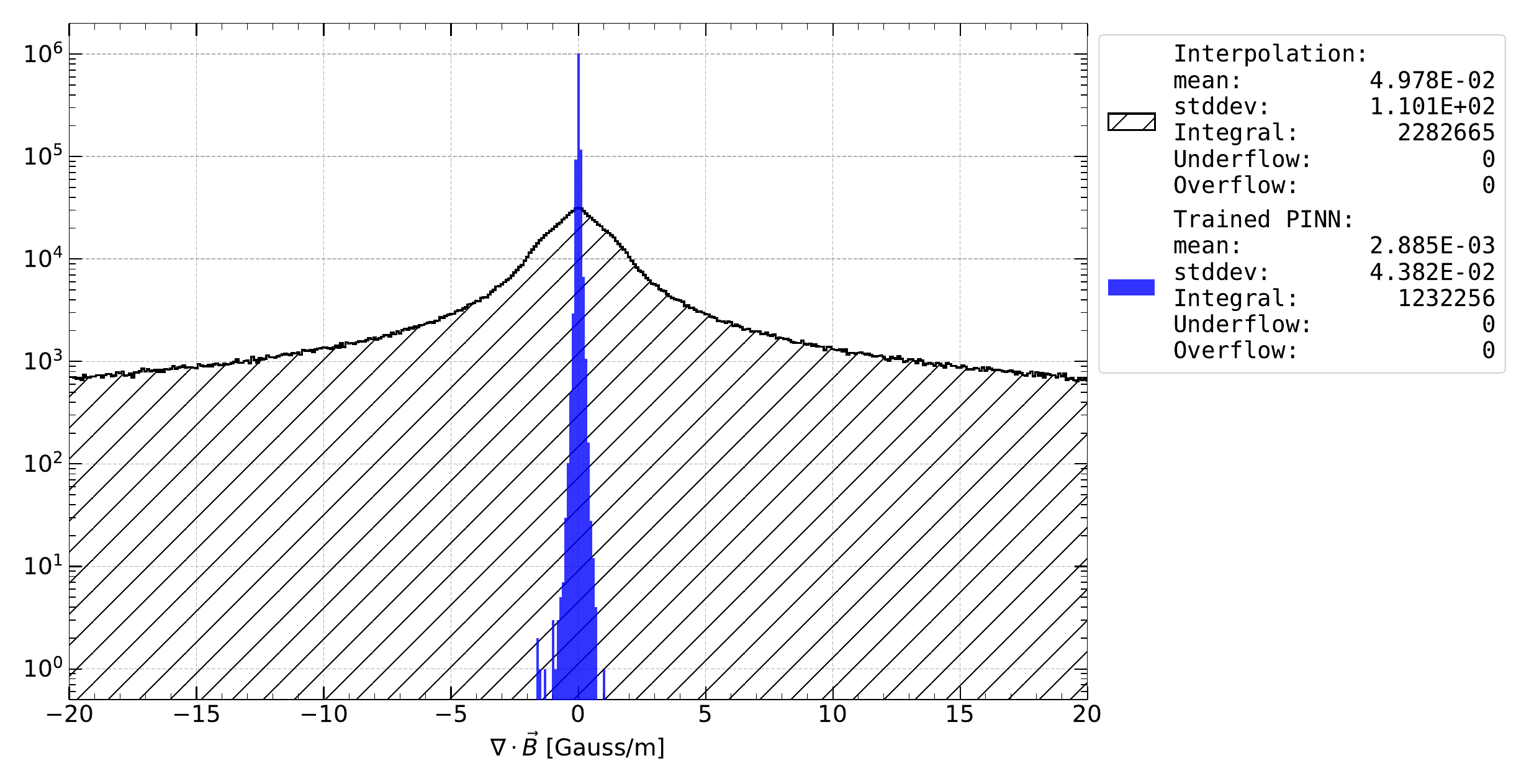}
\end{minipage}
\vspace{1em} 
\begin{minipage}[t]{0.45\textwidth}
\centering
\includegraphics[trim=0cm 0cm 0.0cm 0.0cm, width=1.0\textwidth]{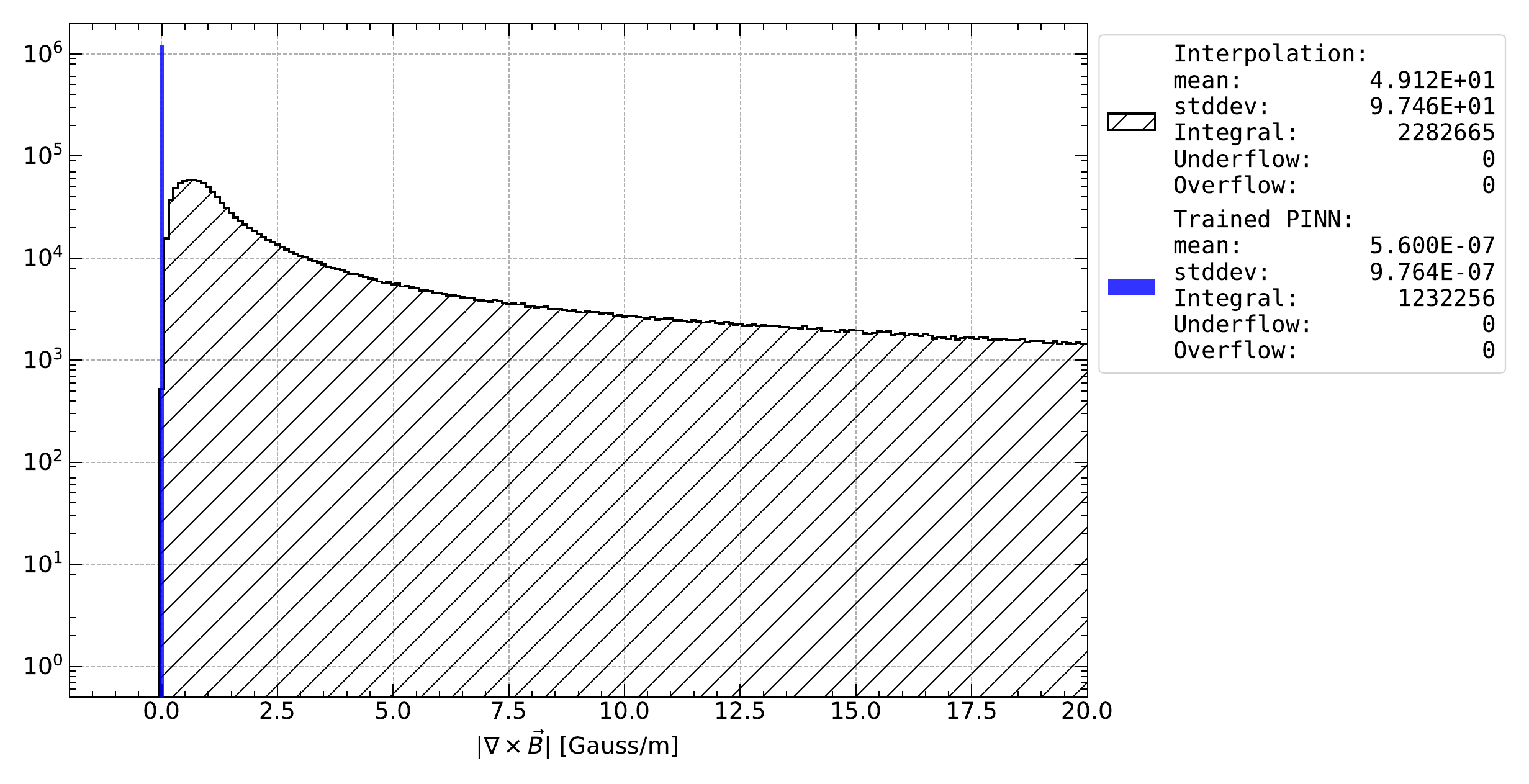}
\end{minipage}
\caption{Plots of divergence (top) and magnitude of the curl (bottom) of the trained PINN (blue bars) compared to derivatives introduced by linearly interpolating the magnetic field using the original field calculation evaluated at the test sample points (gray hatches). The former is evaluated at the test sample field points using automatic differentiation, while the latter is evaluated at locations sampled uniformly throughout the mapped volume. For clarity the $x$ axis of each plot is restricted to magnitudes less than $20$~Gauss/m, but the interpolation derivatives extend to approximately $\pm\numprint{1000}$~Gauss/m.}
\label{fig:div_curl_comparison}
\end{figure}

To quantify the consistency of the trained network with Maxwell's equations, we compare the divergence and the magnitude of the curl of the trained PINN evaluated at the test sample points to the divergence and curl induced by interpolating the same test dataset at randomly sampled points throughout the volume (Figure~\ref{fig:div_curl_comparison}). This interpolation scheme is used for physics simulations and analysis needed to interpret data from Mu2e. The standard deviation and extent of tails in the PINN are several orders of magnitude smaller than those introduced by interpolation. As the interpolation has been demonstrated to be sufficient for producing high quality physics simulations, we can conclude that the trained PINN with $\lambda_{\mathrm{max}}=0.1$ is well behaved physically.

\begin{figure}
    \centering
    \begin{minipage}[t]{0.45\textwidth}
        \centering
        \includegraphics[width=\linewidth]{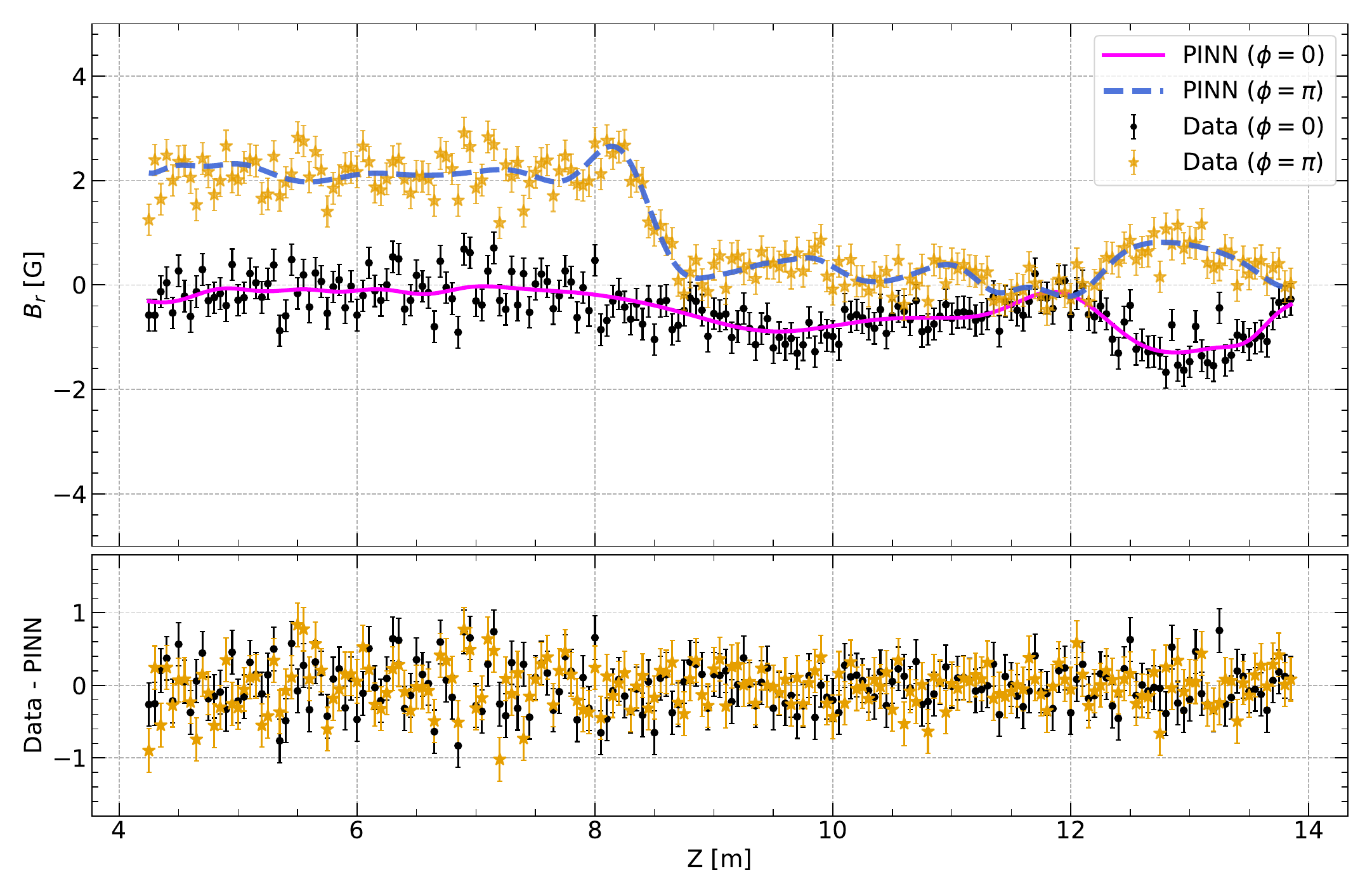}
    \end{minipage}
    
    \vspace{-0.5em} 
    
		\begin{minipage}[t]{0.45\textwidth}
        \centering
        \includegraphics[width=\linewidth]{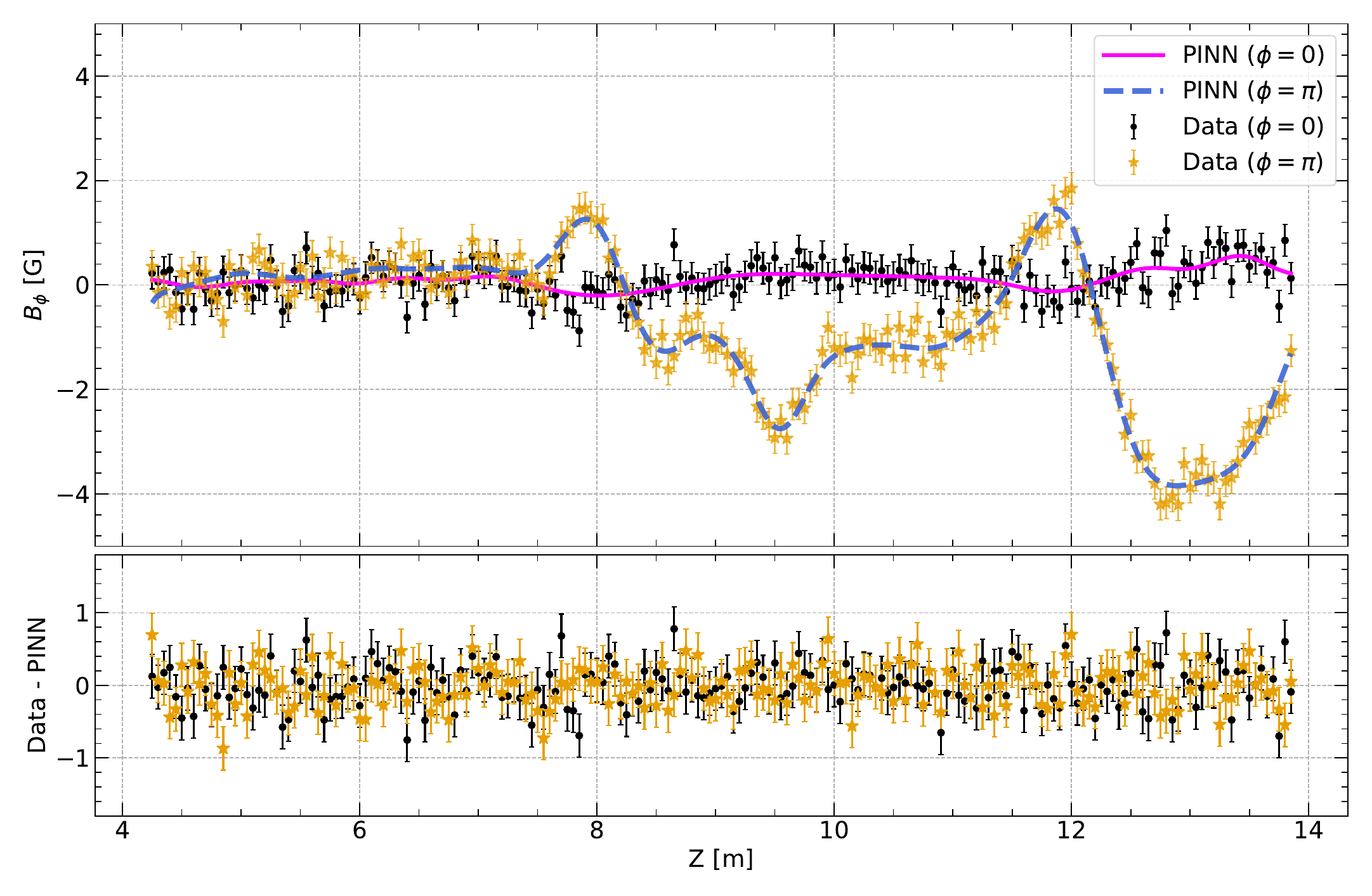}
    \end{minipage}
    
    \vspace{-0.5em} 

    \begin{minipage}[t]{0.45\textwidth}
        \centering
        \includegraphics[width=\linewidth]{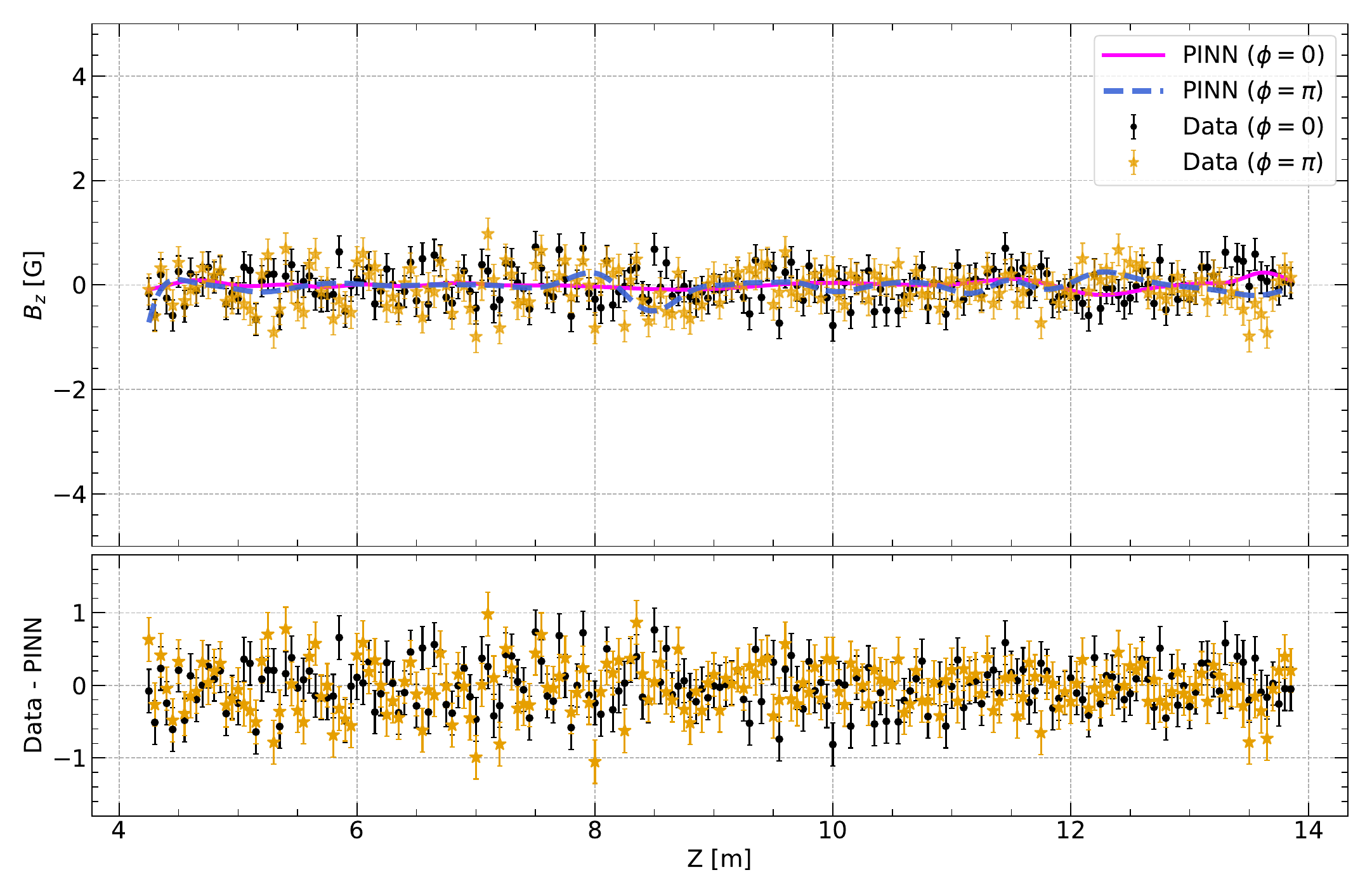}
    \end{minipage}
    \caption{Axial profiles at $\cylAzi=0,\pi$~rad, $r=0.8$~m comparing the PINN input and training evaluation. The three field components $B_r$, $B_{\cylAzi}$, and $B_z$ are displayed on the top, middle, and bottom, respectively. Each plot contains two panels. The top panel shows the least-squares residual (black and gold points) that are supplied as input and the trained PINN evaluation (magenta and blue lines). The bottom panel shows the residual of the trained PINN with the input.}
    \label{fig:PINN_profile_nominal}
\end{figure}

The behavior of the PINN is highlighted for the profiles at $\cylAzi=0,\pi$~rad at a radius $r=0.8$~m for the three cylindrical field components (Figure~\ref{fig:PINN_profile_nominal}). The PINN learns Gauss-level features in $B_r$ and $B_\cylAzi$ which have some sub-dominant oscillatory structures. The residual panels show unstructured residuals with a scale consistent with the measurement noise. There are very small features in the PINN for the axial component $B_z$, which is already well modeled by the least-squares fit.

In the following sections the prediction of the PINN is denoted as $\vec{B}_{\mathrm{PINN}}$.

\subsubsection{Least-Squares Iteration}
\label{subsubsec:results_leastsquares_iter}

After the PINN has been trained, the modeling is completed by a final iteration of the least-squares fitting. In this final iteration, the input data are prepared by subtracting the trained PINN evaluation from the original measurement:
\begin{equation}
\vec{B}_{\mathrm{res, NN}} = \vec{B}_{\mathrm{meas}} - \vec{B}_{\mathrm{PINN}}\ .
\end{equation}
This second residual field should be well modeled by the least-squares fit and may find small improvements in convergence since the structures of the field that the least-squares model originally struggled to reconstruct in the first least-squares fit are captured by the PINN. The iterated fit uses an identical set of parameters to the first fit, but are initialized to their best-fit values.

The resulting fits to $\vec{B}_{\mathrm{res, NN}}$, which we define as $\vec{B}_{\mathrm{LSQ,2}}$, yield a final fit quality statistic of ${\chi^2_{\mathrm{reduced}} = 1.034\pm0.005}$. The value for the toy included in the figures is ${\chi^2_{\mathrm{reduced}} = 1.031}$. The final field model is defined as the sum of the iterated least-squares fit and the evaluation of the trained PINN ($\vec{B}_{\mathrm{PINN}}$):
\begin{equation}
\vec{B}_{\mathrm{model}} = \vec{B}_{\mathrm{LSQ,2}} + \vec{B}_{\mathrm{PINN}}\ .
\end{equation}

The condensed picture of the fit quality is presented in histograms of field component residuals across all points of the measurement sample in Figure~\ref{fig:residuals_df_meas_nominal}. The first least-squares fit does a modest job of reconstructing the measured field. Training the PINN brings the residuals in line with the underlying measurement noise model. Small gains are seen in the final least-squares fit iteration.

While studying fit residuals are instructive and help us understand modeling behavior, what we ultimately care about is the fidelity of the final model with respect to the true, underlying magnetic field. In other words, we want to know how well the model predicts the field in the absence of measurement noise. The component residuals evaluated on the test sample provide this information (Figure~\ref{fig:residuals_df_test_nominal}). Due to the strong physical constraints in the model as well as strong ability to generalize, the RMS of the residuals on the test sample are approximately five times smaller than the measurement noise. Recalling that efforts in recent experiments have resulted in modeling accuracies at the $10^{-3}$ level, we conclude our modeling procedure can outperform alternative methods applied to other large bore solenoidal field by several orders of magnitude.

The remainder of Section~\ref{sec:results} is dedicated to robustness tests of this modeling technique.

\subsection{Systematic Effect: Hall Probe Bias}
\label{subsec:results_HP_bias}

To consider a realistic systematic effect in the data, we consider an overall field strength scaling systematic applied independently to each Hall probe. We consider the case where the scale error is uncorrelated between probes; each scale factor is drawn from a Gaussian distribution with a mean of one and a standard deviation of $1\times10^{-4}$. This choice is a reflection of the requirements set by physics goals of the Mu2e experiment~\cite{FMS2018} and is expected to be achieved in the calibration.
Each scale factor $s_i$ is applied to the points in the measurement sample that are collected from Hall probe $i$ to create the modified measurements $\tilde{\vec{B}}$:
\begin{equation}
\tilde{\vec{B}}_{\mathrm{meas}, i} = s_i\ \vec{B}_{\mathrm{meas}, i}
\end{equation}
while the point measurements in the test sample are left unmodified. A summary of the scale factors from the particular draw for this example are provided in Table~\ref{tab:HPC_sfs}. The non-zero sample mean $\overline{s_i - 1} = 2.2 \times 10^{-5}$ should result in a small bias in the final model. If all steps of modeling are on firm ground, there should be no major consequence of the calibration scale factors.

\begin{figure*}
    \centering
    \begin{minipage}[t]{0.9\textwidth}
        \centering
        \includegraphics[width=\linewidth]{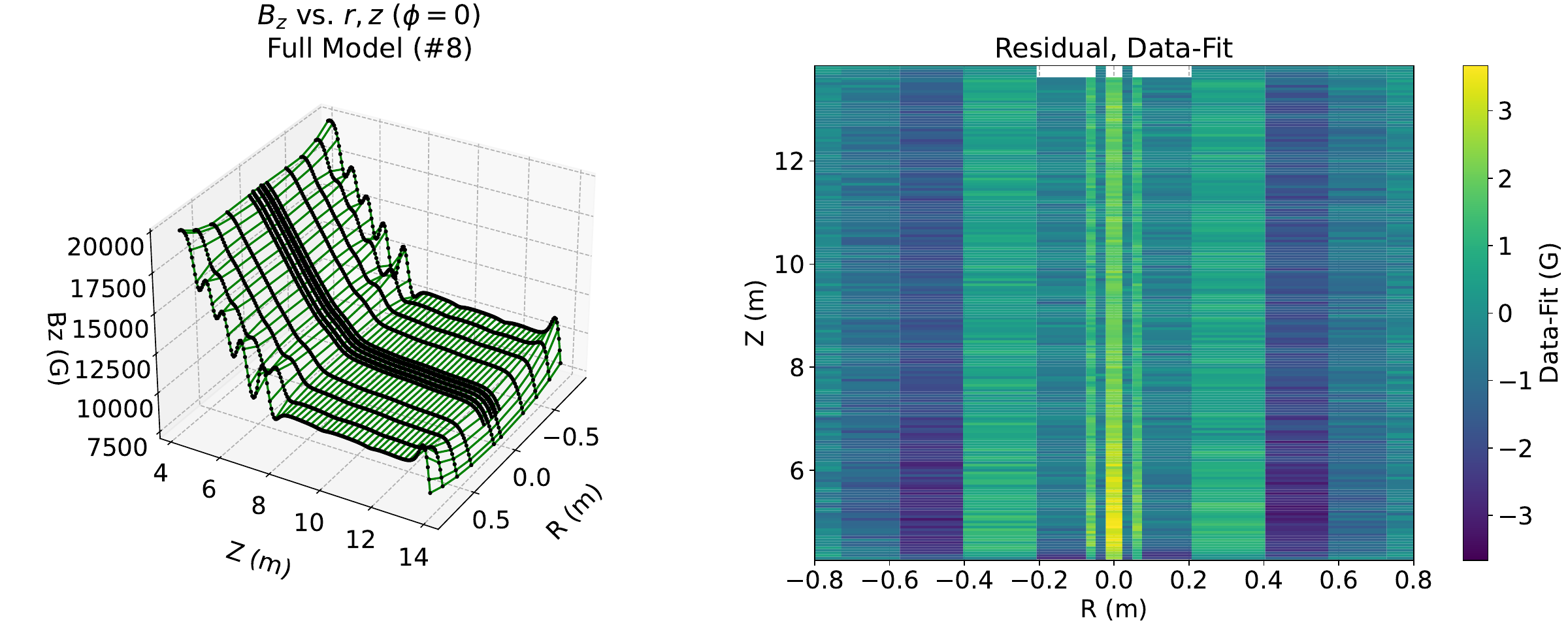}
    \end{minipage}
    \caption{Results of the full modeling applied to the Hall probe calibration systematic test case for $B_z$. The left panel shows the data (black points) and fit evaluated at the measurement points (green wireframe) for the $\cylAzi=0$~rad plane of the measurement sample. The right panel uses color to indicate the residual for the same plane of data. The residual pattern indicates a resistance of the model to fitting the bias in the probes.}
    \label{fig:FullModel_fit_HPC}
\end{figure*}

\begin{figure}
    \centering
    \begin{minipage}[t]{0.45\textwidth}
        \centering
        \includegraphics[width=\linewidth]{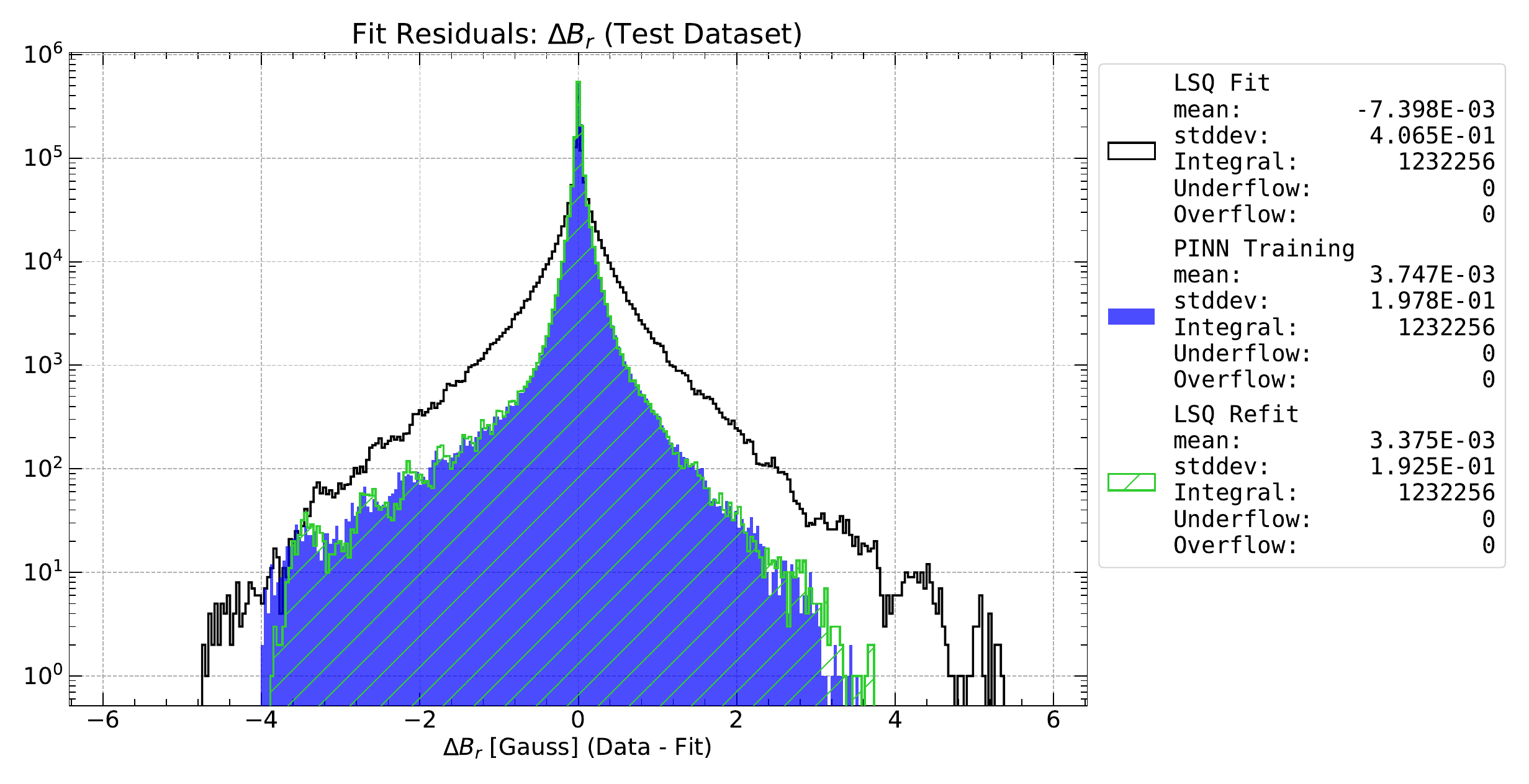}
    \end{minipage}
    
    \vspace{-0.5em} 
    
		\begin{minipage}[t]{0.45\textwidth}
        \centering
        \includegraphics[width=\linewidth]{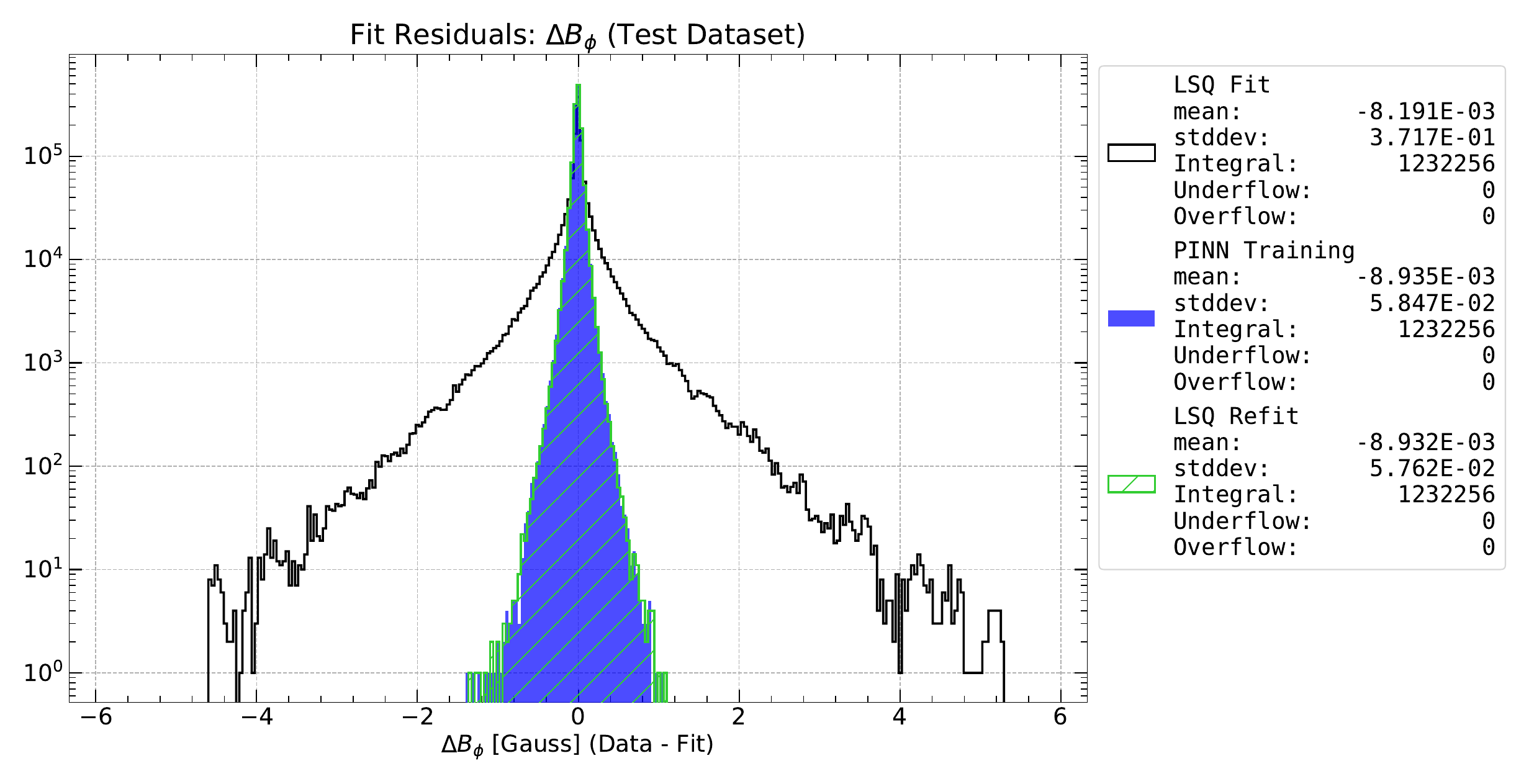}
    \end{minipage}
    
    \vspace{-0.5em} 

    \begin{minipage}[t]{0.45\textwidth}
        \centering
        \includegraphics[width=\linewidth]{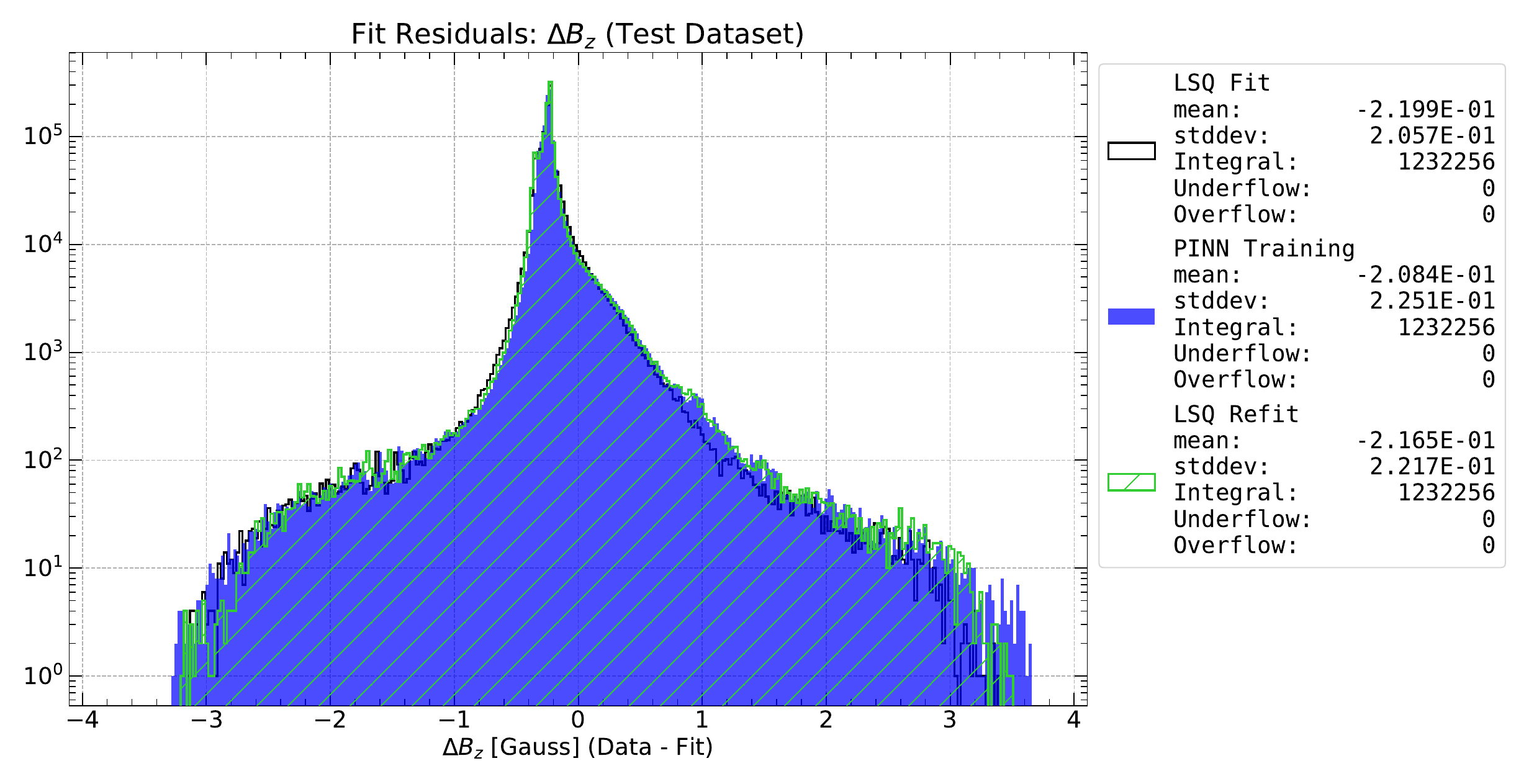}
    \end{minipage}
    \caption{Residual plots, data - fit, for the three field components $B_r$ (top), $B_\cylAzi$ (middle), $B_z$ (bottom) evaluated at all points in the test sample for the Hall probe calibration systematic test. The residuals after each step of the modeling are included: least-squares fit (black outline), PINN training (blue filled), and least-squares iteration (green hatch).}
    \label{fig:residuals_df_test_HPC}
\end{figure}

Figures~\ref{fig:FullModel_fit_HPC}~and~\ref{fig:residuals_df_test_HPC} summarize the result of the full modeling procedure and include a plot of $B_z$ in the $\cylAzi=0$~rad plane and a plot of residuals on the test samples, respectively. The initial least-squares fit results in a fit statistic ${\chi^2_{\mathrm{reduced}} = 8.3}$, and ${\chi^2_{\mathrm{reduced}} = 7.3}$ for the final model. The increased $\chi^2$ is dominated by the model successfully resisting the bias, which appears most strongly in $B_z$. The residual histograms for the test sample points indicate a bias in $B_z$, as expected. Overall the model is still of high quality; neither $B_r$ nor $B_\cylAzi$ are biased, and the RMS of the residuals are below the noise level of the probes.

\begin{table}
\center
\caption{Hall probe radial locations and the deviations of scale factors $s_i$ from unity for the systematic measurement
error test case. ``BP'' refers to Hall probes on the Big Propeller, while ``SP'' refers to Hall probes on the Small Propeller\cite{FMS2018}.}
\bgroup
\def\arraystretch{1.3}
\begin{tabular}{l|c|r}
\textbf{Probe Label} & \textbf{Radius [m]} & \multicolumn{1}{c}{$\mathbf{s_i - 1}$} \\\hline
SP1 &  $0.000$ & $2.15\times 10^{-4}$\\
\hline
BP1 &  $0.044$ & $-9.24\times 10^{-6}$\\
\hline
SP2 &  $0.054$ & $1.39\times 10^{-4}$\\
\hline
SP3 &  $0.095$ & $-2.07\times 10^{-5}$\\
\hline
BP2 &  $0.319$ & $7.34\times 10^{-5}$\\
\hline
BP3 &  $0.488$ & $-1.44\times 10^{-4}$\\
\hline
BP4 &  $0.656$ & $-6.63\times 10^{-5}$\\
\hline
BP5 &  $0.800$ & $-1.01\times 10^{-5}$
\end{tabular}
\egroup
\label{tab:HPC_sfs}
\end{table}

\subsection{Reduction in Number of Measurements}
\label{subsec:results_reduced_meas}

To assess the value added by the PINN, we pose the question: what unique benefits does the PINN offer compared to traditional methods? Drawing once again upon insights from the Nyquist-Shannon limit and basic Fourier analysis, we observe that certain features in the Mu2e DS field are inherently difficult to reconstruct using a truncated cylindrical harmonic series. One such example, visible in the Figure~\ref{fig:PINN_profile_nominal}, is a shoulder feature in $B_r$ at $\cylAzi=\pi$~rad, $8 \leq z \leq 9$~m, which does not appear at $\cylAzi=0$ and is not captured by the least-squares fit. However, the PINN is able to model this feature successfully.

This observation raises the possibility that broader and subtler structures -- those that are smooth or slowly varying in space -- may still require a large number of harmonic terms to represent accurately in a truncated cylindrical harmonic series. This is analogous to the classic difficulty of approximating a step function with a truncated Fourier series; although the feature appears coarse, the sharp, localized spatial transition demands many basis functions to resolve. As a result, efforts to reconstruct such features using a truncated least-squares model may be inefficient. The PINN, by contrast, can learn these structures directly from data without relying on an explicit basis, making it more effective at capturing such behavior with fewer learned parameters. This strength in turn could reduce the required spatial sampling density.

This hypothesis has practical implications; a reduction in required measurements translates directly into a faster measurement campaign.

We construct a reduced dataset with half as many mapper set points in both azimuthal angle $\cylAzi$ and axial position $z$. We ensure the least-squares model still satisfies the Nyquist-Shannon limit by setting $m_{\mathrm{max}} = 54$ and $n_{\mathrm{max}} = 2$. The PINN is trained using the same hyperparameter values optimized for the nominal case.

\begin{figure}
    \centering
    \begin{minipage}[t]{0.45\textwidth}
        \centering
        \includegraphics[width=\linewidth]{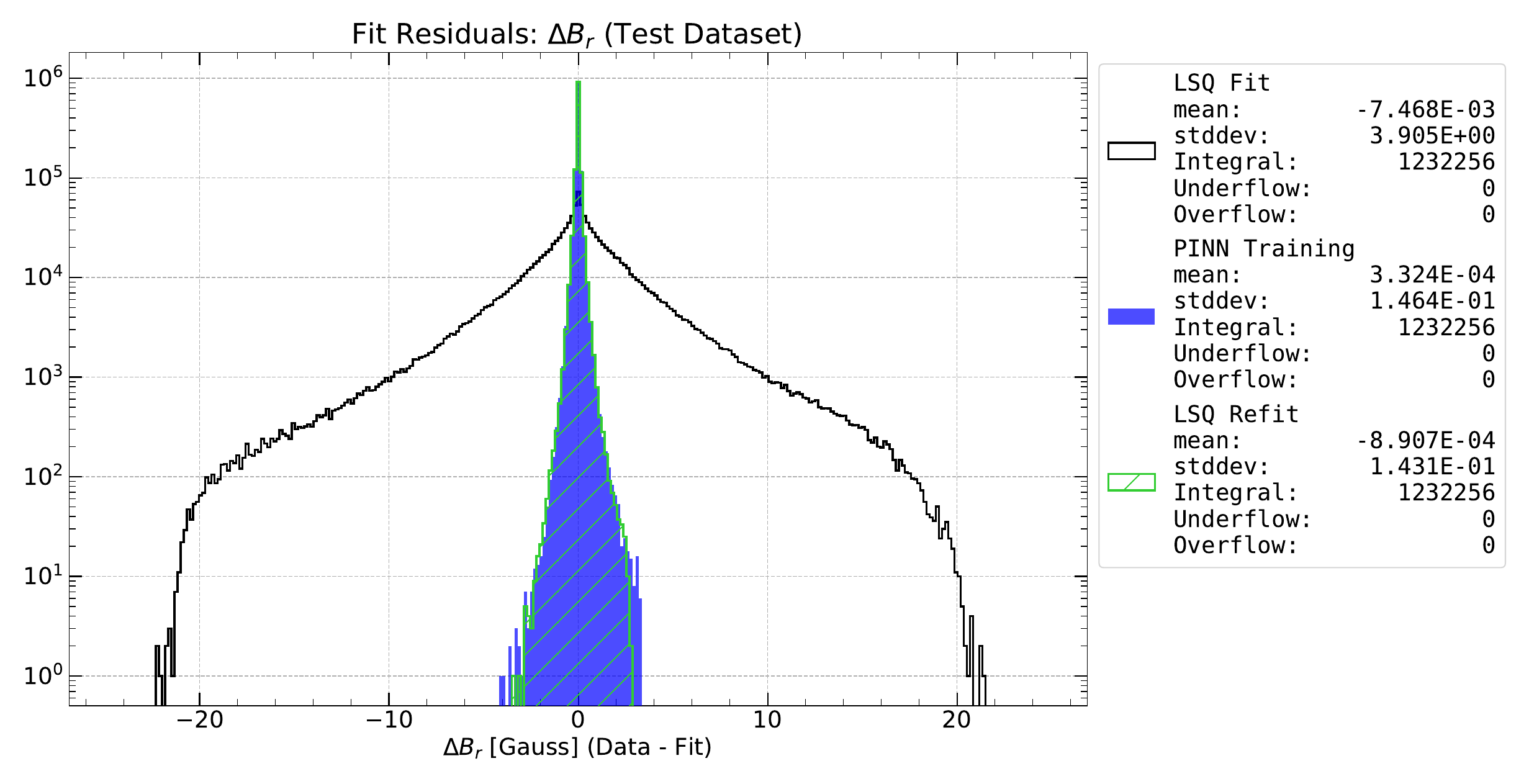}
    \end{minipage}
    
    \vspace{-0.5em} 
    
		\begin{minipage}[t]{0.45\textwidth}
        \centering
        \includegraphics[width=\linewidth]{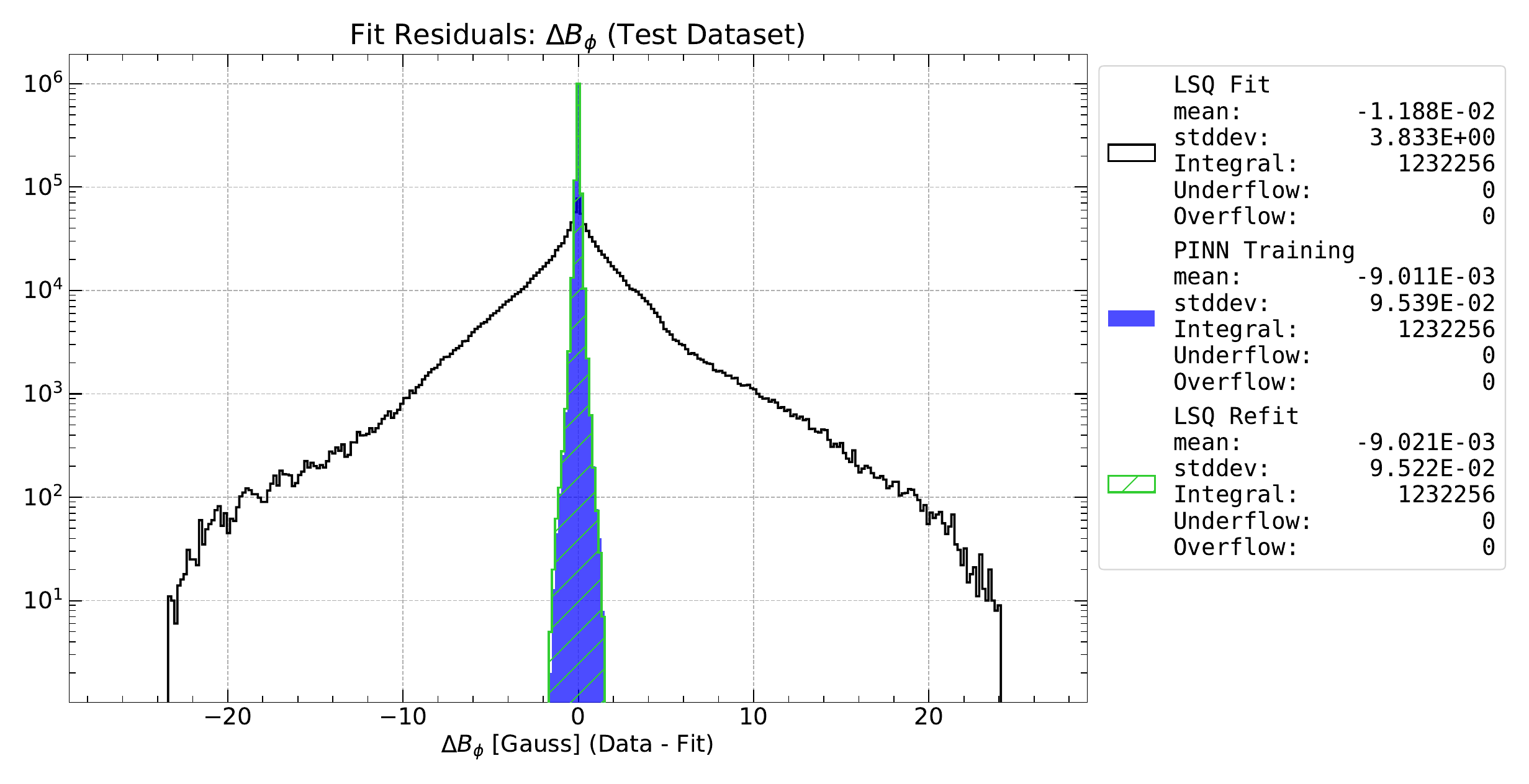}
    \end{minipage}
    
    \vspace{-0.5em} 

    \begin{minipage}[t]{0.45\textwidth}
        \centering
        \includegraphics[width=\linewidth]{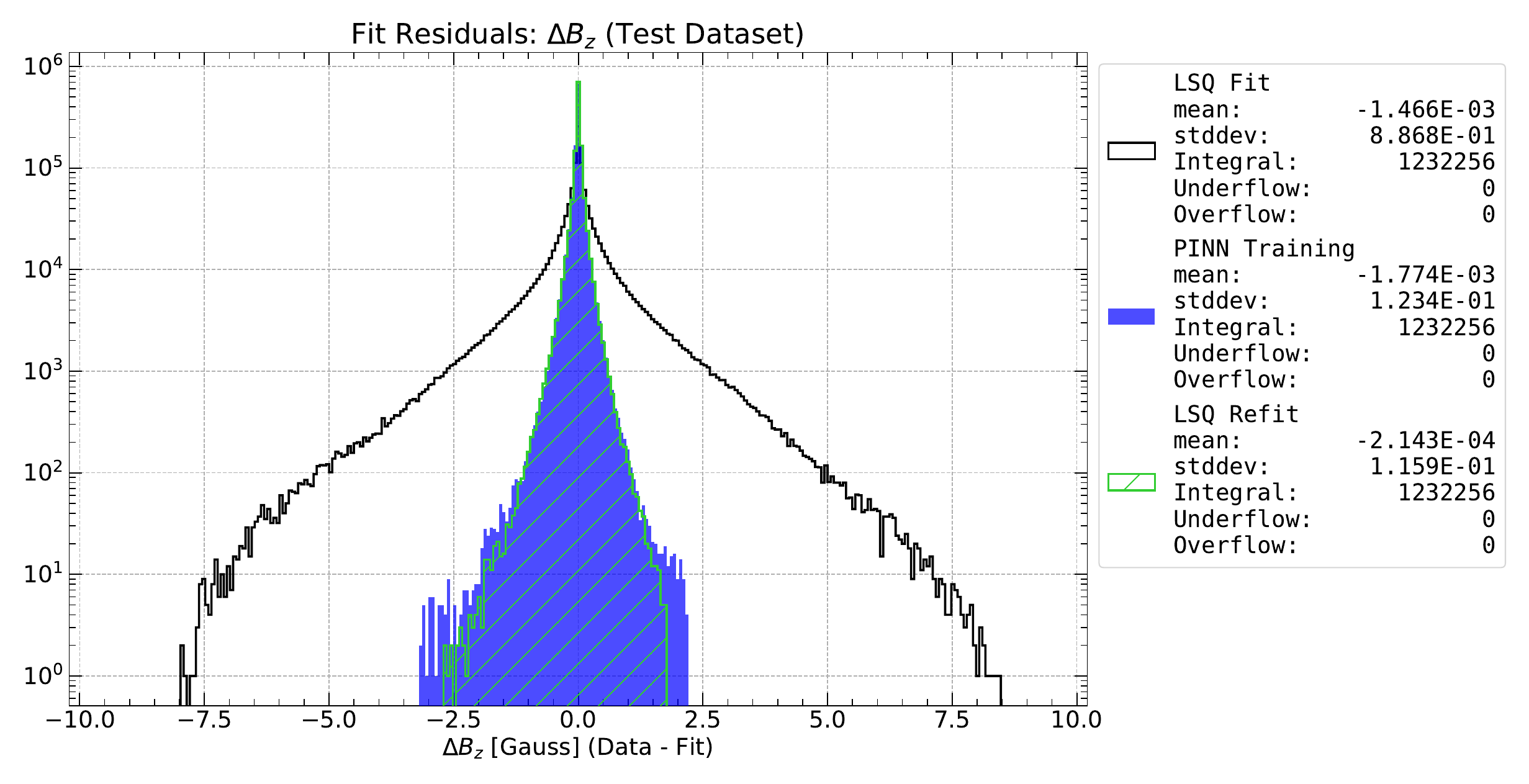}
    \end{minipage}
    \caption{Residual plots, data - fit, for the three field components $B_r$ (top), $B_\cylAzi$ (middle), $B_z$ (bottom) evaluated at all points in the test sample when fitting with a sparse measurement sample. The residuals after each step of the modeling are included: least-squares fit (black outline), PINN training (blue filled), and least-squares iteration (green hatch).}
    \label{fig:residuals_df_test_sparse}
\end{figure}

Figure~\ref{fig:residuals_df_test_sparse} shows the residuals on the test sample under this reduced measurement sample scenario. The initial least-squares fit yields a poor fit quality with ${\chi^2_{\mathrm{reduced}} = 77}$, but after the PINN training, the final model achieves ${\chi^2_{\mathrm{reduced}} = 1.12}$. The residual histograms show that while the model is slightly weaker than in the nominal case, the RMS of the residuals remains well below the measurement noise, and the tails are very minor (typically within a few Gauss).

\begin{figure}
    \centering
    \begin{minipage}[t]{0.45\textwidth}
        \centering
        \includegraphics[width=\linewidth]{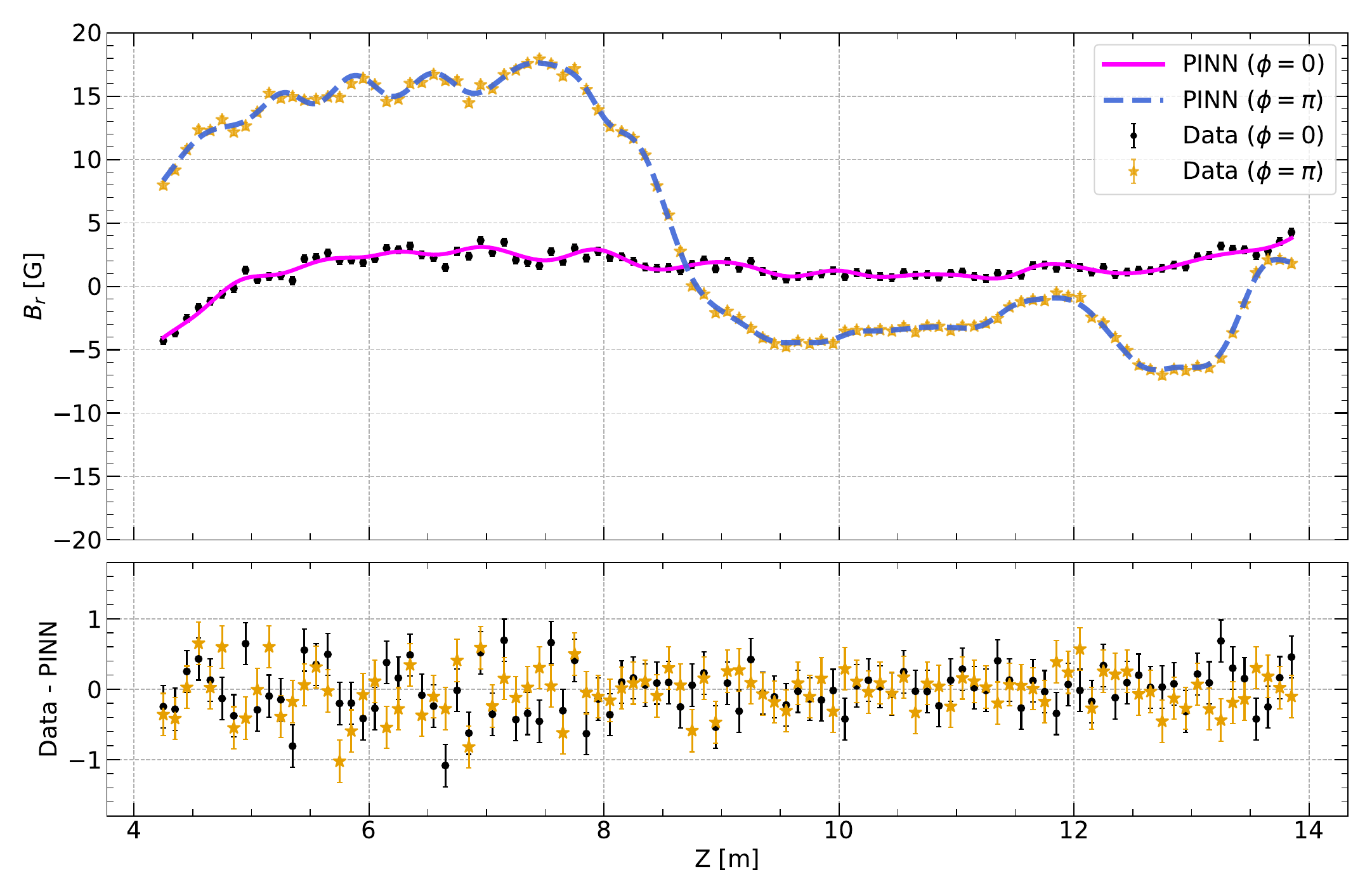}
    \end{minipage}
    
    \vspace{-0.5em} 
    
		\begin{minipage}[t]{0.45\textwidth}
        \centering
        \includegraphics[width=\linewidth]{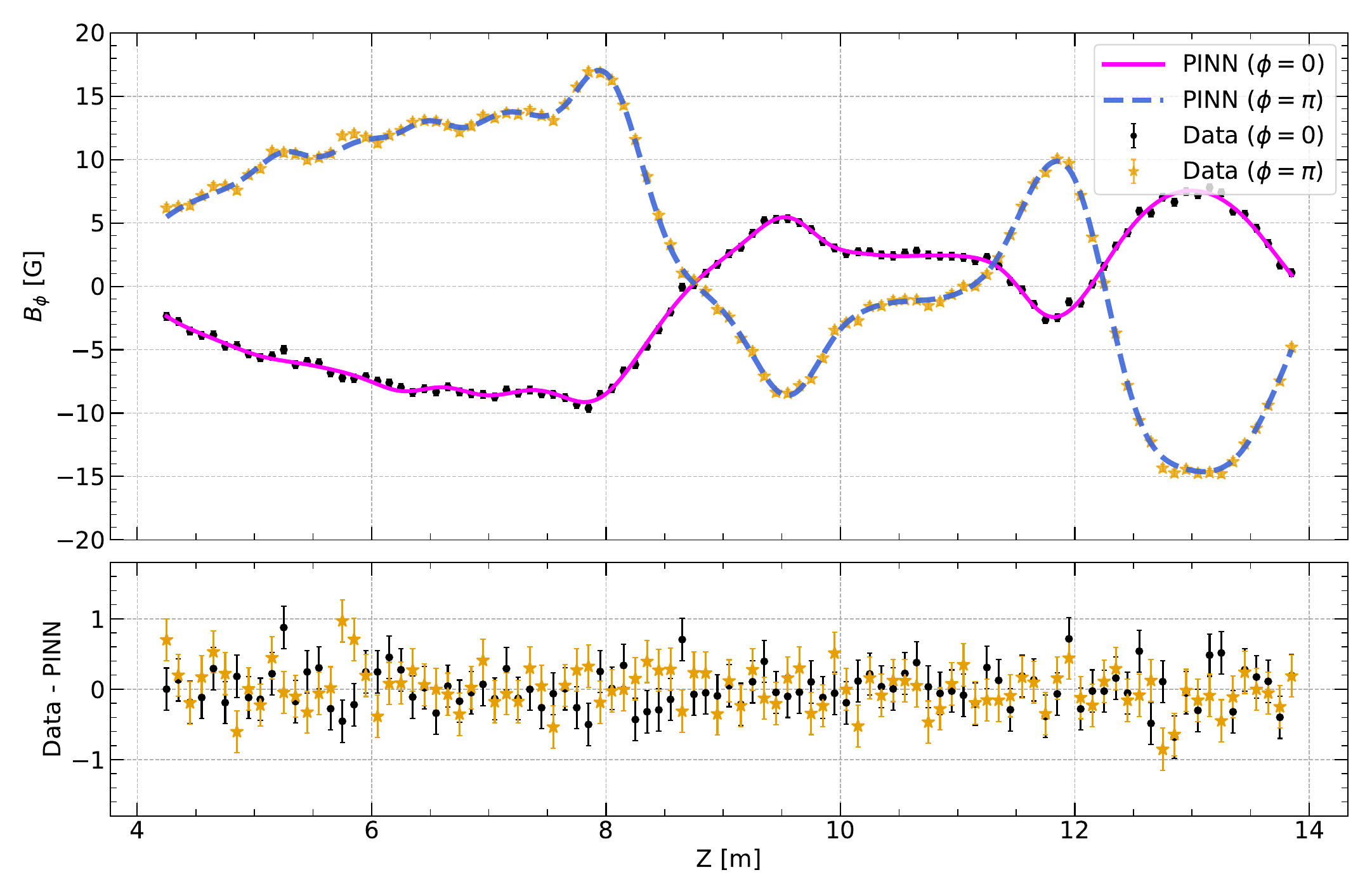}
    \end{minipage}
    
    \vspace{-0.5em} 

    \begin{minipage}[t]{0.45\textwidth}
        \centering
        \includegraphics[width=\linewidth]{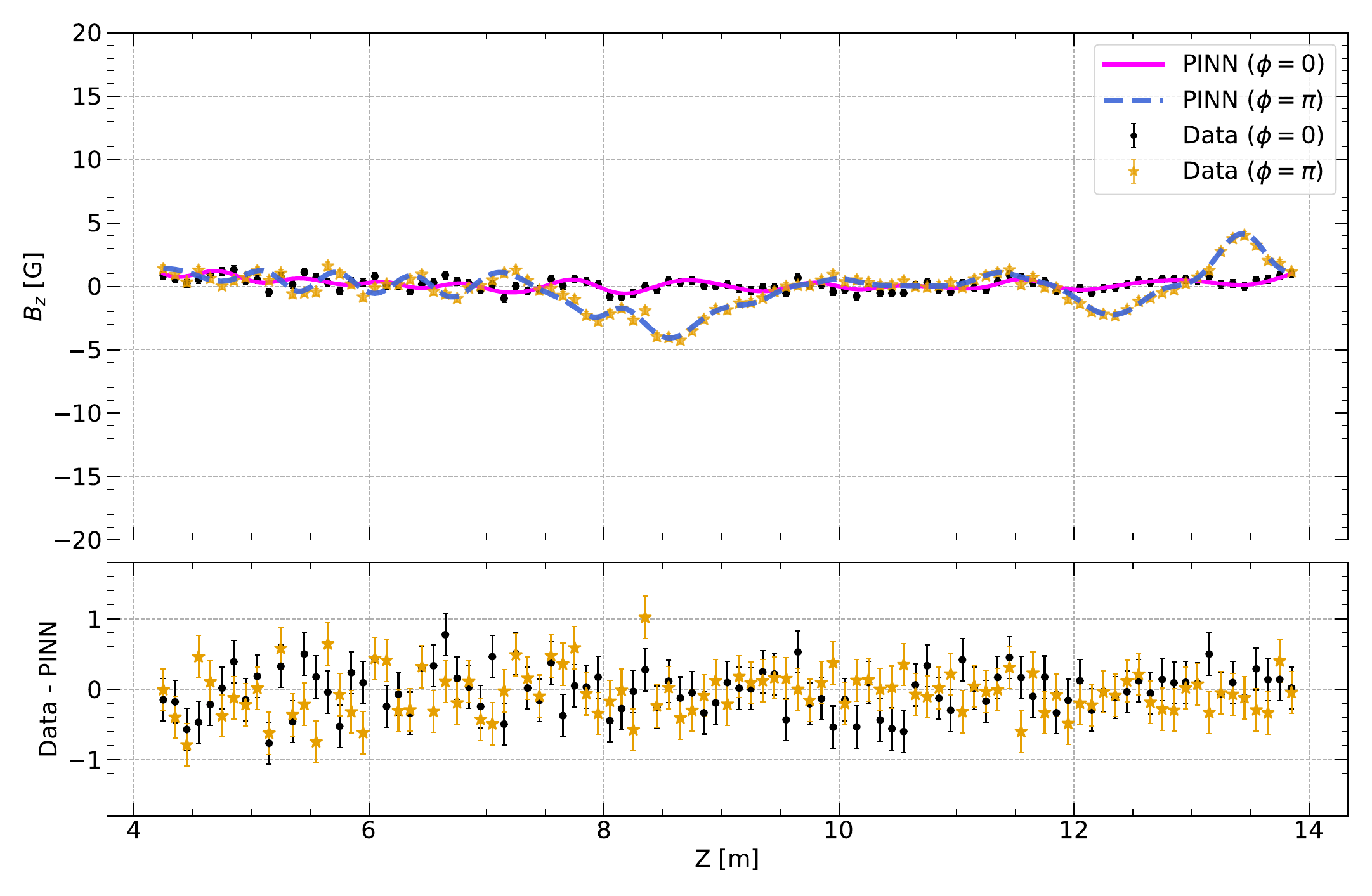}
    \end{minipage}
    \caption{Axial profiles at $\cylAzi=0,\pi$~rad, $r=0.8$~m comparing the PINN input and training evaluation when the PINN is trained on sparse data. The three field components $B_r$, $B_{\cylAzi}$, and $B_z$ are displayed on the top, middle, and bottom, respectively. Each plot contains two panels. The top panel shows the least-squares residual (black and gold points) that are supplied as input and the trained PINN evaluation (magenta and blue lines). The bottom panel shows the residual of the trained PINN with the input.}
    \label{fig:PINN_profile_sparse}
\end{figure}

To further highlight the PINN’s strength, Figure~\ref{fig:PINN_profile_sparse} shows the $r=0.8$~m profiles at $\cylAzi=0,\pi$~rad, allowing direct comparison to the nominal model (Figure~\ref{fig:PINN_profile_nominal}). The PINN successfully captures broader, semi-periodic field features across all components. Notably, the shoulder in $B_r$ is successfully modeled despite a larger magnitude and more complex structure; its amplitude increases by over an order of magnitude and exhibits sub-dominant oscillations. A comparable sawtooth-like structure is visible in $B_\cylAzi$, underscoring the PINN's ability to easily reconstruct the same feature that would require many more terms in the least-squares model.

In practice, the choice of spatial sampling density requires robust studies of any potential downsides of reducing the number of measurements, e.g. increasing impact of systematic effects. We have neglected that here and instead completed a basic demonstration using the nominal measurement without considering systematics.

\section{Summary and Conclusions}
\label{sec:summary}

We have demonstrated a hybrid approach for modeling the magnetic field inside the Mu2e DS bore, combining a truncated cylindrical harmonic least-squares model with a PINN trained on the residuals. This approach results in a final model that is both physically consistent and highly accurate, achieving residuals well below the injected measurement noise and a $\chi^2_{\mathrm{reduced}}$ near unity. The standard deviation of the residuals of the model evaluation compared to the true field is orders of magnitude better than the requirements driven by Mu2e's physics goals.

The PINN architecture, which incorporates both hard and soft constraints from Maxwell’s equations, proves essential in capturing asymmetric and localized field features that are poorly described by the cylindrical harmonic series alone. The DELTAsnake activation function, introduced here, enhances the PINN’s ability to learn structured, oscillatory residuals while maintaining training stability and physical consistency.

Robustness studies show that the model resists calibration systematics and can maintain accuracy even when trained on a substantially reduced dataset.

While this study focuses on the Mu2e DS as a test case, the techniques developed here should be broadly applicable to solenoidal field environments where measurement sparsity, geometric complexity, and physical consistency are key concerns over large spatial volumes. The authors also note the generic applicability of the iterative training, wherein either modeling technique alone cannot sufficiently model the data but the iteration of the two is especially powerful.

\section*{Supplementary Material}
\label{sec:supplementary_material}
See supplementary material for additional details relating to: the \texttt{helicalc} package; the homogeneous magnetic material model; details of the least-squares model and hyperparameter optimization; the PINN hyperparameter optimization; the estimate of the GDF of the PINN.

\section{Acknowledgments}
\label{sec:acknowledgments}

We are grateful for the vital contributions of the Fermilab staff and the technical staff of the participating institutions. This work was supported by the US Department of Energy; the Istituto Nazionale di Fisica Nucleare, Italy; the Science and Technology Facilities Council, UK; the Royal Society, UK; the Leverhulme Trust, UK; the Ministry of Education and Science, Russian Federation; the National Science Foundation, USA; the National Science Foundation, China; the Helmholtz Association, Germany; the Julian Schwinger Foundation, USA; the EU Horizon 2020 Research and Innovation Program under the Marie Sk\l{}odowska-Curie Grant Agreement Nos.\ 858199, 101003460, 101006726 and 101230211; and the Horizon Europe Research and Innovation Program under the Marie Sk\l{}odowska-Curie Grant Agreement No.\ 101234557. This document was prepared by members of the Mu2e Collaboration using the resources of the Fermi National Accelerator Laboratory (Fermilab), a U.S.\ Department of Energy, Office of Science, Office of High Energy and Nuclear Physics HEP User Facility. Fermilab is managed by FermiForward Discovery Group, LLC, acting under Contract No.\ 89243024CSC000002.

\section*{Author Declarations}
\label{sec:auth_decl}
\subsection*{Conflict of Interest}
The authors have no conflicts to disclose.

\subsection*{Author Contributions}
\noindent
\textbf{Cole Kampa:} Conceptualization (shared); Data curation (shared); Methodology (lead); Formal analysis (lead); Investigation (lead); Visualization (lead); Software (lead); Writing -- original draft (lead); Writing -- review \& editing (shared). \textbf{Susan Dittmer:} Data curation (supporting); Methodology (supporting); Writing -- review \& editing (shared).
\textbf{Henry Glass:} Data curation (shared); Software (supporting); Writing -- review \& editing (shared).
\textbf{Michael Schmitt:} Conceptualization (shared); Supervision (lead); Funding acquisition (lead); Writing -- review \& editing (shared).

\section*{Data Availability}

The data that support the findings of this study are openly available in Zenodo at \\\href{https://doi.org/10.5281/zenodo.21983988}{https://doi.org/10.5281/zenodo.21983988}, Ref.~\cite{Kampa_Zenodo_2026}.

The code that supports the findings of this study are openly available in GitHub at \\
\href{https://github.com/Mu2e/docdb-48750/tree/v1.0}{https://github.com/Mu2e/docdb-48750/tree/v1.0}, Ref.~\cite{docdb_repo}.

\bibliography{refs}

\clearpage

\section*{Supplementary Material}
\section{The \texttt{helicalc} Package}
\label{sec:helicalc_package}

We summarize the mathematical formalism and numerical techniques utilized in the \texttt{helicalc} package. Section~\ref{app:helicalc_math} derives the various integrals that the package can solve. Section~\ref{app:helicalc_numeric} elaborates the numerical integration techniques used.
The \texttt{helicalc} package is available in the \texttt{docdb-48750} repository in the \texttt{FMS\_BFieldModel/helicalc\_package} directory~\cite{docdb_repo}. Once installed, the package is invoked in a Python interpreter or script by importing \texttt{helicalc}. A note of the class or method associated with each section below is included for reference.

\subsection{Mathematical Formalism}
\label{app:helicalc_math}

\subsubsection{Notation}
The following notation is used in the rest of this section:
\begin{itemize}
\item $\vec{x} = (x_p, y_p, z_p)$: field point at which the integral is being calculated
\item $\vec{J} = \vec{J}(x, y, z)$: current density per unit area from some source
\item $\vec{r} = (r_x, r_y, r_z)$: vector from a differential volume of current to the field point, i.e. the vector quantity $\vec{x} - \dif \vec{V}$.
\end{itemize}

\subsubsection{Biot-Savart Law}
The Biot-Savart law describes the magnetic flux density generated by a spatial current density that is constant in time \cite{jackson_classical_1999}. In three-dimensions, the field due to current density $\vec{J}$ (e.g. from a superconducting cable) is:
\begin{equation}
\label{eq:biotsavart3D}
\vec{B}(\vec{x}) = \frac{\mu_0}{4 \pi} \iiint_V \frac{\vec{J}\times \vec{r}}{r^3} \dif V
\end{equation}
where $\mu_0$ is the free space permeability and the volume $V$ contains all of the current in the conductor(s). While $\vec{J}$ in principle can be spatially varying, each conductor element of the Mu2e conductor network is given a constant current density which is calculated using the total current $I$ flowing through the conductor and the conductor's cross-section. The current $I$ and conductor cross-section parameters (width $w$, height $h$) are supplied to \texttt{helicalc}; the calculation of $J$ is done internally. The current is always assumed to flow in a direction normal to the conductor cross-section at any point along the conductor.

The derivations and implementation in \texttt{helicalc} use a local coordinate system to simplify the integrals. In general, a translation vector $\vec{x}_c = (x_c, y_c, z_c)$ and three Euler angles are supplied to the integrator for a given conductor so that the integration in the local coordinates can be transformed to the global (i.e. experiment) coordinates. 
The procedure to calculate the field is as follows:
\begin{enumerate}
\item Transform the global field point to the local coordinates of the conductor: $\vec{x} \rightarrow \vec{x}'$
\item Compute the integral to find the field vector in the local frame: $\vec{B}' = \vec{B}(\vec{x}')$
\item Transform the field vector from the local to the global frame: $\vec{B}' \rightarrow \vec{B}$
\end{enumerate}

The values for $\vec{B} = (B_x, B_y, B_z)$ calculated at a number of field points are saved individually for each conductor and are stored in a pandas DataFrame and saved in the serialized pickle format \cite{reback2020pandas, mckinney-proc-scipy-2010}. A post-processing step combines the contribution from the $N$ conductors:
\begin{equation}
\vec{B}(x, y, z)_{\mathtt{helicalc}} = \sum_{i=1}^N \vec{B}_i (x, y, z)
\end{equation}

\subsubsection{Straight Bus Bars}
Straight segments of superconductor of length $L$ are utilized throughout the bus bar network. The cross-section ($w\ h$) of the bar represents the rectangular cross-section of the superconducting cable.
It is convenient to consider the cross-section of the bus bar centered in $x,y$ plane. The bar extends from $z=0$ to $z=L$ and the current points in the $+\hat{z}$ direction. The position and orientation of the bar in the global frame is specified by $\vec{x}_c$
and three Euler angles to define an intrinsic set of rotations about $z,y,z$.

Using Cartesian coordinates, the current density is:
\begin{equation}
\vec{J}(x,y,z) = \vec{J} = (0, 0, 1) \frac{I}{w \ h}
\end{equation}
which we can use to derive the numerator of the Biot-Savart integrand:
\begin{equation}
\vec{J} \times \vec{r} = \frac{I}{w \ h}[-r_y \hat{i} + r_x \hat{j}] \ .
\end{equation}
Given a field point $\vec{x}'$ in the local coordinate system the vector $\vec{r}$ has components ${r_x = x_p' - x}$, ${r_y = y_p' - y}$, ${r_z = z_p' - z}$ for the integration point $\vec{x}_i = (x, y, z)$ along the bar.

With differential volume $\dif V = \dif x \dif y \dif z$, we can substitute into Equation~\ref{eq:biotsavart3D} to find $\vec{B}_{\mathrm{str}}$, the field due to a straight bus bar:
\begin{equation}
\vec{B}_{\mathrm{str}}(\vec{x}') = \frac{\mu_0 I}{4\pi \ w \ h} \int_0^L \int_{-h/2}^{h/2} \int_{-w/2}^{w/2} \frac{[-r_y \hat{i} + r_x \hat{j}]}{r^3} \dif x \dif y \dif z 
\end{equation}
which is implemented in the \texttt{helicalc.busbar.StraightIntegrator3D} class.

\subsubsection{Circular Arc Bus Bars}
Circular arc segments of superconductor of radius $R_0$ and subtended angle $\phi_{\mathrm{max}}$ are utilized throughout the bus bar network. Arcs are also used as interlayer splices for helical solenoids with more than one layer. The cross-section ($w \ h$) of the bar represents the rectangular cross-section of the superconducting cable; the current flows in the $\phi$ direction (i.e. along the arc).
It is convenient to consider cross-section of the bus bar centered in $x,y$ plane. The center of the arc is located at $x=0, y=R_0$ and the bar extends from the $z=0$ plane towards the $+\hat{z}$ direction. In other words, the circle describing the arc lies in the $y,z$ plane. The current flows from the end of the bar at $z=0$ through the bar. The position and orientation of the arc bar in the global frame is specified by $\vec{x}_c$
and three Euler angles to define an intrinsic set of rotations about $z,y,z$.

Using Cartesian coordinates, the current density is:
\begin{equation}
\vec{J}(x,y,z) = \vec{J} = (0, \sin \phi, \cos \phi) \frac{I}{w \ h}
\end{equation}
where $\phi = 0$ corresponds to the start of the arc ($z=0$) and $\phi = \phi_{\mathrm{max}}$ corresponds to the end of the arc. We can extend this definition of $\phi$ to a cylindrical coordinate system in which $\phi$ and radial coordinate $\rho$ are in the $y,z$ plane and $x$ plays the role of the axial coordinate. We will use this modified coordinate system for the integration variables. The numerator of the Biot-Savart integrand is:
\begin{equation}
\vec{J} \times \vec{r} = \frac{I}{w \ h}\bigg[ (r_z \sin\phi - r_y \cos\phi) \hat{i} + (r_x \cos\phi) \hat{j} + (- r_x \sin\phi) \hat{k} \bigg] \ .
\end{equation}
Given a field point $\vec{x}'$ in the local coordinate system the vector $\vec{r}$ has components ${r_x = x_p' - x_i}$, ${r_y = y_p' - \rho \cos \phi - R_0}$, ${r_z = z_p' - \rho \sin \phi}$ for the integration point (cylindrical coordinates)\linebreak
${\vec{x}_i = (\rho, \phi, x)}$ along the arc bar.

With differential volume $\dif V = \rho \dif \rho \dif \phi \dif x$, we can substitute into Equation~\ref{eq:biotsavart3D} to find $\vec{B}_{\mathrm{arc}}$, the field due to an arc bus bar:
\begin{align}
B_{x, \mathrm{arc}}(\vec{x}') &= \frac{\mu_0 I}{4\pi\ w \ h} \int_{-w/2}^{w/2} \int_{R_0-h/2}^{R_0+h/2} \int_{0}^{\phi_{max}} \frac{[r_z \sin\phi - r_y \cos\phi]}{r^3} \rho \dif \phi \dif \rho \dif x
\\
B_{y, \mathrm{arc}}(\vec{x}') &= \frac{\mu_0 I}{4\pi\ w \ h} \int_{-w/2}^{w/2} \int_{R_0-h/2}^{R_0+h/2} \int_{0}^{\phi_{max}} \frac{[r_x \cos\phi]}{r^3} \rho \dif \phi \dif \rho \dif x
\\
B_{z, \mathrm{arc}}(\vec{x}') &= \frac{\mu_0 I}{4\pi\ w \ h} \int_{-w/2}^{w/2} \int_{R_0-h/2}^{R_0+h/2} \int_{0}^{\phi_{max}} \frac{-[r_x \sin\phi]}{r^3} \rho \dif \phi \dif \rho \dif x
\end{align}
which is implemented in the \texttt{helicalc.busbar.ArcIntegrator3D} class.

\subsubsection{Helical Solenoids}
Thick, helically wound, multilayer solenoids are used to describe the Mu2e DS coils. There are a number of parameters to describe the geometry of a coil. The superconducting cable~\cite{Mu2eTDR} is aluminum-stabilized NbTi; a rectangular cross section of twisted wire containing NbTi and copper is encapsulated in an aluminum sheath. The total cross-section is rectangular. Additional layers of insulating materials are included in the geometric model. We note that the bus bars use a similar cable, but the geometric details of the aluminum sheath and insulating materials are irrelevant for deriving the integrands. The local coordinate system for a coil is constructed with the origin at the center of the coil with the axis of the coil along the $z$-axis. The parameters defining a coil are:
\begin{itemize}
\item Superconductor cross section $w \ h$. The superconductor is encased in aluminum stabilizer; the total cable cross-section (neglecting insulation) is $w_{\mathrm{Al}} \ h_{\mathrm{Al}}$.
\item $t_{\mathrm{CI}}$: thickness of cable insulation that encapsulates the cabling (i.e. surrounds the aluminum stabilizer). The total cable cross-section is $(w_{\mathrm{Al}} + 2 t_{\mathrm{CI}}) \ (h_{\mathrm{Al}} + 2 t_{\mathrm{CI}})$.
\item $t_{\mathrm{GI}}$: thickness of ground insulation that encapsulates an entire coil layer.
\item $t_{\mathrm{IL}}$: thickness of interlayer insulation between successive layers of the coil.
\item Inner radius of the coil $R_i$. This denotes the distance from the axis of the coil to the inner edge of the innermost layer of ground insulation.
\item Total length of the coil $L$. This length includes all insulation.
\item Center of the coil (global coordinates): $(x_c, y_c, z_c)$
\item Three extrinsic Euler angles $x$, $y$, and $z$ to rotate the coil in the global frame, i.e. a pitch, yaw, and roll, respectively. All DS coils have zero pitch, yaw, and roll.
\item $\eta = \pm 1$: coil layer winding helicity. $\eta = +1$ is a ``right-handed'' winding (i.e. positive helicity) while $\eta = -1$ is a ``left-handed'' winding (i.e. negative helicity). Note that by choosing the direction of current always in the same direction azimuthally, i.e. $I \sim \hat{\phi} + \eta \hat{z}$, we can create an axial field in the same direction for both layer helicities. $\vec{B} \approx B_z \hat{z}$ with $B_z > 0$ in the nominal configuration. In the Mu2e DS coils, single-layer coils are wound with $\eta=+1$ and two-layer coils are wound with $\eta=-1$ for the inner layer and $\eta=+1$ for the outer layer. The alternating layer helicities serves two purposes. It is easier to splice layers together in this configuration, and this configuration partially cancels the azimuthal asymmetries induced by the helical nature of the windings. It is also worth noting that a choice is made to integrate $\phi$ in the $\eta$ direction. For $\eta = +1$ the integration over $\phi$ is in the positive direction and $z$ increases during integration. For $\eta = -1$ the integration over $\phi$ is in the negative direction but $z$ still increases during integration.
\item Current $I$. For the Mu2e DS, which is connected in series via a network of bus bars, ${I=6,114}$~A.
\item Number of layers $N_L$
\item Number of windings (i.e. turns) per layer $N_t$
\item $\phi_i$: azimuthal location of the input lead for the coil. $0 \leq \phi_i \leq 2 \pi$. For a single-layer coil the input lead is on the $-z$ end of the coil, while for a two-layer coil the input lead is on the $+z$ end of the coil. For the Mu2e DS, a typical $\phi_i$ value is $\phi_i = 225\degree$.
\item $\phi_f$: azimuthal location of the output lead for the coil. $0 \leq \phi_f \leq 2 \pi$. For single-layer and two-layer coils the output lead is on the $+z$ end of the coil. For the Mu2e DS, a typical $\phi_f$ value is $\phi_f = 87\degree$.
\end{itemize}

\begin{figure}[htbp!]
\centering
\begin{minipage}[b]{0.49\textwidth}
        \centering
        \includegraphics[width=\textwidth]{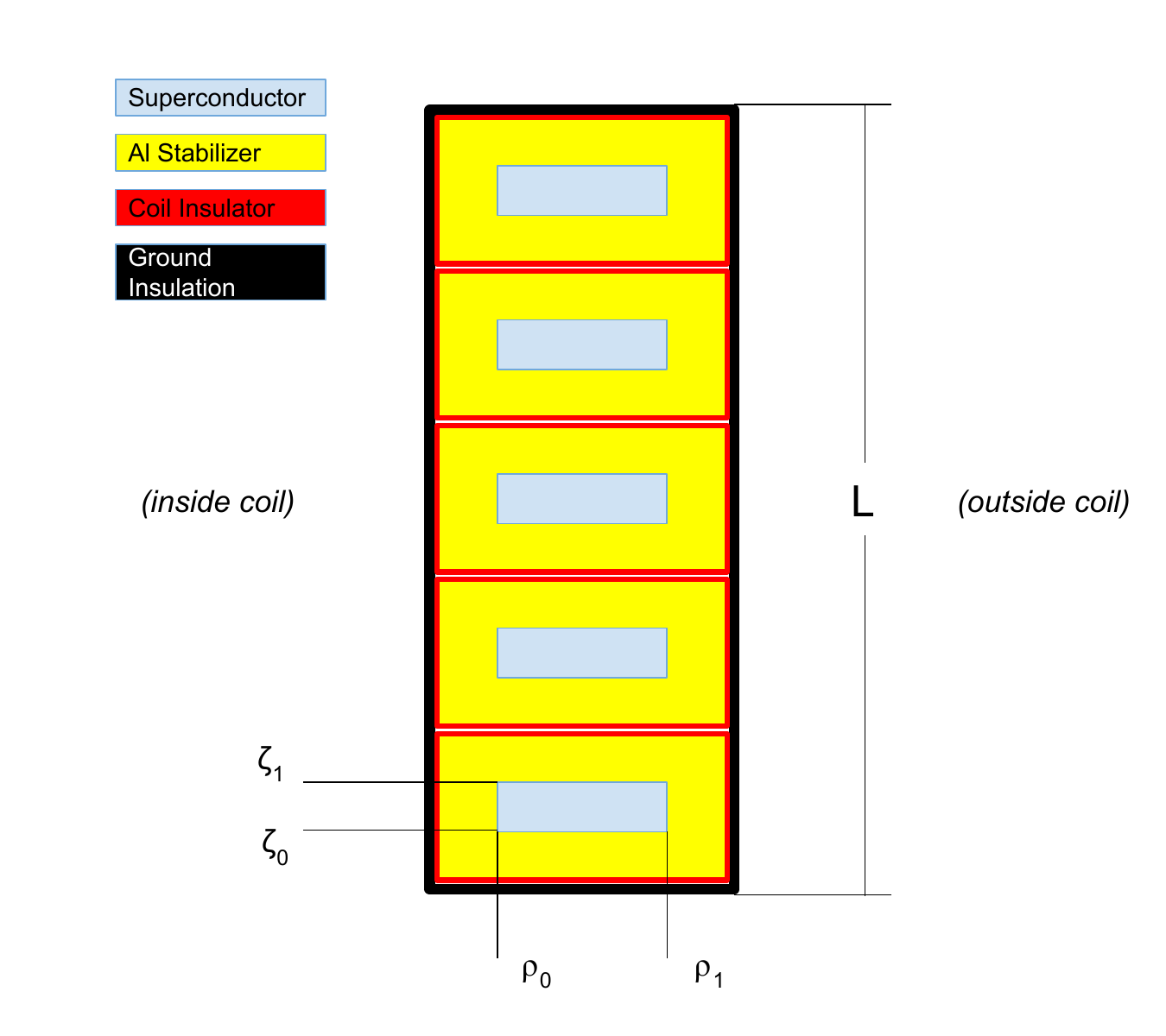}
				\textbf{(a)}\\        
    \end{minipage}
    \begin{minipage}[b]{0.49\textwidth}
        \centering
        \includegraphics[width=\textwidth]{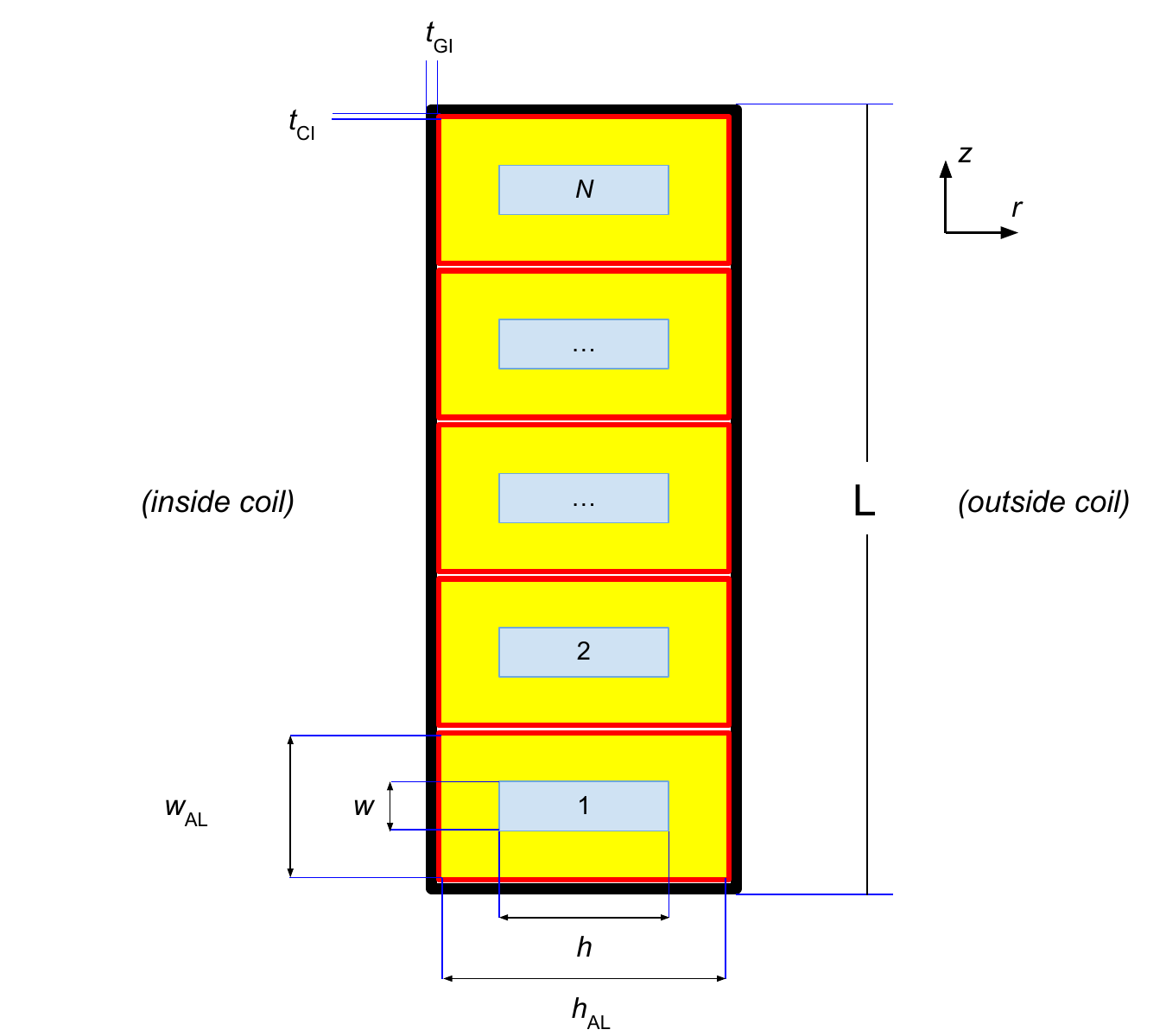}
        \textbf{(b)}\\
    \end{minipage}
    
    \vspace{1em}
    
    \begin{minipage}[b]{0.49\textwidth}
        \centering
        \includegraphics[width=\textwidth]{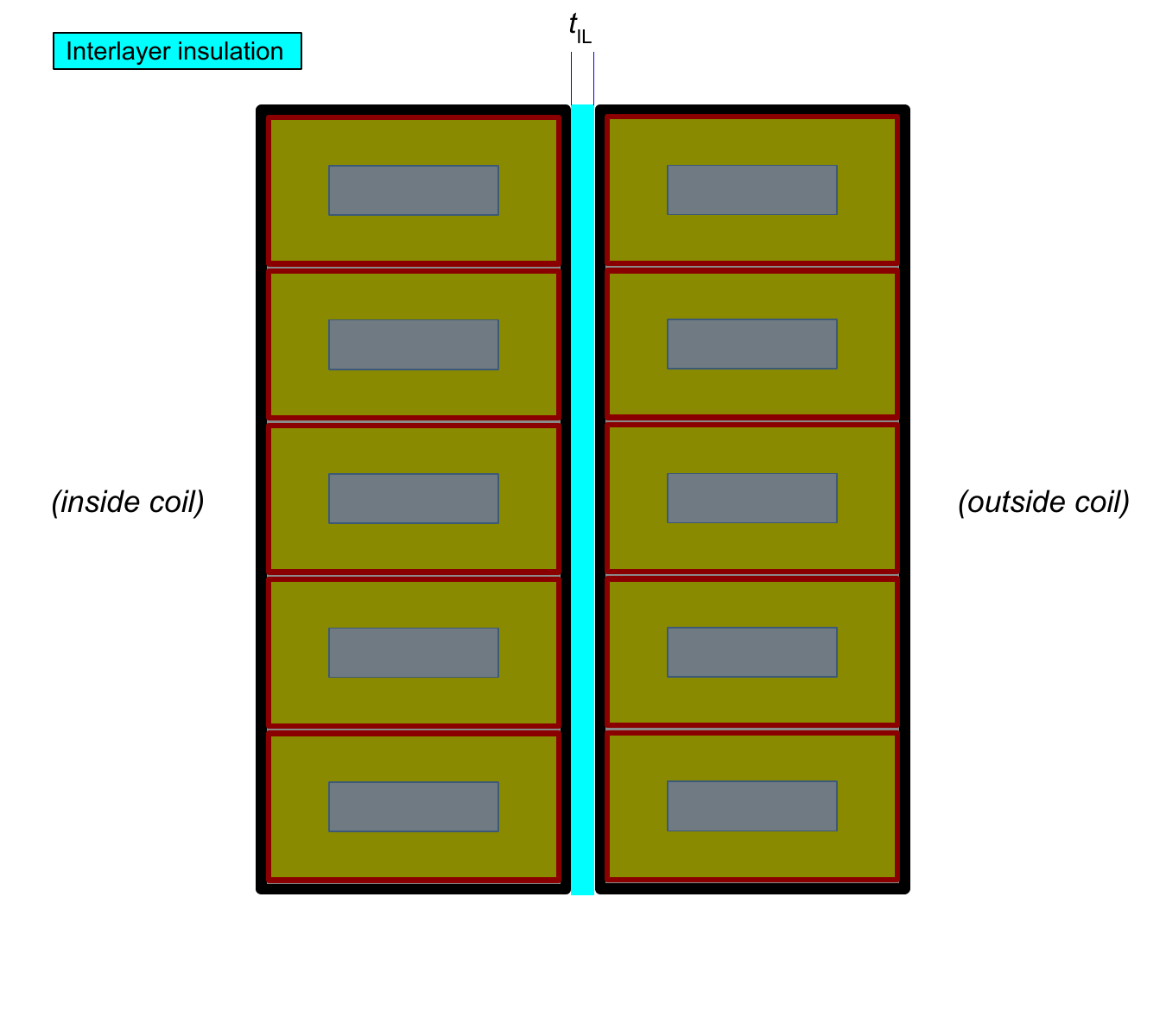}
        \textbf{(c)}\\
    \end{minipage}
    
    \caption{Cross-sections of the helically wound solenoids modeled in \texttt{helicalc}. The cross sections are at $\phi=0$, so the current flows into the page for all superconducting regions of the diagram. (a) A single-layer coil denoting the variables $\rho$ and $\zeta$ which are integrated over. (b) A single-layer coil denoting most of the cable and coil parameters that are used to locate the superconductor. In this coil, the windings are labeled with indexes $1,2,..,N$; the current flows from index $1$ to $N$, so this is a coil with positive helicity. The current flows in the $+\phi$ direction, e.g. from winding $1$ to winding $2$. (c) A two-layer coil, with the interlayer insulation indicated in cyan. In each diagram denotes the superconducting cable with superconductor (NbTi and copper) in gray, aluminum stabilizer in yellow, and cable insulation in red. The outer ground insulation is black.}
		\label{fig:helicalc_geoms}
\end{figure}

Visualizations of single layer and multilayer coil cross-sections are shown in Figure~\ref{fig:helicalc_geoms}. The remainder of the derivation assumes this geometry and refers to integration variables $\rho, \zeta, \phi$. $\rho$ is the radial variable in a standard cylindrical coordinate system. $\phi$ is an accumulating azimuthal angle that is defined to be in the range $[0, 2\pi)$ for the first coil winding.
For $\eta=+1$ increasing $\phi$ increases $z$, while for $\eta=-1$ increasing $\phi$ decreases $z$. Thus to move along an $\eta=-1$ winding one must navigate along $\phi$ in the negative direction. Consider, for example, $\phi = +12 \pi\ \mathrm{rad}$~\mbox{(~$=-12 \pi\ \mathrm{rad}$)}, which corresponds to the same azimuthal position as $\phi = 0$ but in the sixth winding for a coil with $\eta=+1$ ($=-1$). $\zeta$ is fixed to the cable's cross-section at any point in the windings. $\zeta=0$ corresponds to the $-z$ edge of the cable insulation, which has a thickness of $t_{\mathrm{CI}}$. 

Using Cartesian coordinates, the current density is:
\begin{equation}
\vec{J} = (-\sin\phi, \cos\phi, \eta \frac{\Lambda}{2\pi \rho}) \frac{I}{C\ w \ h}
\end{equation}
where $\Lambda = w_{\mathrm{Al}} + 2 t_{\mathrm{CI}}$ is the helical pitch or spacing between windings and $C = C(\rho) = \sqrt{1 + (\frac{\Lambda}{2\pi \rho})^2}$ is a normalization factor that ensures a total current of $I$ flows through the conductor. For the Mu2e coils, $C \approx 1$ ($\rho \gg \Lambda$) and nearly constant ($\Delta \rho = h \ll \rho$); a flag in \texttt{helicalc} allows for a constant $C$ value calculated at the mean value of $\rho$, $\overline{\rho} = \rho_0 + \frac{h}{2}$, for a small computational benefit.

The numerator of the Biot-Savart integrand is:
\begin{align}
\vec{J} \times \vec{r} = \frac{I}{C\ w \ h} &\bigg\{\ \hat{i} (r_z \cos\phi - r_y \eta \frac{\Lambda}{2\pi \rho}) \\
&+ \hat{j} (r_z \sin\phi + r_x \eta \frac{\Lambda}{2\pi \rho}) \notag \\
&+ \hat{k} (-r_y \sin\phi - r_x \cos \phi)\bigg\}\ . \notag 
\end{align}
The vector $\vec{r}$ has components $r_x = x_p' - \rho \cos \phi$, $r_y = y_p' - \rho \sin \phi$, and \linebreak
${r_z = z_p' - [-L/2 + t_{\mathrm{GI}} + \zeta + \frac{h}{2\pi} |\phi - \phi_0|]}$. Note there are small helicity dependent subtleties in $r_z$ not described here.

Using the differential volume $\dif V = (\rho \dif \rho) (\eta \dif \phi) \dif \zeta$ and substituting into Equation~\ref{eq:biotsavart3D} yields the integral to solve to find $\vec{B}_{\mathrm{h.c.}}$, the field due to a single layer of a helical coil:
\begin{align}
B_{x, \mathrm{h.c.}}(\vec{x}') &= \frac{\mu_0 I}{4\pi\ w \ h} \int_{\rho_0}^{\rho_0+h} \int_{\zeta_0}^{\zeta_0+w} \int_{\phi_0}^{\phi_1} \frac{[r_z \eta \rho \cos\phi - r_y \frac{\Lambda}{2\pi}]}{C r^3} \dif \phi \dif \zeta \dif \rho
\\
B_{y, \mathrm{h.c.}}(\vec{x}') &= \frac{\mu_0 I}{4\pi\ w \ h} \int_{\rho_0}^{\rho_0+h} \int_{\zeta_0}^{\zeta_0+w} \int_{\phi_0}^{\phi_1} \frac{[r_z \eta \rho \sin\phi + r_x \frac{\Lambda}{2\pi}]}{C r^3} \dif \phi \dif \zeta \dif \rho
\\
B_{z, \mathrm{h.c.}}(\vec{x}') &= \frac{\mu_0 I}{4\pi\ w \ h} \int_{\rho_0}^{\rho_0+h} \int_{\zeta_0}^{\zeta_0+w} \int_{\phi_0}^{\phi_1} \frac{-\eta \rho [r_y \sin\phi + r_x \cos\phi]}{C r^3} \dif \phi \dif \zeta \dif \rho
\end{align}
which is implemented in the \texttt{helicalc.coil.CoilIntegrator} class. The integration bounds can be derived by studying Figure~\ref{fig:helicalc_geoms}. The radial and axial endpoints are straightforward:
\linebreak
\mbox{$\rho_0 = R_i + (h_{\mathrm{Al}} - h) / 2 + t_{\mathrm{GI}} + t_{\mathrm{CI}}$} and \mbox{$\zeta_0 = t_{\mathrm{CI}} + (w_{\mathrm{Al}} - w) / 2$}. The $\phi$ endpoints are a bit subtler given the varying helicities. $\phi_0 = \phi_i$ starts the integration at the $-z$ end of the coil. If the coil has a single layer $\phi_1 = \phi_0 + \eta [2\pi N_t - |\phi_f - \phi_0|]$; the total number of realized windings is bounded from above by $N_t$.
If the coil has two layers the first (inner) layer winds all $N_t$ turns, $\phi_1 = \phi_0 + \eta [2\pi N_t]$, while the outer layer is bounded in a way equivalent to the single layer coil, ${\phi_1 = \phi_0 + \eta [2\pi N_t - |\phi_f - \phi_0|]}$. While $L$ enters the definition of $r_z$ to define the physical location of the windings, the realized length of the coil may differ from $L$ by up to $w_{\mathrm{Al}}$ for single-layer coils, given the pitch is defined assuming no additional spacing between windings and therefore does not count the final winding which may not span a full $2\pi$~rad.

For two-layer coils, $\phi_0$ ($-z$ end of coil) for the first layer is adjusted to splice the two layers with an arc. For the inner layer $\phi_0 \rightarrow \phi_0 \pm 2 \pi / 10$~rad (minus sign for an inner layer with $\eta=-1$) while for the second layer $\phi_0$ is left unmodified. The two helical termination points at the  $-z$ end of the coil are spliced with a circular arc that spans an angle of approximately $2\pi/10$~rad and has a radius equal to the mean of the two layer radii; the current flows from the first layer (negative helicity) to the second layer (positive helicity). Bus bars are connected at the $+z$ ends of the two-layer coils to bring current in and out.

\subsubsection{Ideal Solenoids}
\begin{figure}[htbp!]
\centering
\includegraphics[width=0.49\textwidth]{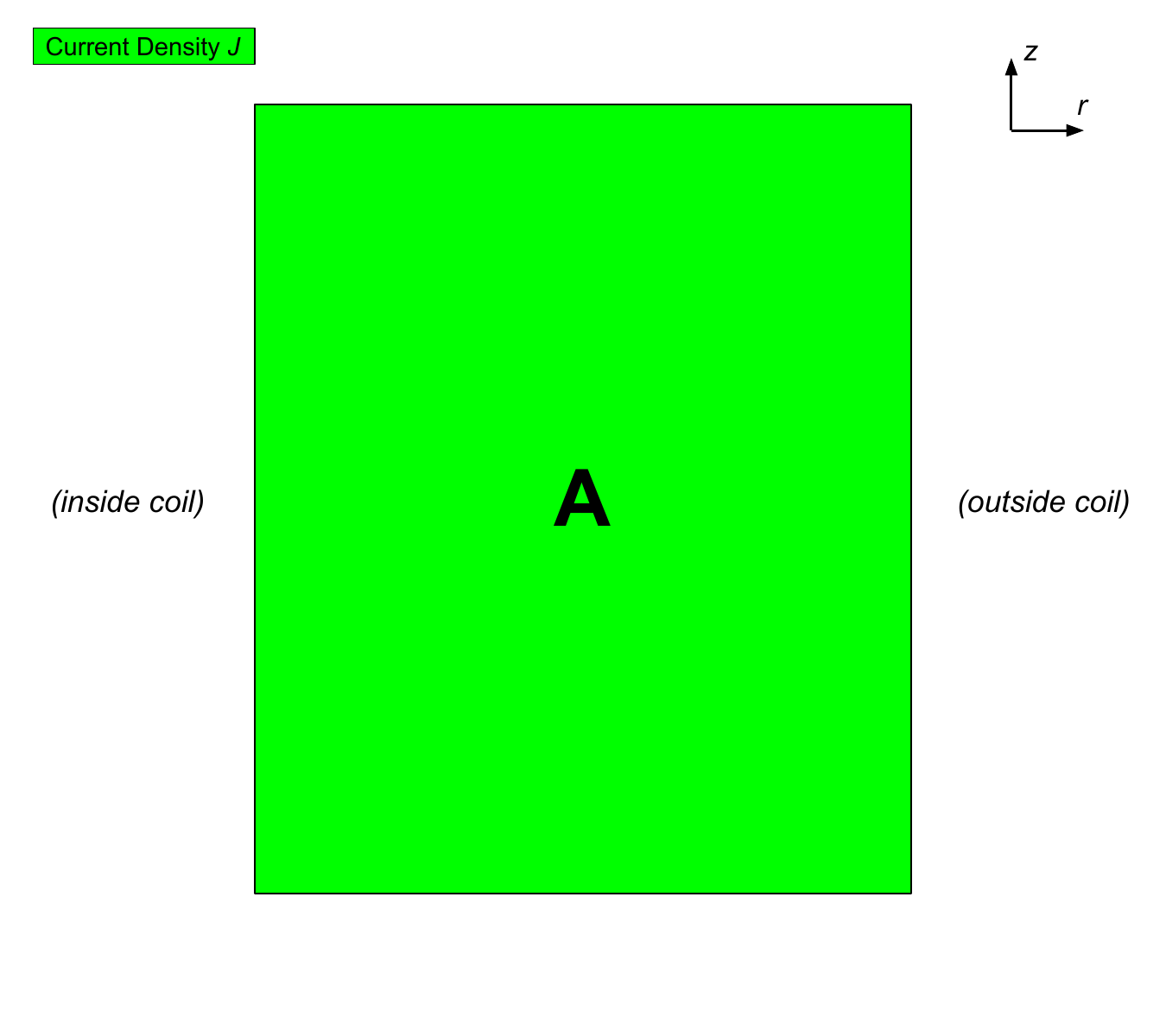}
    \caption{Cross-section of the ideal solenoid modeled in \texttt{helicalc}. The cross sections are at $\phi=0$, so the current flows into the page. Note that the area $A$ contains the entire physical extent of the coil, including superconductor, stabilizer, and insulation layers.}
		\label{fig:solcalc_geom}
\end{figure}

To calculate the field from ideal solenoids, we utilize the same technique as the \texttt{SolCalc} \cite{SOLCALC2018} package, which integrates the exact magnetic field from a filamentary loop current \cite{knoepfelMagFields} over the axial ($z$) and radial ($\rho$) dimensions. Thus, the ideal solenoid representation is a cylindrical shell of current. For a consistent geometric description, many of the parameters use for ideal solenoids match those used in the helical solenoid integrations. The parameters defining an ideal solenoid coil are:
\begin{itemize}
\item Inner radius of the coil $R_i$. This denotes the distance from the axis of the coil to the inner edge of the innermost layer of ground insulation.
\item Outer radius of the coil $R_o$. This denotes the distance from the axis of the coil to the outer edge of the outermost layer of ground insulation.
\item Total length of the coil $L$. This length includes all insulation.
\item Center of the coil (global coordinates): $(x_c, y_c, z_c)$
\item Three extrinsic Euler angles to rotate the coil in the global frame, i.e. a pitch, yaw, and roll, respectively. While PS and DS coils have zero pitch, yaw, and roll, each TS coil has a unique yaw.
\item Current $I$. The Mu2e PS, TS, and DS coils are configured in independent circuits with current values of ${I=9,200}$~A, ${I=1,730}$~A, and ${I=6,114}$~A, respectively.
\item Number of layers $N_L$
\item Number of windings (i.e. turns) per layer $N_t$
\end{itemize}
The current distribution is approximated as being uniformly distributed throughout the entire coil cross-section without consideration of where the current is actually localized (in the superconductor). In other words, the integral is computed over a simple rectangular cross-section of area $A = L \ (R_o - R-i)$. The total current $I_t = I \ N_t \ N_L$ is distributed throughout this cross-section: $\vec{J} = I_t / A \hat{\phi}$. A visualization of the cross-section is given in Figure~\ref{fig:solcalc_geom} and can be compared to the more complex geometry used for helical coils in Figure~\ref{fig:helicalc_geoms}.

Knoepfel derives the magnetic field generated by an infinitesimal filamentary circular loop by first determining the magnetic vector potential $\vec{A} = A_\phi \hat{\phi}$ by integrating over the path of the current. Elliptic integrals $E$ and $K$ are used in to write the solution in closed form. The filamentary field components $B_{r, \mathrm{fil}}$ and $B_{z, \mathrm{fil}}$ are then calculated using the relationship $\vec{B} = \nabla \times \vec{A}$ ({equation~2.3-11} in Ref.~\cite{knoepfelMagFields}). Noting that the differential area $\dif A = \dif \rho \dif z$, we can construct the ideal solenoid coil integrals $B_{r, \mathrm{i.c.}}$ and $B_{z, \mathrm{i.c.}}$ as in Ref.~\cite{SOLCALC2018}:
\begin{align}
B_{\rho, \mathrm{i.c.}}(\rho, z) &= \frac{\mu_0 I}{2\pi\ A} \int_{R_i}^{R_o} \int_{-L/2}^{L/2} \frac{z}{\rho \sqrt{(R + \rho)^2 + z^2}} \bigg[ -K + \big( \frac{R^2 + \rho^2 + z^2}{(R-\rho)^2 + z^2} \big) E \bigg] \dif z \dif r \\
B_{z, \mathrm{i.c.}}(\rho, z) &= \frac{\mu_0 I}{2\pi\ A} \int_{R_i}^{R_o} \int_{-L/2}^{L/2} \frac{1}{\sqrt{(R + \rho)^2 + z^2}} \bigg[ K + \big( \frac{R^2 - \rho^2 - z^2}{(R-\rho)^2 + z^2} \big) E \bigg] \dif z \dif r
\end{align}
where $\rho$ is the filament radius and parameter $k^2$ is defined as:
\begin{equation}
k^2(\rho, z) = \frac{4 R \rho}{(R+\rho)^2 + z^2}
\end{equation}
noting that as in the other derivations, we assume the local coordinate system where the solenoid axis is aligned with $\hat{z}$. The complete elliptic integrals of the first and second kind $K$ and $E$, respectively, are defined as:
\begin{align}
K &= K(k^2) = \int_0^{\frac{\pi}{2}} (1 - k^2 \sin^2 \gamma)^{-1/2} \dif \gamma  \\
E &= E(k^2) = \int_0^{\frac{\pi}{2}} (1 - k^2 \sin^2 \gamma)^{1/2} \dif \gamma
\end{align}
whose tabulated values are used for computational efficiency \cite{scipy2020}. The ideal solenoid integration is implemented in the class \texttt{helicalc.solcalc.SolCalcIntegrator}.

\subsection{Numerical Integrations}
\label{app:helicalc_numeric}
The $2$-dimensional and $3$-dimensional Biot-Savart integrals that are solved in the \texttt{helicalc} packages are well behaved and smooth for field points sufficiently far from the current sources being integrated. All field points considered in the Mu2e DS are approximately $20$~cm from the nearest sources of current. We therefore us an iterative trapezoidal integration technique on a uniform grid. This choice allows a massive computational benefit by running the computations on a GPU, which excels for large matrix operations. We bootstrap the efforts from the machine learning community which has developed a number of useful tools implemented on GPUs. In particular, we use \texttt{pytorch}~\cite{pytorch2019} due to the array structures that closely mimic \texttt{numpy.array} structures \cite{numpy2020}. All conductor geometries implemented in \texttt{helicalc}, with the exception of the ideal solenoids, are implemented to use either a GPU (preferred, \texttt{pytorch}) or a CPU (\texttt{numpy}). When \texttt{helicalc} was developed, \texttt{pytorch} did not have an implementation of the elliptic integrals, which means it cannot be used for the ideal solenoid calculations; the ideal solenoid calculations, which are already fast, are implemented for CPU computations only (\texttt{numpy}).

\section{Opera Field Model}
\label{sec:opera_model}

\subsection{Magnetic Material Model}
\label{subsec:opera_material}
All contributions from passive magnetic materials included in the calculation of the Mu2e DS magnetic field are computed using the Opera magnetostatics simulation software suite. A number of structural elements near the solenoids are included in the calculation, such as reinforcement bar (rebar) in concrete shielding blocks surrounding the solenoids and present in the floor of the experimental hall, as well as steel plates embedded in the floor.

While rebar typically comprises thin cylindrical rods in a complex grid structure within the concrete it is supporting, we use a simplified representation to ease the computational burden. The fractional rebar content $f$ of each concrete block is estimated from engineering drawings. A set of rectangular prisms, or blocks, with homogeneous magnetic properties are defined to represent each volume of rebar. A custom $B-H$ curve is set for each block and is defined as a linear mixture of the materials:
\begin{equation}
B(H)_{\mathrm{mix}} = \epsilon f \ B(H)_{\mathrm{rebar}} + (1 - \epsilon f) \ (\mu_0 H)_{\mathrm{concrete}}
\end{equation}
where $\epsilon$ is an efficiency factor included to account for the structured orientation of the real rebar layout and the expected fringe field that will magnetize that material, or in other words what fraction of the rebar is expected to efficiently transport the flux. For example, the floor rebar contains a grid of parallel rods along $x$ and parallel rods along $z$. The fringe field at the floor should be approximately in the $z$ direction. Thus, an efficiency factor of $\epsilon=0.5$ is applied to the floor blocks. 

\begin{figure*}[h!]
\centering
\includegraphics[trim=0cm 0cm 0.0cm 0.0cm, width=0.8\textwidth]{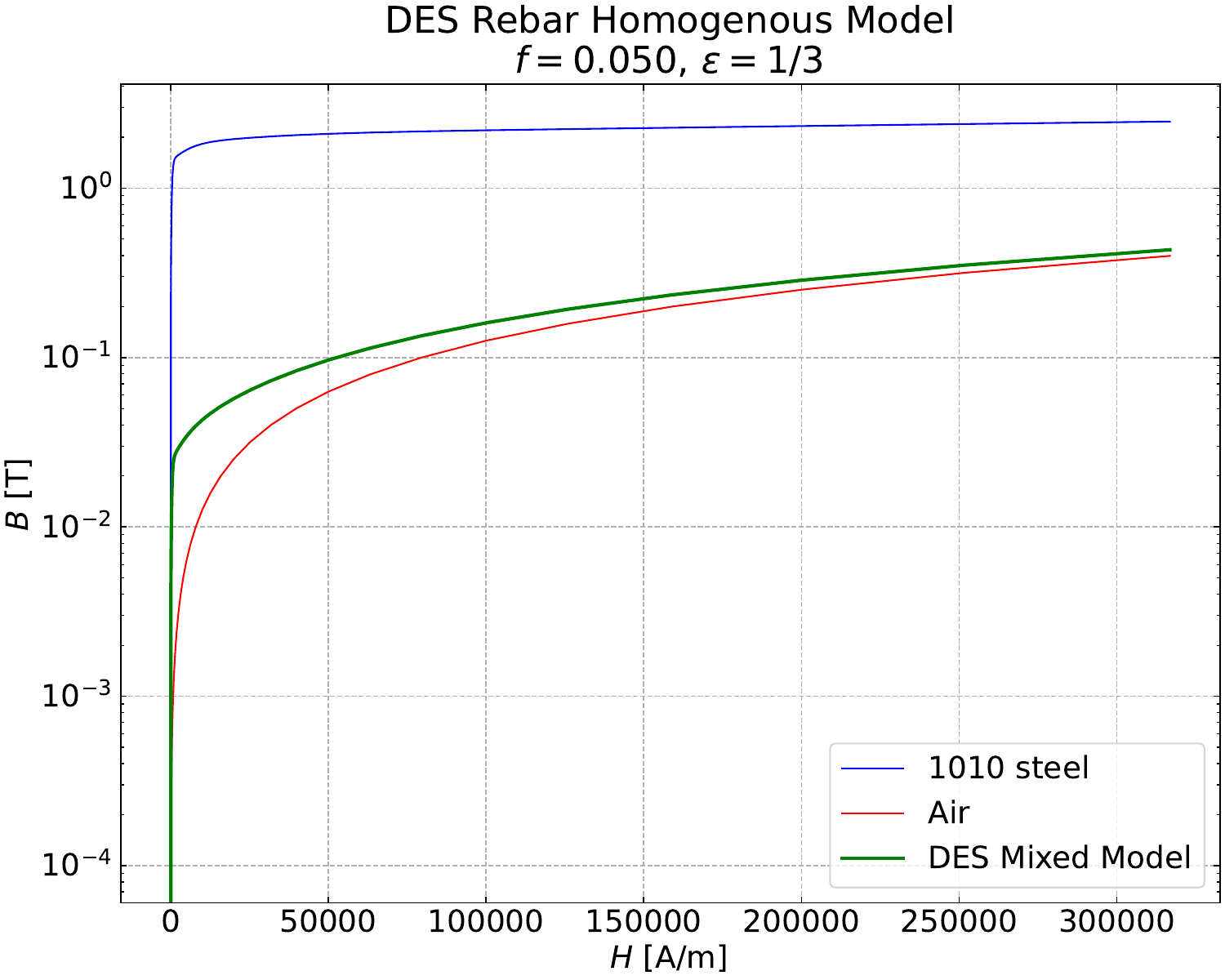}
\caption{The custom $B-H$ curve used to model the DES rebar blocks. This model mixes 1010 steel and air.}
\label{fig:DES_B-H}
\end{figure*}

Including the steel floor plates is simpler, though one can still use the homogeneous treatment with $f=1$, $\epsilon=1$. Table~\ref{tab:opera_params} summarizes the material values used in the Opera model. Figure~\ref{fig:DES_B-H} shows the $B-H$ curve used to model the rebar in the shielding blocks.

\begin{table}[h!]
\centering
\caption{The parameters used to define the different magnetic material elements in the Opera model.}
\begin{tabular}{l|c|c|l}
\textbf{Source}                   & $\mathbf{f}$    & $\mathbf{\epsilon}$ & \textbf{Magnetic Material} \\ \hline
Shielding block rebar & $0.050$ & $1/3$      & 1010 steel        \\ \hline
Floor rebar (top layer)        & $0.033$ & $1/2$      & 1010 steel        \\ \hline
Floor rebar (bottom layer)     & $0.100$ & $1/2$      & 1010 steel        \\ \hline
Floor plates             & $1$     & $1$        & A-36 steel
\end{tabular}
\label{tab:opera_params}
\end{table}

\subsection{Combination with \texttt{helicalc}}
\label{subsec:opera_helicalc}

The field point locations in the Opera maps are slightly offset from those in \texttt{helicalc}, so a trilinear interpolation of each field component of the Opera maps is used to evaluate the field contributions from Opera at the field points locations calculated in \texttt{helicalc}. The Opera calculation is computed on a dense 3D Cartesian grid with a step size in each dimension of $25$~mm and the field is smoothly varying, so interpolation errors are negligible.

\section{Least-Squares Model}
\label{sec:lsq_supp}

\subsection{Modifications to the Model}
\label{subsec:lsq_mods}

In practice, there are a number of non-trivial modifications to the least-squares model functions (see paper Equation~\ref{eq:B_cyl_final}) to improve the usability and prevent free parameter degeneracies. The $n=0$ terms codify azimuthally symmetric contributions to the field, which can be seen in the pre-factor of $n$ which results in a vanishing $B_{\cylAzi,\mathrm{cyl}}$. Additionally, terms with $C_{m, 0}$ and $D_{m,0}$ are zero since $\sin 0 = 0$, thus we fix $C_{m, 0} = D_{m, 0} = 0$.
Finally, $B_{\cylAzi, \mathrm{cyl}}$ has an apparent divergence at $r=0$ due to division by zero. In the limit as $r \to 0$, $B_{\cylAzi, \mathrm{cyl}}$ is well behaved in that it is finite and equal along any azimuthal $\cylAzi$ profile. Therefore, the problem is mitigated near $r=0$ by replacing the radially dependent terms of $B_{\cylAzi, \mathrm{cyl}}$ by a leading-order approximation. To simplify the notation, we isolate radially dependent terms of $B_{\cylAzi, \mathrm{cyl}}$ and denote it by $f_{m,n}(r)$:
\begin{equation}\label{eq:Bphi_f}
f_{m,n}(r) = \frac{n}{r} I_n(k_m r)\ .
\end{equation}
By considering the power series definition of $I_n$ (see DLMF Equation~\href{https://dlmf.nist.gov/10.25.E2}{10.25.2} from Ref.~\cite{NIST:DLMF}):
\begin{equation}\label{eq:I_n_powerseries}
I_n(k_m r) = (k_m r/2)^n \sum^\infty_{j=0} \frac{(k_m^2 r^2/4)^j}{j! \Gamma(n + j + 1)}
\end{equation}
we see that the $1/r$ divergence is artificial for $n>0$, and only appears due to the separate computation of $I_n(k_m r)$ and $1/r$; for any $j$ in Equation~\ref{eq:I_n_powerseries}, the power of $r$ is $\geq 1$ and absorbs the factor of $1/r$ in Equation~\ref{eq:Bphi_f}.
For $n=0$, the prefactor $n$ in Equation~\ref{eq:Bphi_f} instead gives $f_{m,n=0}(r) = 0$.
Given that Equation~\ref{eq:I_n_powerseries} converges rapidly for small $k_m r$ (see Ref.~\cite{TEMME1976}), we approximate $f_{m,n}$ for small $k_m r$ and $n>0$ by including only the first order ($j=0$) term. For $k_m r < \epsilon = 1\times10^{-7}$ we approximate $f_{m,n}(r)$ as:
\begin{equation}
f_{m,n}(r) \approx
\begin{cases}
0, & \text{if } n = 0, \\
\frac{n k_m^n r^{n-1}}{2 \Gamma(n+1)}, & \text{if } n > 0.
\end{cases}
\end{equation}
which is consistent with the small argument asymptotic expression for $I_n$ given in DLMF Equation~\href{https://dlmf.nist.gov/10.30.E1}{10.30.1} (Ref.~\cite{NIST:DLMF}). We validate the approximation by comparing $B_{\cylAzi, \mathrm{cyl}}$ evaluated using the approximation for $f_{m,n}$ and the exact form of $f_{m,n}$ at $k_m r = \epsilon$. Using the nominal least-squares model best-fit parameter values, we find the largest error in $B_{\cylAzi, \mathrm{cyl}}$ is $< 1 \times 10^{-15}$~Gauss.

\subsection{Hyperparameter Optimization}
\label{subsec:lsq_opt}

The least-squares model function, given in paper Equation~\ref{eq:B_cyl_final}, requires choosing only three hyperparameters. The truncation indices $m_{\mathrm{max}}$ and $n_{\mathrm{max}}$ were selected from the Nyquist-Shannon sampling criterion and empirical convergence studies, as discussed in paper Sections~\ref{subsubsec:results_leastsquares}~and~\ref{subsec:results_reduced_meas}. Hyperparameter $L$, the length defining the axial bound used in the boundary conditions, is determined by a grid search. We use a simplified model function with $m_{\mathrm{max}} = 54$ and $n_{\mathrm{max}} = 0$, i.e. a model with only azimuthally symmetric terms, to evaluate many $L$ values in a reasonable period of time.
The scan shows a broad plateau for $L \geq 11.5$~m, with a shallow minimum at $L=12.5$~m.
We validate that this is a reasonable choice for the full model by running the full model fit ($m_{\mathrm{max}}=54$ and $n_{\mathrm{max}}=6$) for a limited set of $L$ values selected from the grid search. The results of the two scans are provided in Table~\ref{tab:LSQ_L_opt}.

\begin{table}[htbp!]
\center
\caption{Hyperparameter scans over length $L$ in the least-squares model. A first scan uses a simplified model with $m_{\mathrm{max}}=54$ and $n_{\mathrm{max}}=0$ to decrease runtime. The value $L=9.6$~m corresponds to the exact axial range of the data used in the fit. A second limited scan with the nominal $m_{\mathrm{max}}=54$ and $n_{\mathrm{max}}=6$ is included for validation. The value $L=12.5$~m is used in all fits in the paper.}
\begin{tabular}{c|c|c|c}
$L$ [m] & $n_{\mathrm{max}}$ & $\chi^2_{\mathrm{reduced}}$ & Runtime [s] \\
\hline
$9.0$ & $0$ & $1.25 \times 10^7$ & $6.5$ \\
$9.6$ & $0$ & $6.15 \times 10^6$ & $8.2$ \\
$10.0$ & $0$ & $1.16 \times 10^6$ & $5.0$ \\
$10.5$ & $0$ & $2.12 \times 10^4$ & $4.1$ \\
$11.0$ & $0$ & $451.20$ & $4.7$ \\
$11.5$ & $0$ & $394.54$ & $3.7$ \\
$12.0$ & $0$ & $394.31$ & $5.4$ \\
\cellcolor{green!30} $12.5$ & \cellcolor{green!30} $0$ & \cellcolor{green!30} $394.30$ & \cellcolor{green!30} $4.1$ \\
$13.0$ & $0$ & $394.32$ & $14.5$ \\
$13.5$ & $0$ & $394.34$ & $14.0$ \\
$14.0$ & $0$ & $394.35$ & $13.5$ \\
$14.5$ & $0$ & $394.37$ & $18.4$ \\
$15.0$ & $0$ & $394.42$ & $18.2$ \\
\hline
$10.0$ & $6$ & $1.19 \times 10^6$ & $364$ \\
\cellcolor{green!30} $12.5$ & \cellcolor{green!30} $6$ & \cellcolor{green!30} $2.15$ & \cellcolor{green!30} $1.22 \times 10^3$ \\
$15.0$ & $6$ & $2.27$ & $2.22 \times 10^3$ \\
\end{tabular}
\label{tab:LSQ_L_opt}
\end{table}

The behavior of the model as $L$ is adjusted can be understood from the boundary condition matching the magnetic potential at $z=\pm L/2$ (see paper Equations~\ref{eq:axial_BC_LSQ}~and~\ref{eq:axial_BC_LSQ_full}). If $L$ is too small, the boundary condition is imposed too close to the data, making the model too rigid near the axial ends of the fitted region and preventing it from capturing the field structure, including asymmetric components. Thus, choosing $L$ equal to the axial extent of the data is not optimal.
On the other hand, if $L$ becomes too large, the finite basis spans an unnecessarily long axial interval. For fixed $m_{\mathrm{max}}$, this reduces the maximum axial spatial frequency represented in the fit region, making the model less efficient.
Large $L$ also significantly increases the runtime.

\section{PINN Hyperparameter Optimization}
\label{sec:PINN_opt}

We complete a number of hyperparameter optimizations for the PINN by training with different hyperparameter configurations using the nominal least-squares fit result.

First, we perform a grid search optimization to determine the optimal network structure. In each case, we use the DELTAsnake activation at every node of every hidden layer and set $f=0.5$, $a=5$, and $G=1$ which were determined in an initial hand tuning. The results of the scan are described in Table~\ref{tab:hyperparam_opt_structure}. While the results are somewhat insensitive to the exact structure near the optimum, we choose to use the deepest and densest network that we can fit in GPU memory. This choice, $N_{\mathrm{layers}} = 8$ and $N_{\mathrm{nodes}} = 64$, not only yields the absolute best result in the nominal case, but also ensures the same network structure can be used throughout the robustness tests.

\begin{table}[htbp!]
\center
\caption{A hyperparameter scan over the number of hidden layers and number of nodese per hidden layer
to determine the optimal scalar PINN network structure for the nominal Mu2e DS dataset.
The hyperparameter $\lambda=0.1$ is set in each test. In each case the DELTAsnake activation function with $f=0.5$, $a=5$, and $D=1$ is
used at each node. The optimal structure is marked by a shaded green box. Any test where the GPU memory was exceeded is marked by a 
shaded red box and is considered unsuccessful. In these cases
an additional test with half the number of collocation points was run to try to mitigate the problem, unsuccessfully.}
\begin{tabular}{l|c|c|c|c}
\backslashbox{Nodes}{Layers} & $2$ & $4$ & $8$ & $10$ \\
\hline
$16$ &  \makecell[r]{ $L_{PINN} = 0.1885$ \\ $L_B = 0.1885$ \\ $L_{\nabla \cdot B} = 0.0001$ \\ $L_{val}$ = 0.1924\\ Runtime$ = 2$ m} &  \makecell[r]{ $L_{PINN} = 0.0977$ \\ $L_B = 0.0946$ \\ $L_{\nabla \cdot B} = 0.0304$ \\ $L_{val}$ = 0.0941\\ Runtime$ = 4$ m} &  \makecell[r]{ $L_{PINN} = 0.0936$ \\ $L_B = 0.0925$ \\ $L_{\nabla \cdot B} = 0.0110$ \\ $L_{val}$ = 0.0924\\ Runtime$ = 8$ m} &  \makecell[r]{ $L_{PINN} = 0.0987$ \\ $L_B = 0.0970$ \\ $L_{\nabla \cdot B} = 0.0176$ \\ $L_{val}$ = 0.0965\\ Runtime$ = 14$ m} \\
\hline
$32$ &  \makecell[r]{ $L_{PINN} = 0.1120$ \\ $L_B = 0.1073$ \\ $L_{\nabla \cdot B} = 0.0467$ \\ $L_{val}$ = 0.1071\\ Runtime$ = 3$ m} &  \makecell[r]{ $L_{PINN} = 0.0911$ \\ $L_B = 0.0907$ \\ $L_{\nabla \cdot B} = 0.0037$ \\ $L_{val}$ = 0.0913\\ Runtime$ = 10$ m} &  \makecell[r]{ $L_{PINN} = 0.0903$ \\ $L_B = 0.0901$ \\ $L_{\nabla \cdot B} = 0.0022$ \\ $L_{val}$ = 0.0910\\ Runtime$ = 20$ m} &  \makecell[r]{ $L_{PINN} = 0.0910$ \\ $L_B = 0.0908$ \\ $L_{\nabla \cdot B} = 0.0019$ \\ $L_{val}$ = 0.0915\\ Runtime$ = 26$ m} \\
\hline
$64$ &  \makecell[r]{ $L_{PINN} = 0.0998$ \\ $L_B = 0.0983$ \\ $L_{\nabla \cdot B} = 0.0153$ \\ $L_{val}$ = 0.0986\\ Runtime$ = 9$ m} &  \makecell[r]{ $L_{PINN} = 0.0900$ \\ $L_B = 0.0898$ \\ $L_{\nabla \cdot B} = 0.0020$ \\ $L_{val}$ = 0.0911\\ Runtime$ = 19$ m} & \multicolumn{1}{>{\columncolor{green!30}}c|}{ \makecell[r]{ $L_{PINN} = 0.0892$ \\ $L_B = 0.0891$ \\ $L_{\nabla \cdot B} = 0.0015$ \\ $L_{val}$ = 0.0919\\ Runtime$ = 25$ m}} &  \cellcolor{red!30} \\
\hline
$96$ &  \makecell[r]{ $L_{PINN} = 0.1127$ \\ $L_B = 0.1110$ \\ $L_{\nabla \cdot B} = 0.0170$ \\ $L_{val}$ = 0.1103\\ Runtime$ = 13$ m} &  \makecell[r]{ $L_{PINN} = 0.0899$ \\ $L_B = 0.0897$ \\ $L_{\nabla \cdot B} = 0.0021$ \\ $L_{val}$ = 0.0911\\ Runtime$ = 19$ m} &  \cellcolor{red!30} &  \cellcolor{red!30} 
\end{tabular}
\label{tab:hyperparam_opt_structure}
\end{table}

In an effort to optimize the activation function hyperparameters and additionally to compare DELTAsnake to alternative activation functions, we perform another set of trainings. In each case we use the optimal network structure, $N_{\mathrm{layers}} = 8$ and $N_{\mathrm{nodes}} = 64$, and modify the activation function used in the hidden layers. For DELTAsnake, we scan over several values of $a$ while fixing $f=0.5$ and $D=1$. For a direct comparison with the snake activation we provide one training using snake with $a=5$; the DELTAsnake with $a=5$ yields a better MSE error and significantly better divergence loss (approximately a factor of $10$). We also complete one training with the $\tanh$ activation, which shows similar performance to the snake training. Table~\ref{tab:hyperparam_opt_activation} summarizes the impact of the choice of activation. We find the best results in a shallow minimum of several $a$ values and select DELTAsnake with $a=5$ for producing the results in the paper. It is worth noting that if $a$ is too large, the training diverges rapidly, typically within a few hundred epochs. It is possible this divergence could be fixed by hand-tuning the activation hyperparameters and training hyperparameters. However, to avoid biasing the values in Table~\ref{tab:hyperparam_opt_activation} by excessively hand-tuning additional hyperparameters and given we achieve high-quality results with smaller $a$, we do not apply such a fix in this demonstration.

\begin{table}[htbp!]
\center
\caption{A hyperparameter scan over the activation used at each node of the hidden layers to determine the optimal scalar 
PINN network structure for the nominal Mu2e DS dataset.
The hyperparameter $\lambda=0.1$ is set in each test, and the number of hidden layers and nodes per hidden layer are set to their
optimal values of $8$ and $64$, respectively. We train using the DELTAsnake activation ($f=0.5$, $G=1$) for various values of $a$. We
run the nominal case of $a=5$ a second time using the standard snake activation ($f=0$, $G=1/a$). 
We use a more convential activation, $\tanh$, for a final comparison. For the results presented in the paper
we use the DELTAsnake activation with $a=5$, which is close to the optimal value in the shallow minimum in the region $a \in [0.5, 10.0]$.
This selection is denoted by the row shaded in green.
Any activation in which the training diverges is shaded red; this only occurs when $a>10$. Considering only the shallow minimum of the
DELTAsnake trials, the DELTAsnake activation outperforms both the snake activation and $\tanh$ activation in the MSE loss in addition to
a $\times 10$ smaller divergence loss in the DELTAsnake trials.}
\begin{tabular}{l|c|c|c||c|c|c|c|c}
Activation & $a$ & $f$ & $G$ & $L_{PINN}$ & $L_B$ & $L_{\nabla \cdot B}$ & $L_{val}$ & Runtime (m) \\\hline
 DELTAsnake &  $0.1$ &  $0.5$ &  $1.0$ & $0.0946$ & $0.0940$ & $0.0059$ & $0.0943$ & $39$\\
\hline
 DELTAsnake &  $0.5$ &  $0.5$ &  $1.0$ & $0.0907$ & $0.0904$ & $0.0027$ & $0.0911$ & $39$\\
\hline
 DELTAsnake &  $1.0$ &  $0.5$ &  $1.0$ & $0.0899$ & $0.0897$ & $0.0021$ & $0.0910$ & $24$\\
\hline
 DELTAsnake &  $2.0$ &  $0.5$ &  $1.0$ & $0.0898$ & $0.0896$ & $0.0018$ & $0.0912$ & $25$\\
\hline
\cellcolor{green!30} DELTAsnake & \cellcolor{green!30} $5.0$ & \cellcolor{green!30} $0.5$ & \cellcolor{green!30} $1.0$ & \cellcolor{green!30}\cellcolor{green!30}$0.0886$ &\cellcolor{green!30} $0.0883$ &\cellcolor{green!30} $0.0023$ &\cellcolor{green!30} $0.0930$ &\cellcolor{green!30} $25$\\
\hline
 DELTAsnake &  $10.0$ &  $0.5$ &  $1.0$ & $0.0905$ & $0.0903$ & $0.0019$ & $0.0912$ & $39$\\
\hline
 \cellcolor{red!30} DELTAsnake &  \cellcolor{red!30} $20.0$ &  \cellcolor{red!30} $0.5$ &  \cellcolor{red!30} $1.0$ &  \cellcolor{red!30} \cellcolor{red!30}$\ $ & \cellcolor{red!30} $\ $ & \cellcolor{red!30} $\ $ & \cellcolor{red!30} $\ $ & \cellcolor{red!30} $\ $\\
\hline
 \cellcolor{red!30} DELTAsnake &  \cellcolor{red!30} $50.0$ &  \cellcolor{red!30} $0.5$ &  \cellcolor{red!30} $1.0$ &  \cellcolor{red!30} \cellcolor{red!30}$\ $ & \cellcolor{red!30} $\ $ & \cellcolor{red!30} $\ $ & \cellcolor{red!30} $\ $ & \cellcolor{red!30} $\ $\\
\hline
 snake &  $5.0$ &  $0.0$ &  $0.2$ & $0.0921$ & $0.0908$ & $0.0125$ & $0.0916$ & $39$\\
\hline
 $\tanh$ &  -- &  -- &  -- & $0.0944$ & $0.0931$ & $0.0139$ & $0.0927$ & $16$\\
\hline
\end{tabular}
\label{tab:hyperparam_opt_activation}
\end{table}

\section{Generalized Degrees of Freedom}
\label{sec:GDF}
To assess the effective number of parameters, or generalized degrees of freedom (GDF), of the PINN, we implement the perturbative Monte Carlo estimation algorithm proposed by Ye~\cite{YeGDF}. We describe the algorithm implementation here but refer the readers to Ye's work for details on the mathematical underpinnings and further details of the method.

We start with the measured field values and nominal least-squares fit predictions from the measurement sample $\vec{B}_i$ and $\vec{B}_{\mathrm{pred},i}$, respectively. In this notation $i$ refers to the field point index in the sample. Additionally, we've dropped all ``meas'' indications, since the algorithm is only applied to the measurement sample.
In a given trial, we perturb the PINN target values (i.e. the residual $\vec{B}_{\mathrm{res},i}= \vec{B}_{i} - \vec{B}_{\mathrm{pred},i}$) by some randomly sampled vector $\vec{\delta}_{i}$.
For each field field point, the perturbation vector is sampled as
\begin{equation}
\vec{\delta}_i \sim \mathcal{N}_3(\vec{\mu}, \Sigma)
\end{equation}
where $\vec{\mu} = (0, 0, 0)$ and $\Sigma = (0.5\sigma)^2 \times \mathbb{I}_3$ and $\sigma=0.3$~Gauss is the known measurement uncertainty. This choice is in accordance with Ye's recommendation.

For $T=20$ perturbation trials, indexed by $t$, we:
\begin{itemize}
\item Generate $\vec{\delta}_{t, i}$ for all field points $i$.
\item Perturb the targets for PINN: $\vec{B}_{\mathrm{res},t,i} = \vec{B}_{i} - \vec{B}_{\mathrm{pred},i} + \vec{\delta}_{t,i}$
\item Train the PINN on $\vec{B}_{\mathrm{res},t,i}$ using the standard training procedure. The resulting predictions from the trained PINN are denoted $\vec{B}_{\mathrm{NN},t,i}$.
\end{itemize}
The perturbed PINN predictions from the $T$ trials are compared to the nominal PINN predictions ($\vec{B}_{\mathrm{NN},i}$) and used to estimate the pointwise sensitivity ($\hat{h}_{i}$) by estimating the slope of the PINN sensitivity for each field component (index $j$):
\begin{itemize}
\item Evaluate the deviation in the estimated PINN predictions for each $t$: $\Delta \vec{B}_{\mathrm{NN},t,i} = \vec{B}_{\mathrm{NN},t,i} - \vec{B}_{\mathrm{NN},i}$.
\item Estimate the slope of $\Delta \vec{B}_{\mathrm{NN},i}$ vs. $\vec{\delta}_{i}$ for each field component $j$ across the $T$ trials.
The estimated slope, $\hat{h}_{i,j}$, is estimated via least-squares regression of the linear equation
\begin{equation}
\Delta B_{\mathrm{NN},t,i,j} = h_{i,j} \delta_{t,i,j} + \alpha_{i,j}
\end{equation}
\item Estimate the pointwise sensitivity for field point:
\begin{equation}
\hat{h}_i = \sum^3_{j=1} \hat{h}_{i,j}\ .
\end{equation}
\end{itemize}
The total estimate of GDF is the sum of the pointwise sensitivities:
\begin{equation}
\widehat{\mathrm{GDF}} = \sum_{i}^{N_{\mathrm{meas}}} \hat{h}_i\ .
\end{equation}

To assess a statistical uncertainty for the estimate of the GDF, we apply the bootstrap method~\cite{EfronBootstrap} with $M=5\times10^4$ bootstrap replicas. In each replica, we sample $T$ complete perturbation trials with replacement and recompute the GDF using the same procedure.
The standard deviation of the bootstrap distribution is taken as the uncertainty on the GDF estimate.

We estimate the GDF is:
\begin{equation}\label{eq:GDF_result_Ye}
\widehat{\mathrm{GDF}} = \numprint{1156} \pm 57
\end{equation}
which we alternately denote as $N_{\mathrm{PINN}}$ in the paper.

To validate this estimate, we estimate the GDF in an alternative way using the same set of $T$ trials. Instead of estimating the pointwise sensitivity ($\hat{h}_{i}$) and summing them, we estimate the global sensitivity in a given trial $t$ over all field points ($i$) and components ($j$) by regressing the slope $\widehat{\mathrm{GDF}}_t / N_{\mathrm{meas}}$ from the linear equation
\begin{equation}
\Delta B_{\mathrm{NN},t,i,j} = \frac{\mathrm{GDF}_t}{N_{\mathrm{meas}}} \delta_{t,i,j} + \alpha_{t}\ .
\end{equation}
The mean and standard error of the mean of the $\widehat{\mathrm{GDF}}_t$ estimators are taken as estimates of the GDF and the uncertainty on the estimate, respectively. With this alternative method we find:
\begin{equation}
\widehat{\mathrm{GDF}} = \numprint{1144} \pm 67
\end{equation}
which is in good agreement with the estimate using Ye's method. We quote the estimate given in Equation~\ref{eq:GDF_result_Ye} in the paper.

\end{document}